\documentclass[aps,prb,twocolumn,superscriptaddress,showpacs,floatfix]{revtex4-2}
\usepackage{bm}
\usepackage{amssymb}   % for math
\usepackage{amsmath}
\usepackage{wasysym}
\usepackage[colorlinks=true,citecolor=blue,linkcolor=magenta, breaklinks=true]{hyperref}
\usepackage{graphicx}
\usepackage{braket}
\usepackage{natbib}
\usepackage [latin1]{inputenc}
\usepackage{blkarray, bigstrut}
\usepackage[dvipsnames]{xcolor}
\usepackage{comment}
\usepackage{physics}

\newcommand{\be}{\begin{equation}}
\newcommand{\ee}{\end{equation}}
\newcommand{\bk}{{{\bf{k}}}}

\newcommand{\beal}{\begin{align}}
\newcommand{\eeal}{\end{align}}

\newcommand{\upa}{\uparrow}
\newcommand{\dna}{\downarrow}

\newcommand{\pdg}{{\phantom\dagger}}

\def\a{\alpha}
\def\b{\beta}

\def\d{\delta}

\def\g{\gamma}

\def\i{i}

\def\m{\mu}
\def\n{\nu}

\def\s{\sigma}

\def\w{\omega}

\def\mh{{\mathcal{H}}}

\def\bcao{BaCo$_2$(AsO$_4$)$_2$}
\def\ncto{Na$_2$Co$_2$TeO$_6$}
\def\bcpo{BaCo$_2$(PO$_4$)$_2$}
\def\ncso{Na$_3$Co$_2$SbO$_6$}
\def\fm{FM}
\def\zz{ZZ}
\def\sl{{$\widetilde{\mathrm{SL}}$}}
\def\chixx{\chi{''}_{\!\!\!xx}}

\begin{document}

\title{Proximate Dirac spin liquid in honeycomb lattice $J_1$-$J_3$ XXZ model: Numerical study and
application
to cobaltates}

\begin{abstract}
Recent theoretical and experimental work suggest that the honeycomb cobaltates, initially proposed as candidate 
Kitaev quantum magnets, are in fact described by a pseudospin-$1/2$ easy-plane 
spin Hamiltonian with nearest neighbor ferromagnetic (FM) exchange $J_1$ being frustrated by antiferromagnetic 
third-neighbor exchange $J_3$ and weaker compass anisotropies.
Using exact diagonalization and density-matrix renormalization 
group (DMRG) calculations, we show that this model exhibits FM order at small $J_3/J_1$
and zig-zag (ZZ) order at large $J_3/J_1$, separated by an intermediate phase, which  we label as {\sl}. This
{\sl} phase is shown to exhibit spin-liquid-like correlations in DMRG, although we cannot preclude weak 
broken symmetries, e.g. weak Ising type N\'eel order, given the limits on our 
explored system sizes.
%We find that the DMRG correlations appear very different in the two different cylinder geometries we have studied.
Using a modified parton mean field theory and variational Monte Carlo on
Gutzwiller projected wavefunctions, we show that the optimal FM and ZZ orders as well as the intermediate {\sl} state 
are proximate to a `parent' Dirac spin liquid (SL). This
Dirac SL is shown to capture the broad continuum in 
the temperature and magnetic field dependent
terahertz spectroscopy of {\bcao}, and the reported low temperature metallic thermal conductivity in {\ncto} and {\bcao} upon 
incorporating disorder induced broadening.
%suggesting that it might be a useful parent SL.
\end{abstract}

\author{Anjishnu Bose}
\thanks{These authors contributed equally to this work.}
\affiliation{Department of Physics, University of Toronto, 60 St. George Street, Toronto, ON, M5S 1A7 Canada}
\author{Manodip Routh}
\thanks{These authors contributed equally to this work.}
\affiliation{Department of Condensed Matter Physics and Materials Science, S.N. Bose National Centre for Basic Sciences, Kolkata 700098, India.}
\author{Sreekar Voleti}
\thanks{These authors contributed equally to this work.}
\affiliation{Department of Physics, University of Toronto, 60 St. George Street, Toronto, ON, M5S 1A7 Canada}
\author{Sudip Kumar Saha}
\affiliation{Department of Condensed Matter Physics and Materials Science, S.N. Bose National Centre for Basic Sciences, Kolkata 700098, India.}
\author{Manoranjan Kumar}
%\email{manoranjan.kumar@bose.res.in}
\affiliation{Department of Condensed Matter Physics and Materials Science, S.N. Bose National Centre for Basic Sciences, Kolkata 700098, India.}
\author{Tanusri Saha-Dasgupta}
%\email{tanusri@bose.res.in}
\affiliation{Department of Condensed Matter Physics and Materials Science, S.N. Bose National Centre for Basic Sciences, Kolkata 700098, India.}
\author{Arun Paramekanti}
%\email{arun.paramekanti@utoronto.ca}
\affiliation{Department of Physics, University of Toronto, 60 St. George Street, Toronto, ON, M5S 1A7 Canada}
\affiliation{Department of Condensed Matter Physics and Materials Science, S.N. Bose National Centre for Basic Sciences, Kolkata 700098, India.}
\affiliation{International Centre for Theoretical Sciences, Bengaluru 560089, India}
\date{\today}
% \pacs{75.25.aj, 75.40.Gb, 75.70.Tj}
\maketitle

Quantum spin liquids (SLs) are exotic magnetic liquids featuring fluctuating singlet correlations and fractionalized spin excitations,
%While initially proposed in the context of the one-dimensional (1D) Heisenberg model,
which may be fruitfully described using the language of gauge-matter theories \cite{QSLgauge_XGWenBook,QSLgauge_Lee2014,QSLreview_Savary_2017,QSLreview_Broholm2020,QSLreview_Knolle2019}. While 
%quantum SLs were 
initially proposed as interesting magnetic ground states formed by a quantum superposition of a large number of 
classically frustrated spin or singlet dimer configurations \cite{rvb_anderson1973,rvb_moessner_prl2001},
interest in SLs exploded with the idea that doping such insulators might lead to high temperature superconductivity
\cite{hightc_anderson_science1987,rvb_rokhsarkivelson_prl1988}. 
For gapped SLs, their robust topological order \cite{QSLgauge_XGWenBook} and the absence of local order parameters 
in SLs has also rendered them as platforms of potential interest 
for topologically protected quantum memories \cite{KitaevModel_Kitaev2006}.
%These enigmatic phases thus serve as a bridge connecting condensed matter physics, high energy physics,
%and quantum information. 

The search for quantum SLs has unearthed several promising candidates in materials
with strong geometric frustration such as kagom\'e \cite{kagome_neutron_Lee2007,kagome_review_mendels2016,kagome_diracvmc_hermele2007,kagome_diracvmc_iqbal2013,kagome_diracdmrg_pollmann2017},
hyperkagom\'e \cite{hyperkagome_Takagi_PRL2007,hyperkagome_Lawler_PRL2008,hyperkagome_YiZhou_PRL2008}, and pyrochlore \cite{Pyrochlore_Ross_PRX2011,pyrochlore_review_gingras2014,Pyrochlore_Hermele_PRL2014,Pyrochlore_Fennell_NatPhys2018} magnets, as well as in quantum simulators using 
Rydberg atoms \cite{Rydberg_Lukin_Science2021}. A significant effort
has also been devoted to exploring the physics of the exactly solvable Kitaev quantum SL with Majorana excitations
\cite{KitaevModel_Kitaev2006,Kitaev_review_Hermanns2018,Kitaev_review_Motome2020} in pseudospin-$1/2$ honeycomb magnets
such as iridates \cite{KitaevModel_Jackeli2009,iridates_kimchi2014,KitaevReview_Takagi2019,Iridates_review_Trebst2022},
$\alpha$-RuCl$_3$ \cite{RuCl3_SOC_Kim2014,RuCl3_Raman_Burch2015,RuCl3_neutron_Nagler2017}, 
and their higher spin generalizations \cite{KitaevSpinS_Baskaran2008,KitaevSpinS_Kee2019}. While many of
these quantum magnets have ordered ground states, their dynamical response is nevertheless expected to
provide a window into SL physics at intermediate energy scales.

Recently, it was proposed that the
honeycomb cobaltates, with $d^7$ Co$^{2+}$ ions in an octahedral crystal field environment, 
provide an alternative venue to realize the Kitaev model \cite{CoKitaev_Liu2018,CoKitaev_Liu2020}.
This has led to a large body of
experimental work on a variety of these materials including
{\bcao} \cite{BCAO_regnault1977,BCAO_Regnault2018,BCAO_Zhong2019,Armitage2022,BCAO_Broholm2023,BCAO_thermalcond_Li2022}, 
{\bcpo} \cite{BCPO_Nair2018}, {\ncto} \cite{ncto_simonet2016,ncto_ncso_stock2020,Kappaxy_NCoTeO_Park2022,LANL_ncto2022,kappa_NCTO_Sun_PRB_2023}, 
and {\ncso} \cite{ncso_mcguire2019,ncto_ncso_stock2020}. While most of these cobaltates display collinear
or spiral magnetic ground states \cite{BCAO_regnault1977,BCAO_Regnault2018,BCPO_Nair2018,BCAO_Zhong2019,CTO_Yuan2020,CTO_Coldea_Ncomm2021,Armitage2022,Anisotropy_NCoSb_Yuan2022,BCAO_Broholm2023},
or proposed triple-Q orders \cite{tripleQ_Janssen2022},
there are tantalizing signatures in {\bcao} of a spin-liquid-like broad continuum
in the terahertz (THz) response \cite{Armitage2022}. Experiments on {\ncto} and {\bcao}
have also reported a residual metallic thermal conductivity $\kappa \propto T$ 
\cite{kappa_NCTO_Sun_PRB_2023,BCAO_thermalcond_Li2022}
which hints at mobile fermionic spinons; however, phonon contributions to $\kappa(T)$ remain to be understood \cite{hong2023phonon}.
% thermal Hall effect whose origin remains unclear \cite{Kappaxy_NCoTeO_Park2022}

The THz and thermal conductivity
results suggest that although the ground state of the cobaltates might exhibit magnetic order, 
a parent gapless quantum SL could be lurking in close proximity to these ordered phases.
While bond-anisotropic interactions are relevant to some of the observations
\cite{Anisotropy_NCoSb_Rachel2022,Anisotropy_NCoSb_Yuan2022,Kappaxy_NCoTeO_Park2022},
{\it ab initio} and exact diagonalization (ED) calculations 
of the magnetic exchange interactions 
\cite{Abinitio_Das2021,Abinitio_Streltsov2022,Abinitio_HSKim2022}, 
as well as detailed fits to 
high resolution inelastic neutron scattering data on {\bcao} \cite{BCAO_Broholm2023}, 
%instead 
suggest that a more appropriate starting point is
an XXZ ferromagnet, with easy-plane anisotropy, and frustration arising from 
third-neighbor antiferromagnetic (AFM) exchange.
Could such XXZ models in the $J_1$-$J_3$ parameter space with subdominant bond-anisotropic exchanges
support quantum SLs distinct from the Kitaev SL? If so, can such SLs explain the dynamical THz spin response \cite{Armitage2022} and the 
unusual remarkable metallic thermal transport observed in a
Mott insulator \cite{BCAO_thermalcond_Li2022,kappa_NCTO_Sun_PRB_2023}? Addressing these questions can open up new lines of 
investigation into these exotic magnetic fluids.

The impact of further-neighbor frustration in honeycomb magnets has been previously explored with competing
second-neighbor exchange in $J_1$-$J_2$ models. For classical Heisenberg models, the ferromagnetic or 
N\'eel order of the nearest-neighbor Heisenberg magnet gets destroyed with increasing $J_2$ \cite{j1j2j3_rastelli1979,j1j2j3_Fouet2001}, 
leading to coplanar spirals which support a Bose surface of spin wave 
excitations \cite{j1j2_ganesh2010,j1j2_ganesh2011}.
In the quantum spin-$1/2$ Heisenberg antiferromagnet, a regime of this fluctuating
spiral order gives way to valence bond crystals \cite{j1j2j3_Fouet2001,j1j2_ganesh2010,j1j2_peps_Boninsegni2012,j1j2_DQCP_ganesh2013,j1j2_DQCP_pujari2013,j1j2_dmrg_sheng2013}. 
In the strong easy-plane limit, this model may realize a gapped (possibly chiral) spin liquid 
\cite{j1j2_rigol2013,j1j2_galitski2014,j1j2_Ting2021}.

Here, we study the spin-$1/2$ $J_1$-$J_3$ model on the honeycomb lattice, with FM $J_1$ 
and frustrating AFM third-neighbor exchange $J_3$,
in the strong easy-plane limit \cite{Abinitio_Das2021,spinwaveXXZ_chernyshev_PRB2022,BCAO_Broholm2023}, 
as relevant to the cobaltates, and also explore the impact of
weaker bond-dependent exchange couplings. Our main result,
obtained using exact diagonalization (ED) and DMRG, is evidence of a distinct intervening phase, between the FM phase at small
$J_3/J_1$ and the zig-zag phase at large $J_3/J_1$. 
%This intervening phase persists with
%weak bond-dependent exchange anisotropy.
Given the spin-liquid-like characteristics of this
intermediate phase, we label it {\sl}. Our numerics, however, do not definitively rule out weak 
symmetry breaking orders in the proposed {\sl} state.
%and weak applied in-plane Zeeman field. 
We present a parton theory and Gutzwiller wavefunction study which suggests that this {\sl}
state descends from a parent gapless Dirac SL with large low energy spinon density of states. While the 
Dirac SL state may be ultimately be unstable to {\sl} via confinement or spinon pairing, it
is nevertheless shown to provide an excellent starting point to describe the 
experimental THz spin response and thermal conductivity in the honeycomb cobaltates.

\section{Model Hamiltonian}

\begin{figure}[t]
    \centering
    \includegraphics[width=0.45\textwidth]{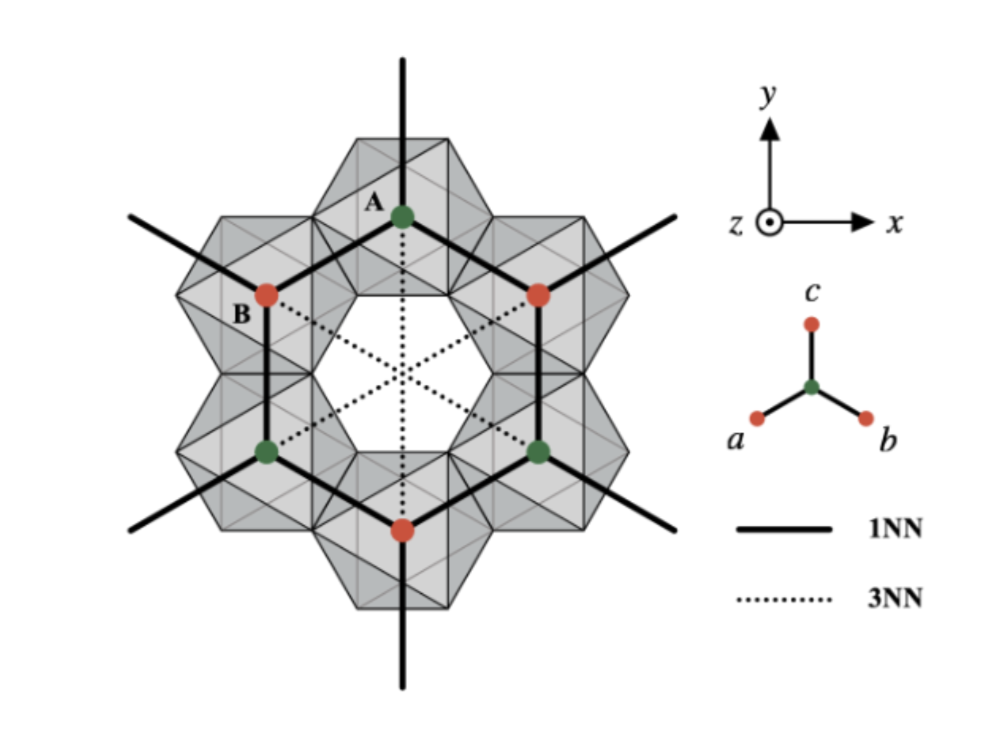}
   \caption{Schematic view of honeycomb lattice showing pseudospin-$1/2$ Co$^{2+}$ ions enclosed in edge-sharing oxygen octahedra. 
    Solid lines connecting A-B sublattices represent first-nearest neighbors (1NN), while 
    dotted lines represent third-nearest neighbors (3NN); bonds are labelled $(a,b,c)$ as shown on the right. We
    also show crystallographic $xyz$ coordinate system which is used for the Hamiltonian.}
\label{fig:lattice}
\end{figure}

The honeycomb cobaltates have the structure depicted in Fig.\ref{fig:lattice}. Their
magnetism is described by an easy-plane XXZ spin model, with weaker bond-dependent anisotropies, which was
derived using {\it ab initio} and ED studies \cite{Abinitio_Das2021,Abinitio_HSKim2022,Abinitio_Streltsov2022}. 
In CoTiO$_3$ \cite{CTO_Yuan2020,CTO_Coldea_Ncomm2021}, a dominant
FM $J_1$, with weaker interlayer AFM exchange, leads to in-plane 
ferromagnetic order staggered from one layer to the next. Small compass anisotropies pin the moment
orientation and can induce a weak gap to the Goldstone mode \cite{Abinitio_Das2021,CTO_Coldea_Ncomm2021}. 
By contrast, in {\bcao}, {\bcpo}, {\ncto},
and {\ncso}, the larger spacing between layers renders the system more two-dimensional (2D), while the pnictogen (P, As, Sb) 
or chalcogen (Te) mediates significant frustrating antiferromagnetic $J_3$ \cite{Abinitio_Das2021}
1as also deduced from fits to the neutron scattering data \cite{BCAO_Broholm2023}.
%the combination of these two factors 
%appears to be responsible for the observed spin-liquid like responses in these cobaltates. 
Motivated by this, we consider the 
2D honeycomb lattice XXZ model, with the Hamiltonian
\begin{eqnarray}
    \label{bond term in H}
        \mh_1 &=&  \frac{1}{2} \sum_ {i,j} \;  J_{ij} \left(S_i^x S_j^x + S_i^y S_j^y + \lambda S_i^z S_j^z\right),
\end{eqnarray}
written in the global $xyz$ basis, where $\hat{z}$ is perpendicular to the honeycomb plane as shown in 
Fig.\ref{fig:lattice}. The exchange couplings $J_{ij}$ are chosen to be ferromagnetic $J_{ij}\!=\! -J_1$
for the first neighbor bond, and frustrating antiferromagnetic $J_{ij}\!=\!+J_3$ for the 
third neighbor bond, with $J_1, J_3 \!>\! 0$. 
The Heisenberg limit of this model, $\lambda=1$, has been partially explored earlier \cite{j1j2j3_rastelli1979,j1j2j3_Fouet2001}
inspired by old experiments on {\bcao} \cite{BCAO_regnault1977}. 
Here, we focus on the easy-plane regime of interest $0 \!<\! \lambda \!\ll\!1$.
We supplement this Hamiltonian with weaker nearest-neighbor compass anisotropy terms
\begin{eqnarray}
\mh_2 &=& \sum_{\langle i,j\rangle} 
\left[ \left( K_1  \cos{\phi_{ij}} + K_2 \sin{\phi_{ij}} \right)   \left( S_i^x S_j^x - S_i^y S_j^y \right) \right. \nonumber  \\ 
        &&- \left. \left( K_1  \sin{\phi_{ij}} - K_2 \cos{\phi_{ij}} \right)  \left(S_i^x S_j^y + S_i^y S_j^x  \right) \right].
\end{eqnarray}
The compass terms have couplings $K_1,K_2$, with the angles $\phi_{ij}\!=\! 0 , 2\pi/3 , 4\pi/3$ corresponding 
respectively to the three nearest-neighbor honeycomb bonds $c,a,b$ centered on the A-sublattice (see Fig.\ref{fig:lattice}).
Motivated by {\it ab initio} studies \cite{Abinitio_Das2021} and fitting to neutron data \cite{BCAO_Broholm2023},
we set $K_1\!=\! -K_2\!=\! K$, with $K\!>\!0$ and work in the regime $K \!\ll \! J_1$ as relevant to 
the cobaltates. The total Hamiltonian we study is thus $\mh = \mh_1 + \mh_2$.

\section{Exact diagonalization study}

\begin{figure}[t]
    \centering
     \includegraphics[width=0.46\textwidth]{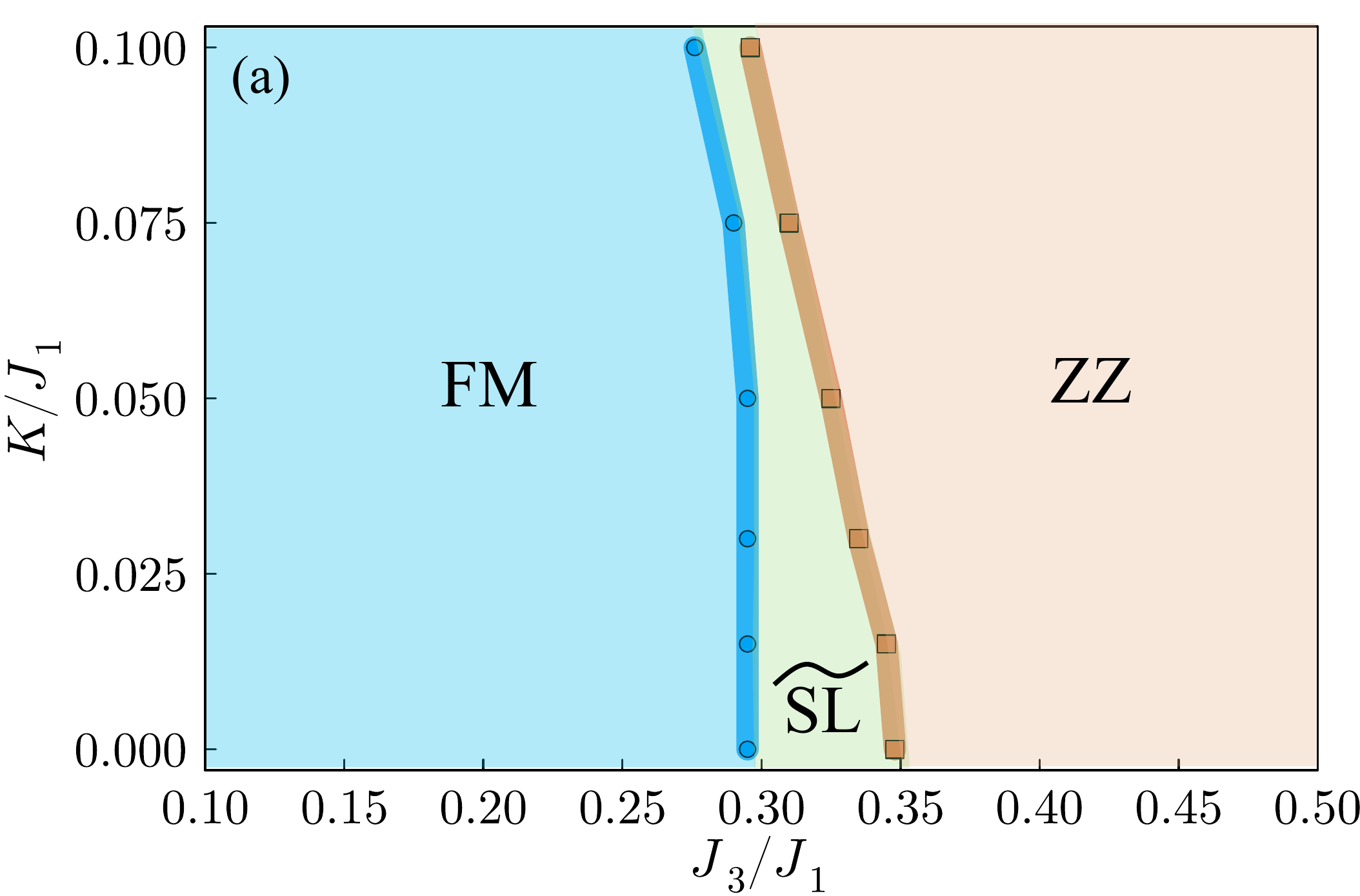}
    \includegraphics[width=0.44\textwidth]{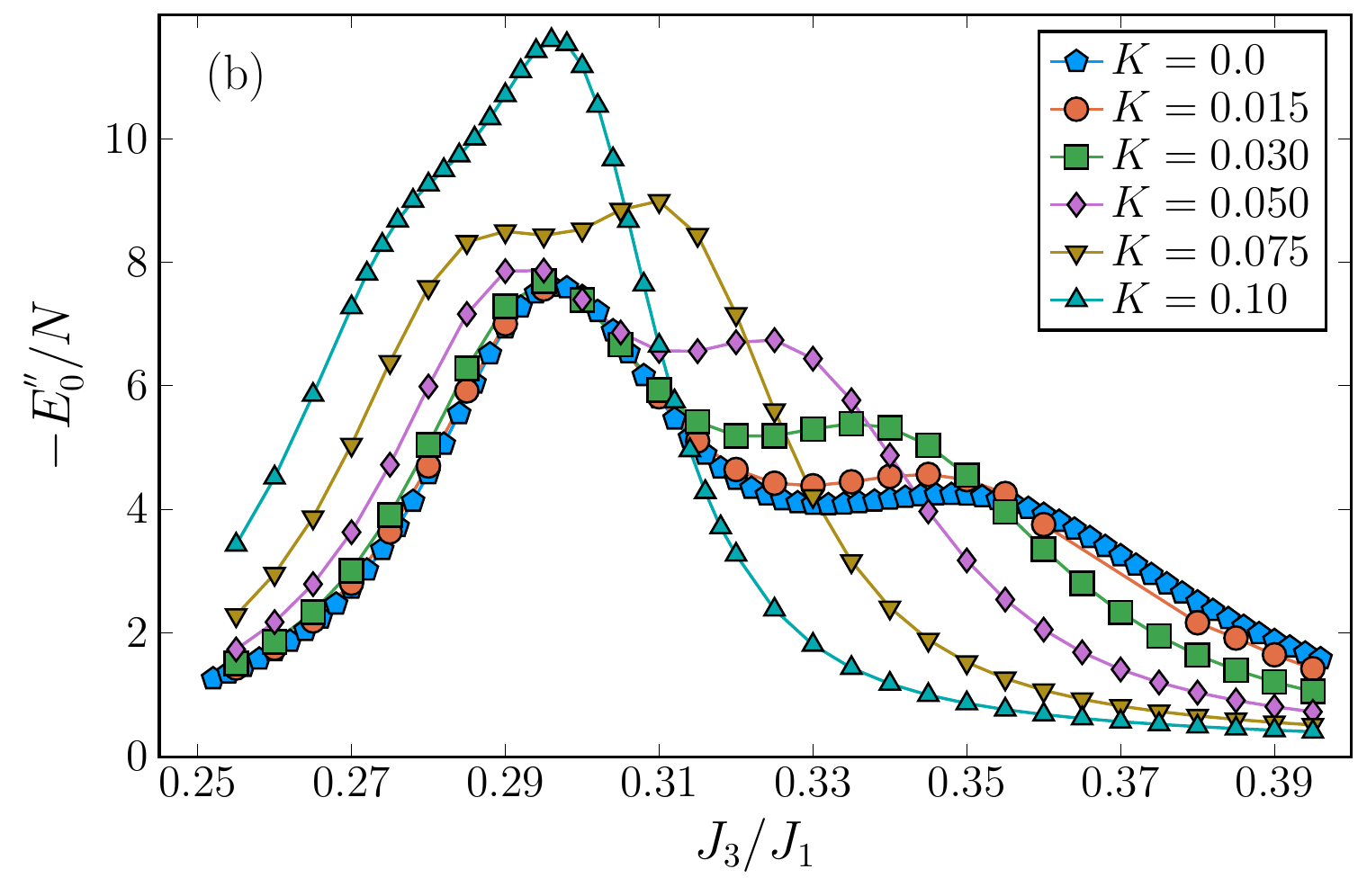}
   \caption{(a) Phase diagram of the model Hamiltonian
    $\mh$ (see text) for fixed $\lambda=0.25$ as a function of third-neighbor exchange 
    $J_3/J_1$ and compass anisotropy $K/J_1$. Based on ED and DMRG spin correlations, we identify 
    three phases: an easy-plane ferromagnet {\fm}, spin-liquid-like phase {\sl},
    and a zig-zag phase {\zz}. The phase boundaries are identified by peaks in ED of
    $(-\partial^2 E/\partial J_3^2)$. (b) Cuts through the phase diagram in showing
    $-E_0''/N = (1/N) (-\partial^2 E/\partial J_3^2)$ computed using ED on $24$-site torus, 
    as a function of $J_3/J_1$ for various compass anisotropies $K/J_1$. The two peaks in $-E_0''/N$ 
   serve to delineate the FM-{\sl} and {\sl}-ZZ phase boundaries.
    }
\label{fig:EDphasediag}
\end{figure}

We have carried out extensive ED calculations  
on this model for system sizes ranging from $N=18$ to $N=30$ spins (using the Davidson algorithm \cite{DAVIDSON197587, davidson1993monster, MURRAY1992382}).
Fig.~\ref{fig:EDphasediag}(a) shows the zero field phase diagram of this model as we vary $J_3/J_1$ and 
$K/J_1$ for fixed $\lambda=0.25$. Based on studying spin
correlations we find (i) an easy-plane {\fm} state for small $J_3/J_1$, (ii) a
zig-zag phase {\zz} for large $J_3/J_1$, and (iii) a small window of an intervening phase
where spin correlations appear to exhibit liquid-like behavior \cite{suppmat}; we characterize this phase 
further below using DMRG calculations on larger system sizes. 
This intermediate state
could arise from quantum disordering the incommensurate spiral order of the classical spin model \cite{j1j2j3_Fouet2001}.
Due to the spin-liquid-like nature of this
intermediate phase, we term it as {\sl}.
The phase boundaries in Fig.~\ref{fig:EDphasediag}(a) are obtained from peaks in
the second derivative of the ground state energy with respect to $J_3$, as shown in
in Fig.\ref{fig:EDphasediag}(b).
%See Supplemental Material \cite{suppmat} for details and additional data at $\lambda=0.0$. 
With increasing compass anisotropy, the FM-{\sl} and {\sl}-ZZ
phase boundaries shift such that the {\zz} phase expands as seen from Fig.~\ref{fig:EDphasediag}(a) 
while the FM and {\sl} phases shrink.
Nevertheless, the intermediate {\sl} persists over a window of $K/J_1$ in this extended phase diagram.

\section{Density matrix renormalization group results}

\begin{figure}[t]
    \centering
    \includegraphics[width=0.48\textwidth]{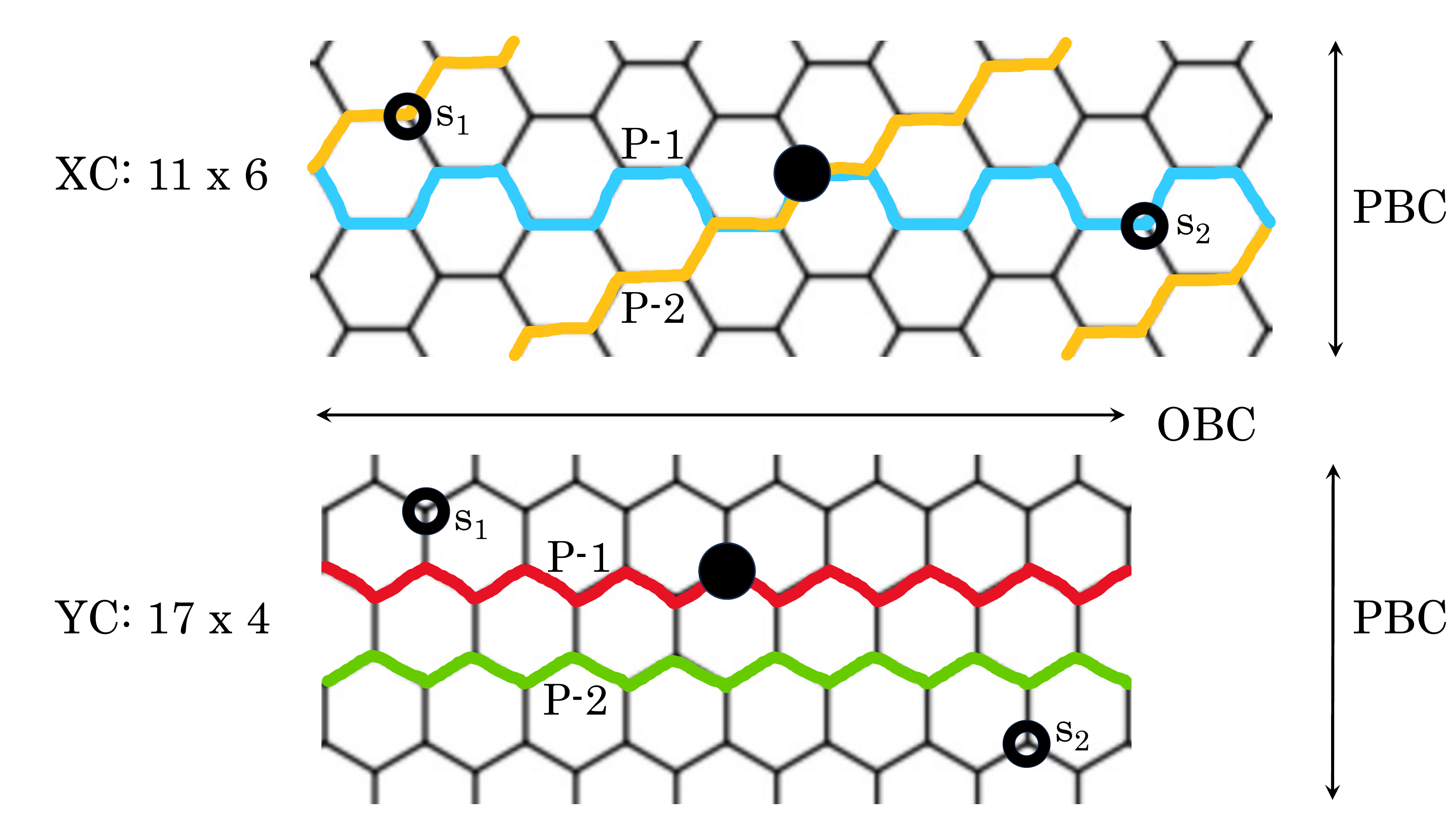}
    \caption{Cylinder clusters XC (top) and YC (bottom) employed in DMRG calculations with indicated maximum system sizes. 
    We use periodic boundary conditions (PBC) along $y$ and open boundary conditions (OBC) along $x$.
    The spin correlations are computed with respect to the reference site (black filled circle) along two indicated paths for
    each cluster (XC path P-$1$, blue; XC path P-$2$, yellow; YC path P-$1$, red; YC path P-$2$, green). 
    For finite size scaling of the order parameters we use the spin correlation of sites $s_1$ and $s_2$ (open circles)
    with respect to reference site.}
    \label{fig:DMRGcluster}
\end{figure}

\begin{figure}[t]
    \centering
    \includegraphics[width=0.35 \textwidth]{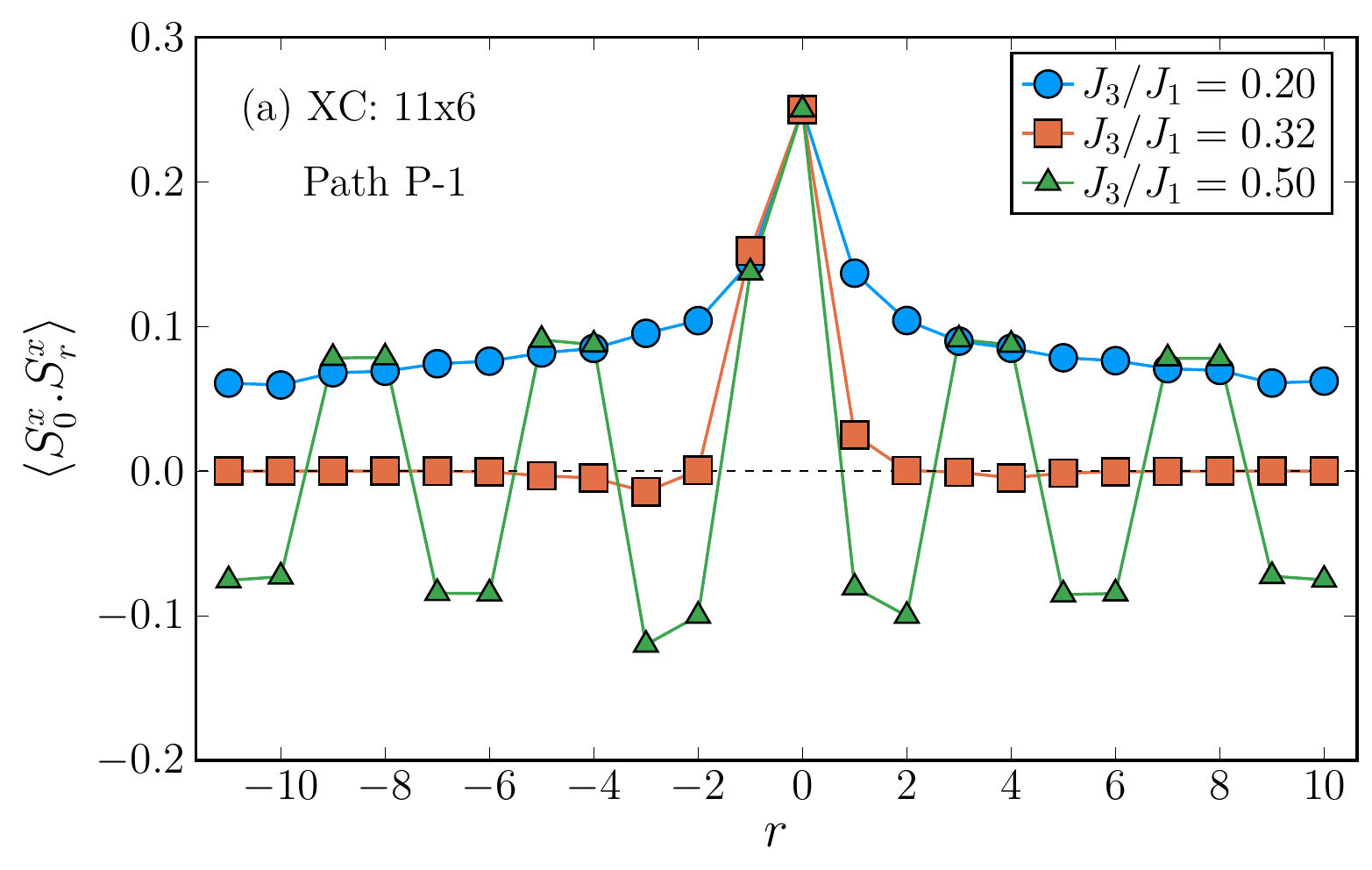}
    \includegraphics[width=0.35 \textwidth]{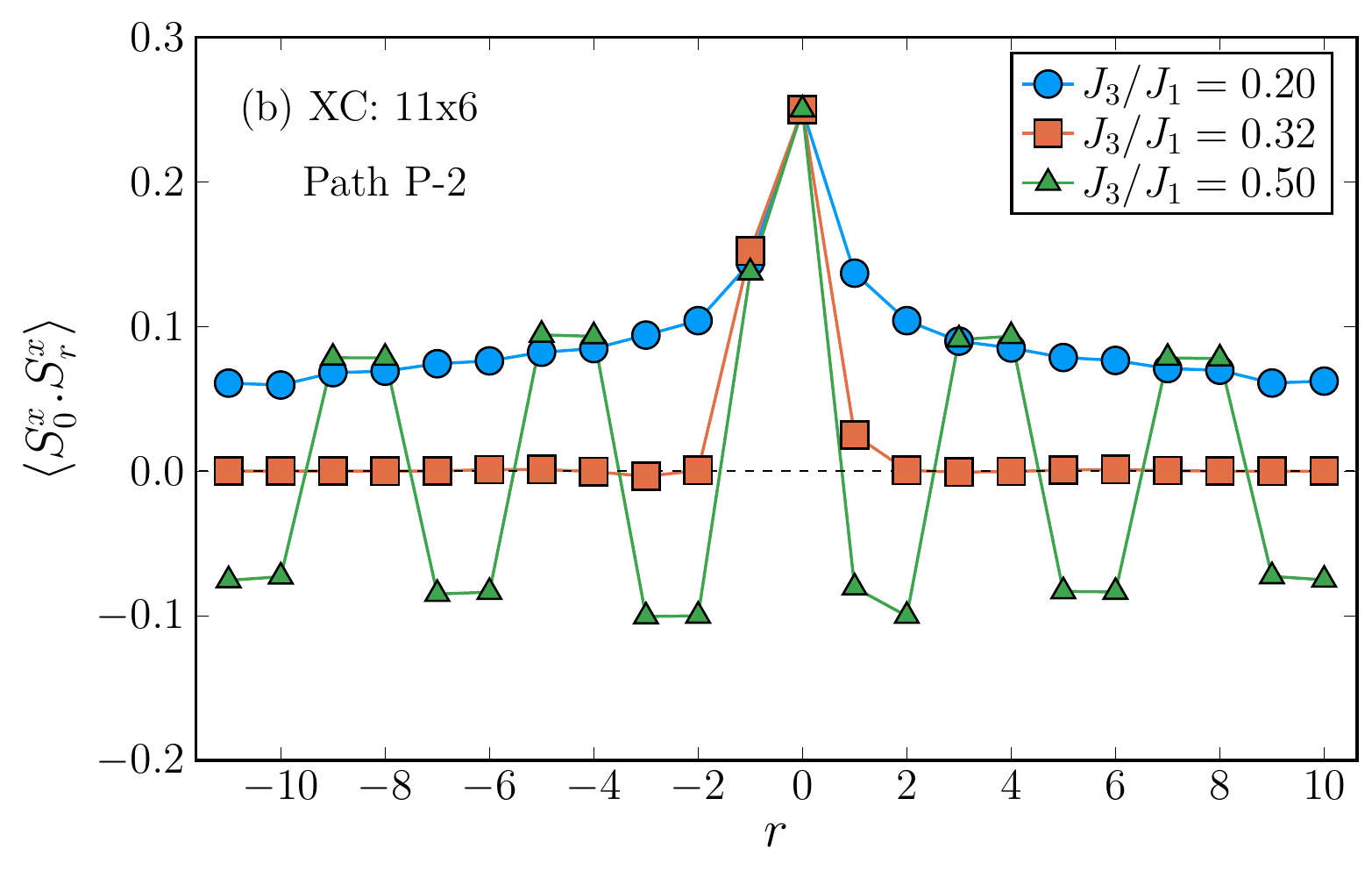}
    \includegraphics[width=0.35 \textwidth]{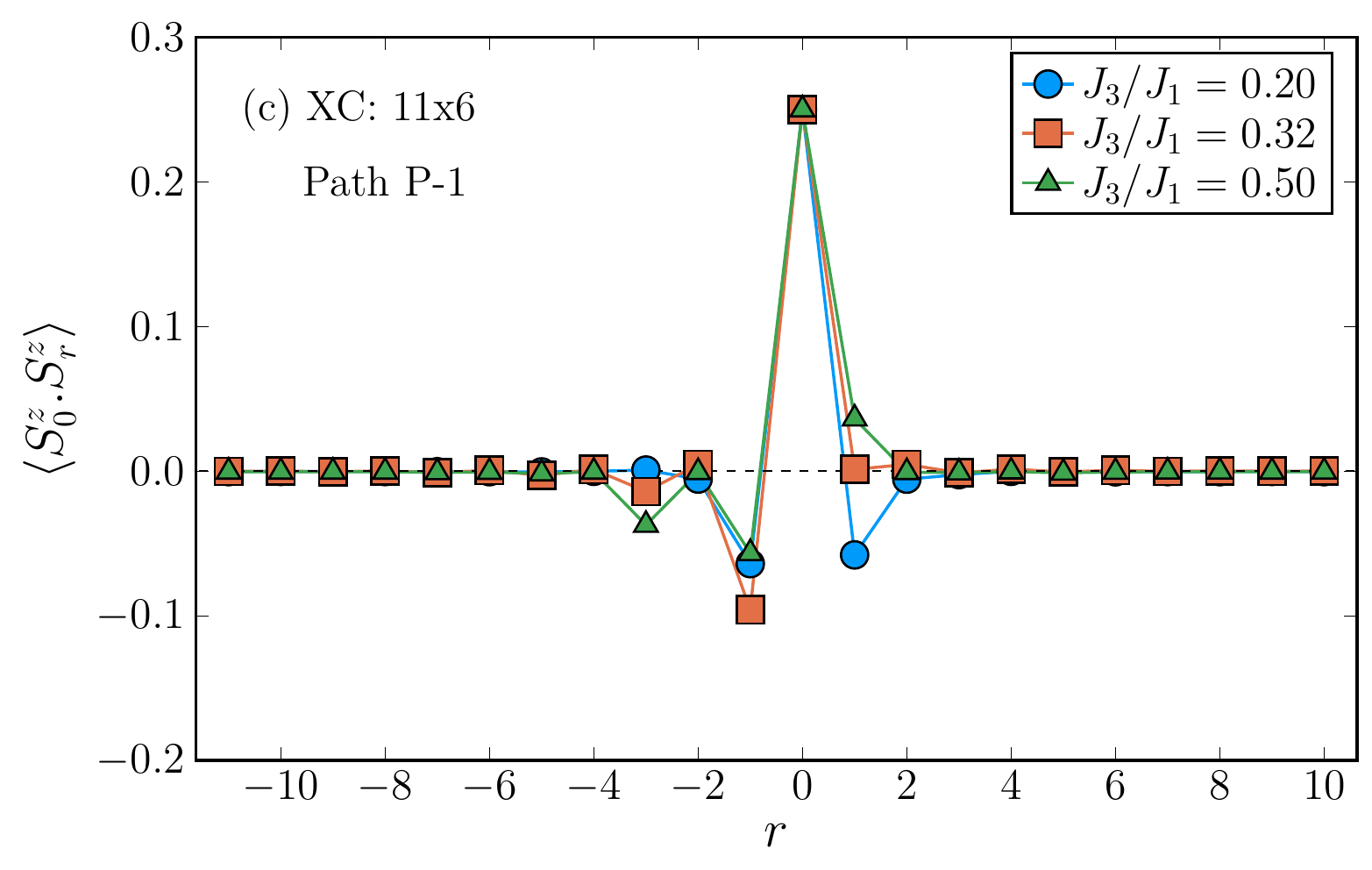}
    \includegraphics[width=0.35 \textwidth]{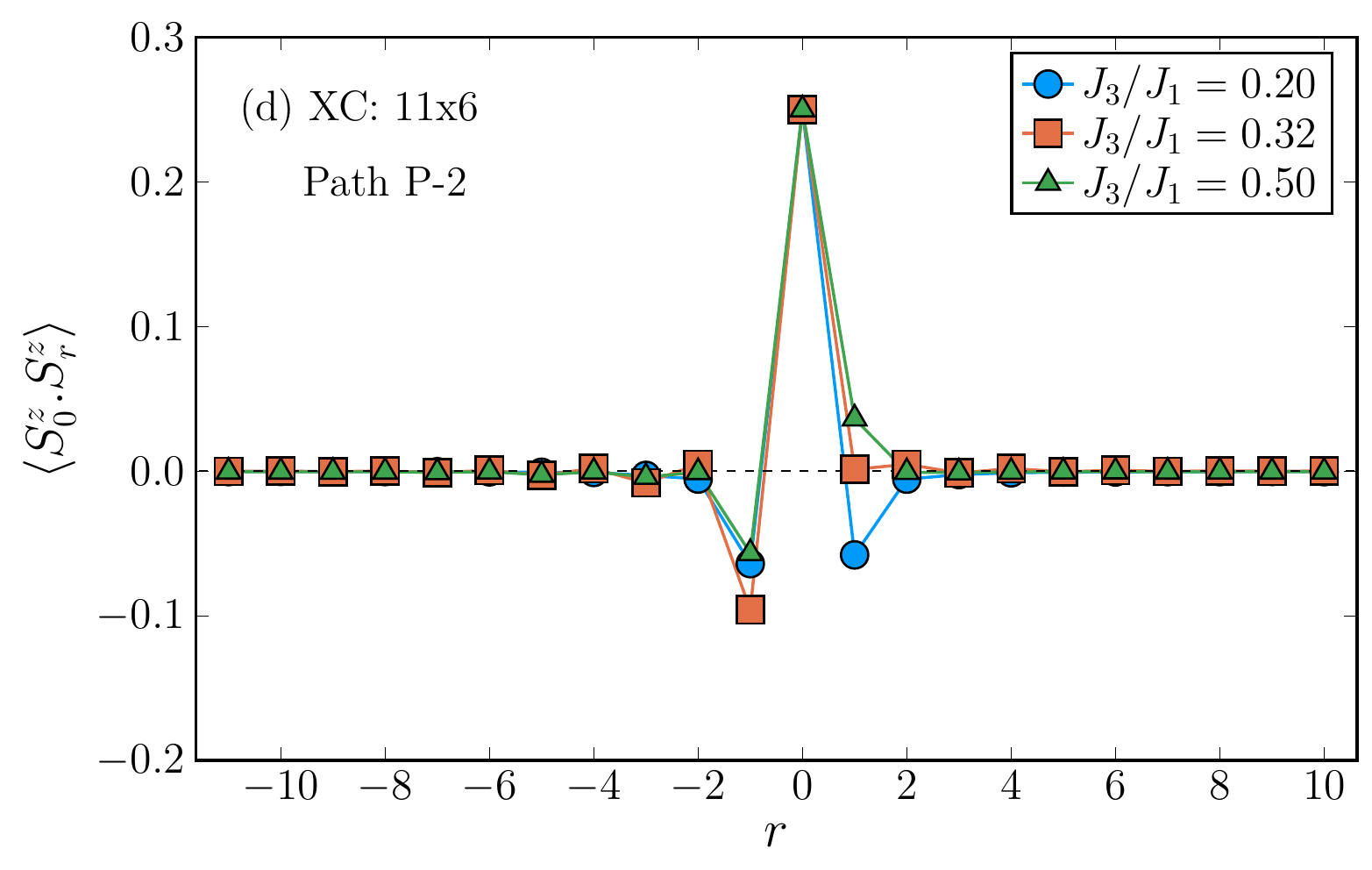}
    \caption{DMRG spin correlations on XC cluster with respect to the reference site shown
    in Fig.~\ref{fig:DMRGcluster} for $(K\!=\!0,\lambda\!=\!0.25)$. (a) $S^x$ correlations along path P-$1$ showing FM behavior at $J_3/J_1 \!=\! 0.2$, ZZ order
    at $J_3/J_1\!=\! 0.5$, and spin-liquid-like short range correlations at $J_3/J_1 \!=\! 0.32$. (b) Similar
    to (a) but for path P-$2$. (c) and (d) $S^z$ correlations showing that they rapidly decay
    to zero for both paths in all phases.}
    \label{fig:DMRG1}
\end{figure}

To further characterize the phases found in ED,
we have used the modified DMRG algorithm \cite{white1992density,white1993density,Mkumar2010,schollwock2005density} 
to calculate the ground state of $\mh$ on cylinders,
studying system sizes upto $66$ sites (XC cylinder) and $68$ sites (YC cylinder); the clusters are shown in 
Fig.~\ref{fig:DMRGcluster} (see Supplemental Material \cite{suppmat} for details about the DMRG and for comparison
with ED results). We have computed
spin correlations, in the three different phases identified in ED, going along different paths as indicated
in Fig.~\ref{fig:DMRGcluster}. These include the labelled zigzag and armchair paths on the honeycomb lattice, with 
sites indexed by steps $n$ and spin correlations computed with respect to the reference point marked by a 
filled circle.
%The presence of the {\zz}, {\fm},  and {\sl} phases are confirmed by examining the spin-spin correlation functions for each component between 
%sites i and j, given by $<S^x_iS^x_j>=<0|S^x_iS^x_j|0>$, $<S^y_iS^y_j>=<0|S^y_iS^y_j|0>$, and $<S^z_iS^z_j>=<0|S^z_iS^z_j|0>$, where $|0>$ is the ground-state wave function.
Since the intermediate {\sl} phase identified in ED is most visible at $K=0$,
we focus on the parameter set $K/J_1\!=\!0$ and $\lambda=0.25$. In this case, the $S^x$ and $S^y$ correlations are identical
but they differ from the $S^z$ correlations.

Fig.~\ref{fig:DMRG1}(a) shows the $S^{x,y}$ correlations on the $66$-site XC cluster along path P-$1$ (blue armchair path in
Fig.~\ref{fig:DMRGcluster}). The 
spin correlations for $J_3/J_1 = 0.2$ and $J_3/J_1 = 0.5$ respectively indicate FM and period-$2$ ZZ order, while they
decay rapidly in the intermediate {\sl} phase. The spin correlations along path P-$2$ (yellow zigzag path 
Fig.~\ref{fig:DMRGcluster}) of the XC-cluster show qualitatively similar results as displayed in Fig.~\ref{fig:DMRG1}(b).
By contrast, the $S^z$ correlations decay rapidly to zero in all three
phases as seen for both path P-$1$ and path P-$2$ as seen from Figs.~\ref{fig:DMRG1}(c) and ~\ref{fig:DMRG1}(d).
We find no evidence for any tendency to incommensurate correlations or other wavevectors in our study.

\begin{figure}[t]
    \centering
    \includegraphics[width=0.33 \textwidth]{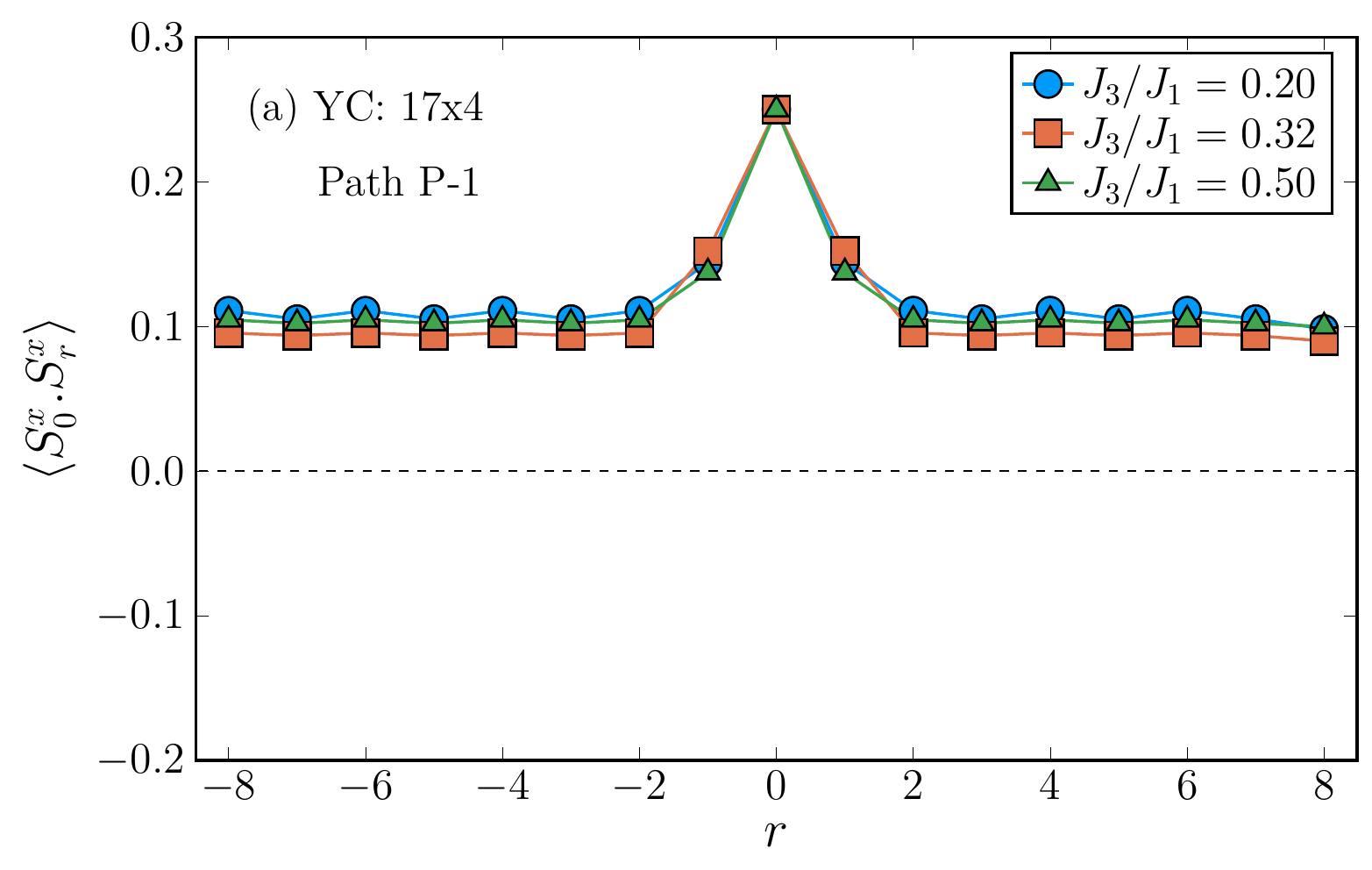}
    \includegraphics[width=0.33 \textwidth]{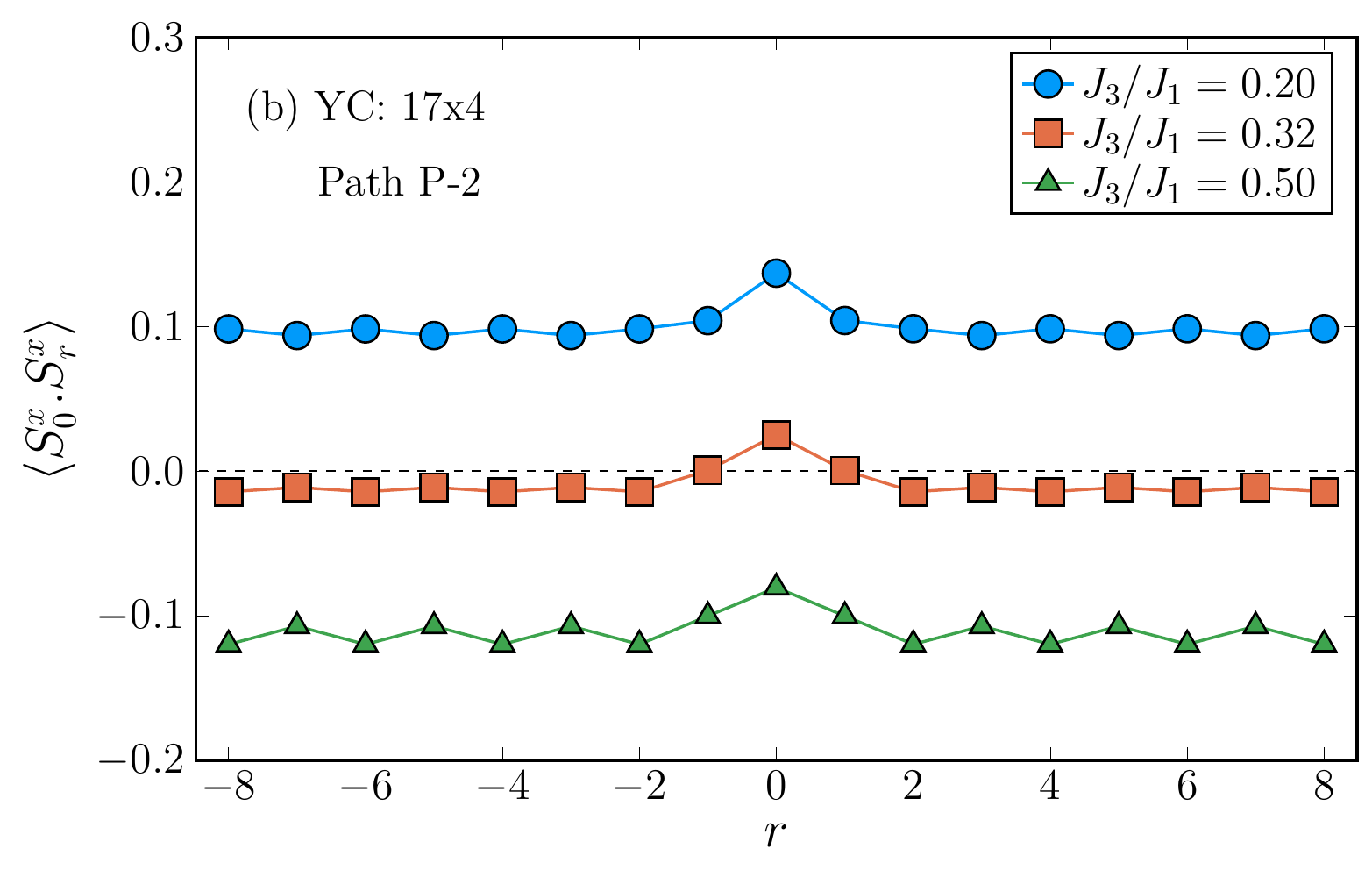}
    \includegraphics[width=0.33 \textwidth]{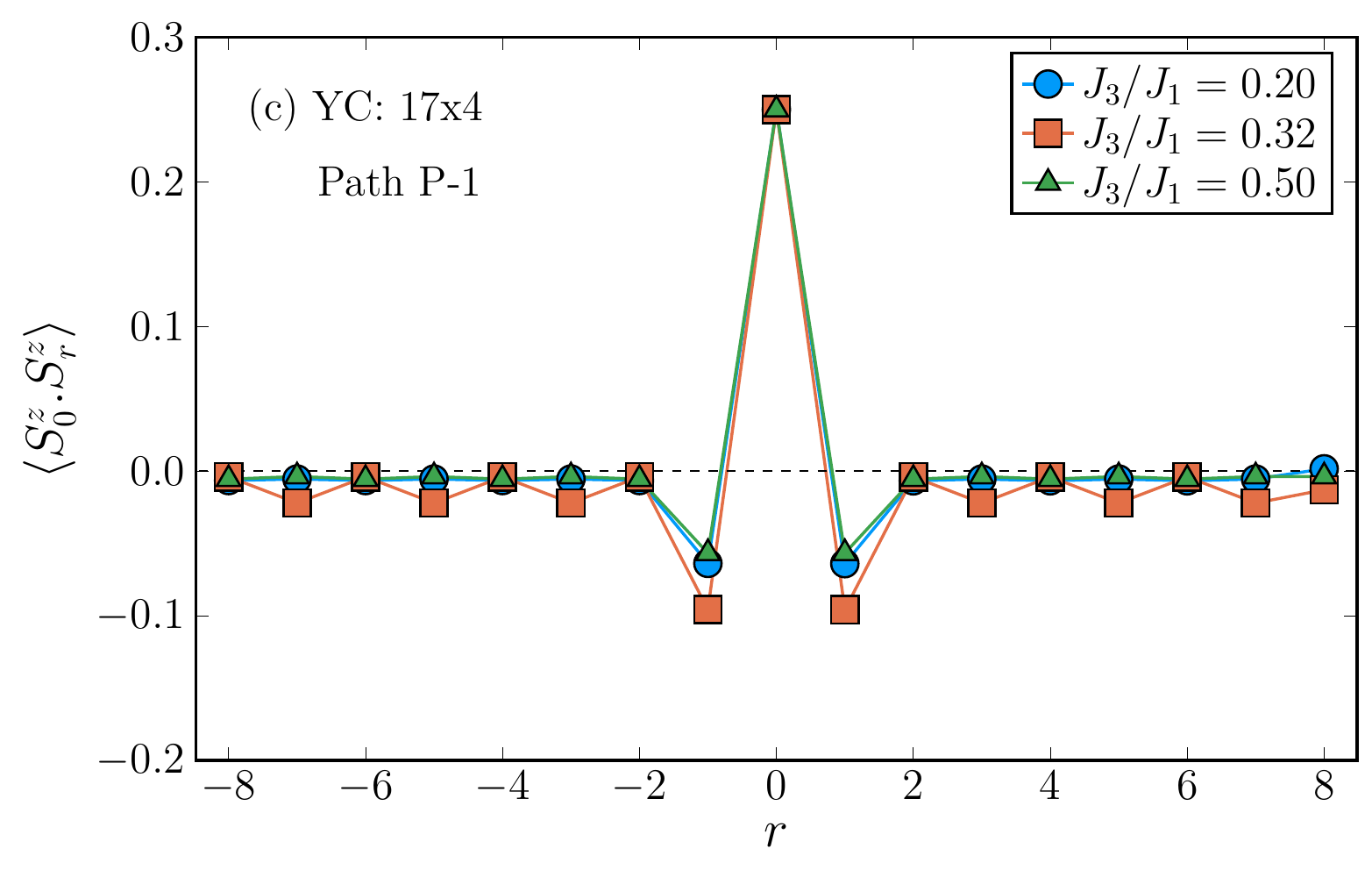}
    \includegraphics[width=0.33 \textwidth]{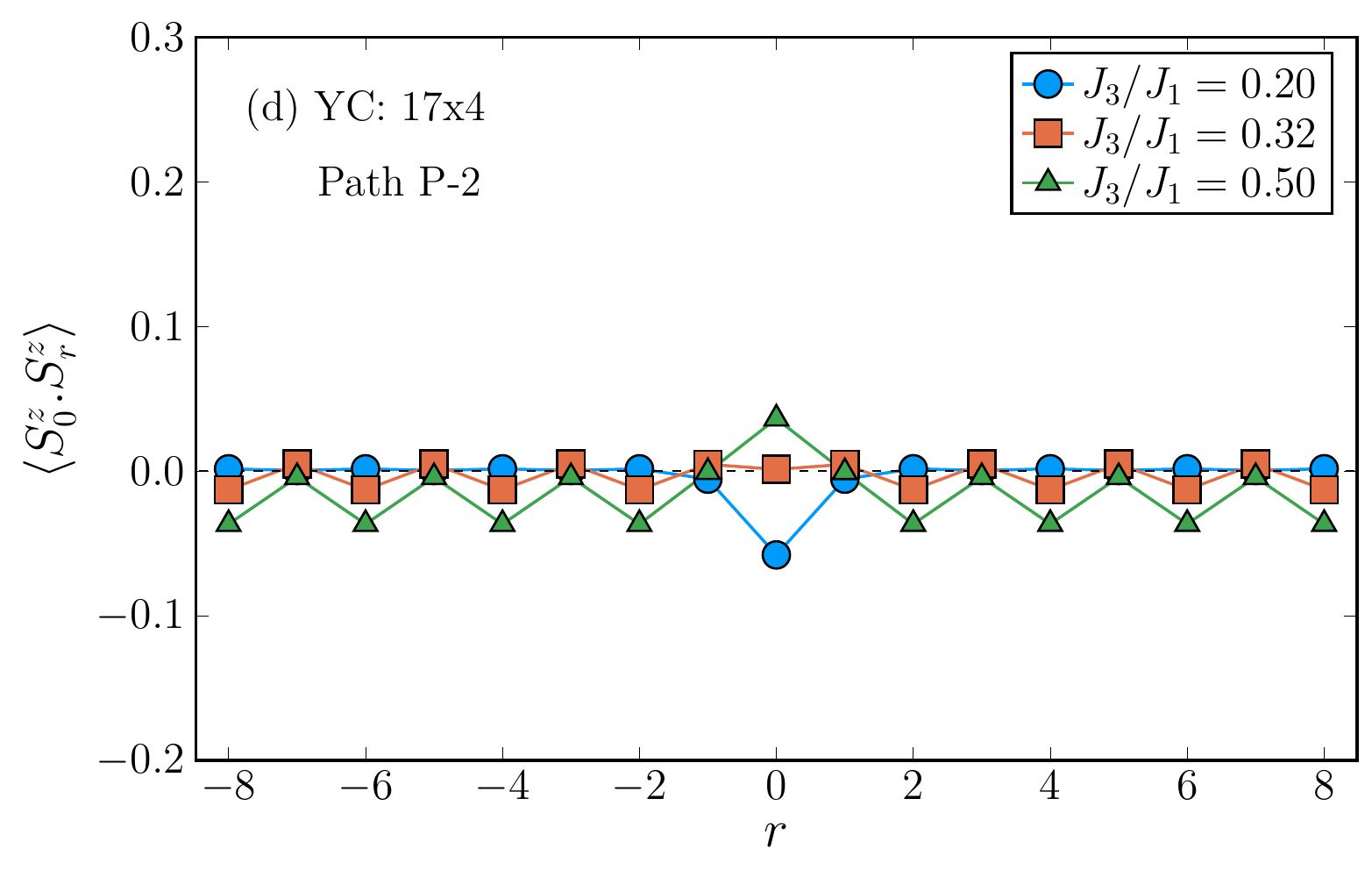}
     \caption{DMRG spin correlations on YC cluster computed for $(K\!=\!0,\lambda\!=\!0.25)$. 
     (a) $S^x$ correlations along path P-$1$ showing apparently non-decaying correlations in all phases. 
     (b) $S^x$ correlations along path P-$2$ showing FM, ZZ, and strongly suppressed correlations
     for indicated $J_3/J_1$ values. (c) and (d) show $S^z$ correlations which exhibit
     weak oscillatory component indicative of $z$-N\'eel order for $J_3/J_1=0.32$ in the {\sl} phase,
     but these oscillations are more pronounced in the ZZ phase for path P-$2$ on this YC cluster.}
    \label{fig:DMRG2}
\end{figure}

Turning to the YC-cluster, we find somewhat ambiguous results. Fig.~\ref{fig:DMRG2}(a) and Fig.~\ref{fig:DMRG2}(c) 
show that the $S^{x,y}$ long ranged spin correlations
in the FM and ZZ phases along path P-$1$ and path P-$2$ are qualitatively consistent with our findings on the XC-cluster. 
However, in the intermediate {\sl} phase, the $S^{x,y}$ correlations along
path P-$1$ seem to indicate long-range order while it is nearly vanishing along path P-$2$. This striking spatial anisotropy ---
relatively long correlations along the same zigzag chain but strongly suppressed correlations with the neighboring zigzag chain ---
is somewhat reminiscent of sliding Luttinger liquid phases \cite{SLL_Emery_PRL2000,SLL_Vishwanath_PRL2001,CSLL_Mukhopadhyay_PRB2001}. 
Such sliding phases have also
been proposed to occur, on intermediate energy scales, in certain 
crossed spin-chain models \cite{CSLL_Starykh_PRL2002,CSLL_Sindzingre_PRB2002}. This
spatial anisotropy of correlations observed in the {\sl} phase on the YC-cluster might reflect the physics of
a strongly fluctuating and partially melted incommensurate spiral order which decouples adjacent zigzag chains.
In this YC geometry, the $S^z$ correlations are ambiguous.
From Fig.~\ref{fig:DMRG2}(b) and Fig.~\ref{fig:DMRG2}(d), $S^z$ correlations along path-$1$ are oscillatory, 
indicative of $S^z$ N\'eel order ($z$-N\'eel) on the honeycomb lattice. However, Fig.~\ref{fig:DMRG2}(d) shows that 
such oscillatory $S^z$ correlations are also found in the ZZ phase (but not in FM) even for our largest system size, 
suggesting that these $S^z$ correlations might be strongly influenced by the YC cylinder geometry and are 
not necessarily a distinguishing feature of {\sl}.

\begin{figure}[t]
    \centering
    \includegraphics[width=0.45 \textwidth]{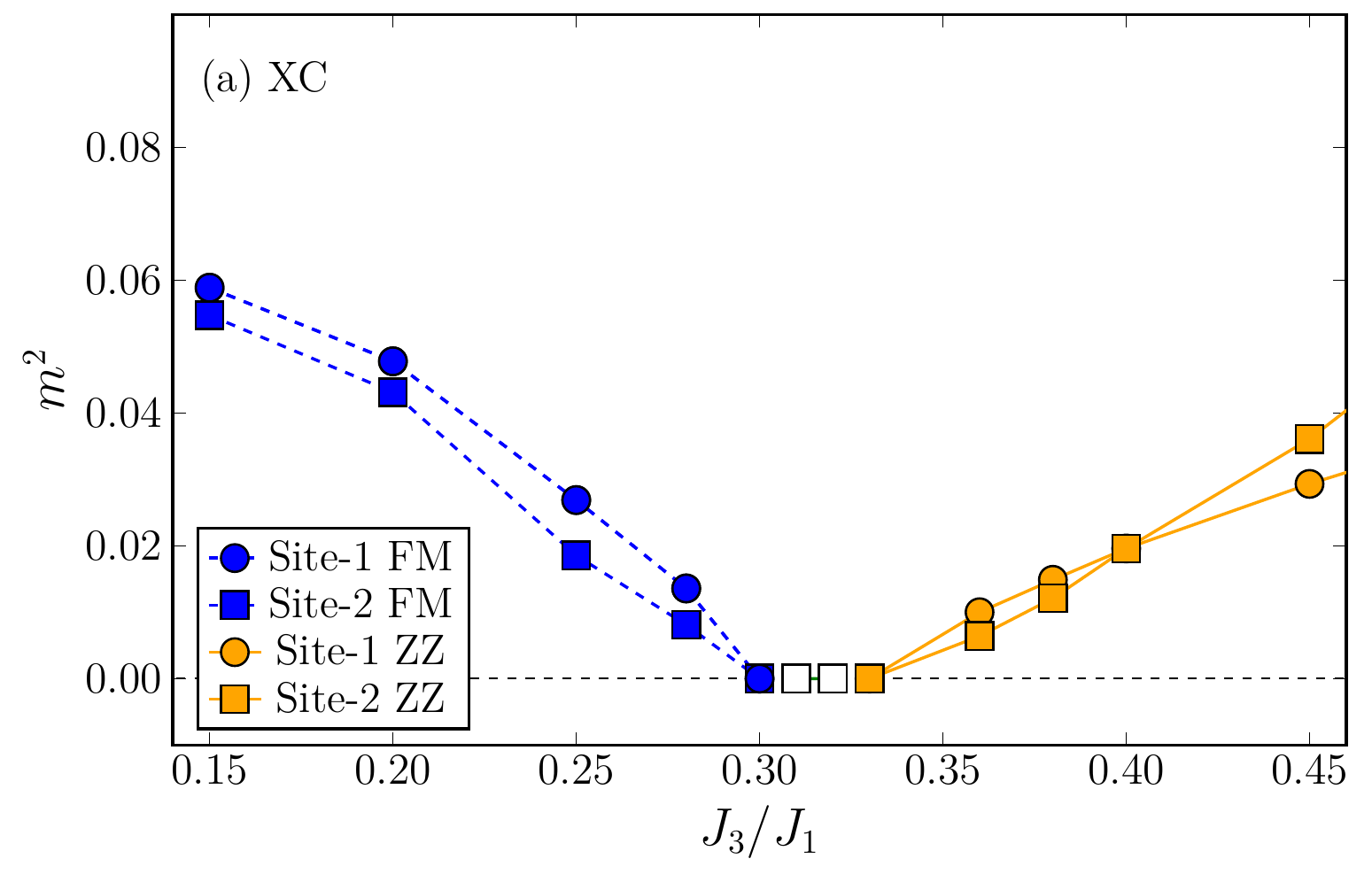}
    \includegraphics[width=0.45 \textwidth]{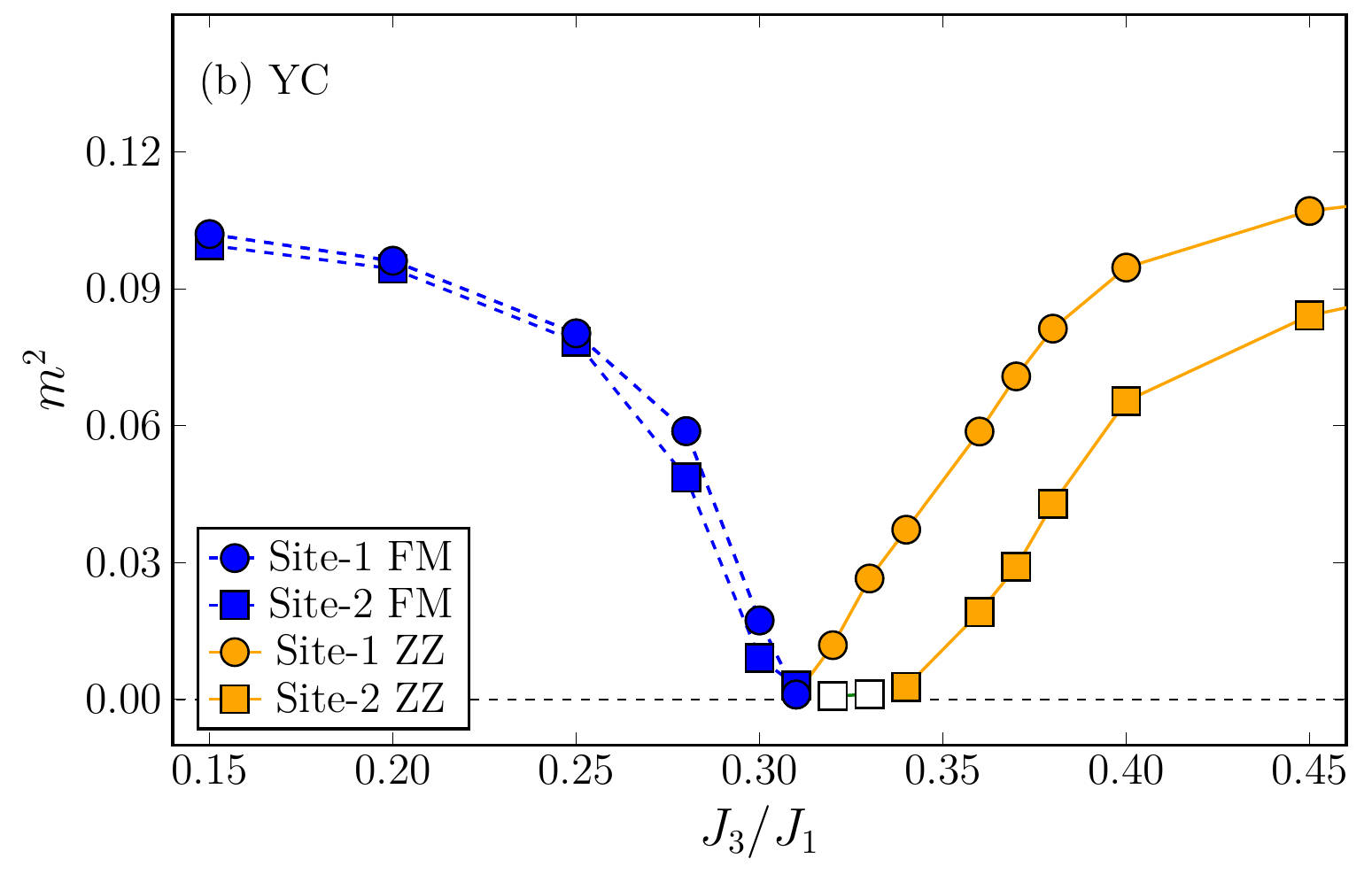}
    \caption{Extrapolated $\langle S^x_0 S^x_r \rangle$ 
    correlations at distant points $s_1$ and $s_2$ with respect to the
    reference site in the limit of large cylinder sizes for $(K\!=\!0,\lambda\!=\!0.25)$. The dashed (FM) and solid (ZZ) lines are a guide to the eye.
    (a) XC cluster and (b) YC cluster. These extrapolated correlations indicate an
    intermediate phase (open squares) which is neither FM nor ZZ. This window is similar in three of
    the four cases, $0.3\lesssim J_3/J_1 \lesssim 0.33$, 
    except the result from extrapolating the $s_1$ correlation in the YC cluster which shows either a 
    very narrow {\sl} window or even a direct FM-ZZ transition. We denote this extrapolated correlation
    as $m^2$ since it is expected to be a measure of the square of the order parameter (for the FM or ZZ
    orders). Open squares correspond to nearly vanishing $m^2$ which is the {\sl} phase.}
    \label{fig:DMRGorderparameter}
\end{figure}

We have studied the the finite size scaling of $S^{x},S^{y}$ spin correlations on the XC and YC clusters 
at distant points $s_1$ and $s_2$
with respect to 
the reference site (see Fig.~\ref{fig:DMRGcluster}). These points are chosen to be one layer
away from the boundary to decrease edge effects. We computed these spin correlations extrapolated to the
large cylinder size limit as a measure of $m^2$, the square of the magnetic
order parameter. As shown in Fig.~\ref{fig:DMRGorderparameter}(a) for the XC cluster, 
we find nonzero extrapolated FM order at small $J_3/J_1$ (blue circles, blue squares), 
and ZZ order at large $J_3/J_1$ (orange circles and orange squares).
This leaves a window $0.3 \lesssim J_3/J_1 
\lesssim 0.33$ where such magnetic orders appear to vanish as indicated by the
open symbols. For the YC cluster, we find similar results as seen in 
Fig.~\ref{fig:DMRGorderparameter}(b); although the extrapolated
correlation for site $s_1$ suggests a very narrow intermediate
phase or a direct transition, there is a clear intermediate phase
from studying correlations for site $s_2$.

We have checked that, together with no clear evidence of magnetic orders,
there is also no evidence of Kekul\'e type
valence bond orders in the intermediate window with no FM and ZZ order. 
Our results are thus strongly suggestive of an 
intermediate SL phase, distinct from the FM and ZZ phases. 
Nevertheless, given the
ambiguity in our DMRG spin correlation results between the XC and YC clusters, we continue to label this phase as {\sl} 
to highlight its spin-liquid-like character.

We emphasize that our results
do not definitively rule out weak broken symmetries in the {\sl} phase
which might become apparent on larger system sizes.
Indeed, after the initial report of our work \cite{bose2023proximate}, a more recent DMRG study of this model
by Jiang, White, and Chernyshev \cite{j1j3_jiang2023quantum}, has reported evidence for an intervening phase consistent
with our results. However, they conclude from studying larger cylinder system sizes
that this intermediate phase, which we call {\sl}, is a symmetry broken phase with 
weak $z$-N\'eel order. Crucially, they deduce this order using finite-size scaling of the induced
order in the presence of an additional strong $z$-N\'eel
boundary pinning field. Whether our results differ from their work due to this difference
in methodology, or simply due to a difference in system sizes, remains to be clarified. We leave this as an open issue for further
studies. Moreover, even if we take into account the larger system DMRG results which
argue that this {\sl} state hosts $z$-N\'eel order, the deduced order parameter they deduce is
very small, $\lesssim 0.1$, suggesting extremely strong quantum fluctuations in this {\sl} state. This places
it in proximity to a genuine SL state and
well beyond the purview of any linear spin-wave theory description. A pseudo-fermion functional renormalization
group study (pf-FRG) of this model \cite{j1j3_watanabe2022frustrated} has also found a small intervening state with
$z$-N\'eel order, but the quantitative reliability of this method for identifying phases of spin-$1/2$ 
models is unclear.

To conclude this section, our DMRG results for the spin-$1/2$ $J_1$-$J_3$ XXZ model on the honeycomb lattice finds evidence for 
the {\sl} phase, with spin-liquid-like characteristics, at intermediate $J_3/J_1$.
Whether or not the {\sl} hosts
weak symmetry breaking (e.g., weak $z$-N\'eel order), it is important to understand
the nature of the proximate SL state from which this phase descends,
and to explore its implications for the phase diagram and for experiments on the honeycomb cobaltates. We
next turn to this issue using parton mean field theory and a numerical variational Monte Carlo study of Gutzwiller 
projected parton wavefunctions.

\section{Gutzwiller projected parton
wavefunctions}

To identify a candidate for the {\sl} phase, we turn to a parton theory of the spin Hamiltonian,
using Schwinger fermions to represent spins via
\(S_i^{\m} = \frac{1}{2} f^{\dagger}_{i\g} \s^{\m}_{\g\d} f^{\vphantom\dagger}_{i\d}\), 
where \(\s^{\m}\) are the Pauli matrices. The spin Hamiltonian $\mh$ is then expressed as
a quartic interaction in terms of partons, and we have to impose
$\sum_{\g} f^{\dagger}_{i\g} f^{\vphantom\dagger}_{i\g}\!=1\!$ as a local Hilbert space constraint
to faithfully represent spin configurations.

\subsection{Parton mean field theory}
We can simplify $\mh$ expressed in terms of partons using
mean field theory where we allow nonzero
expectation values for parton bilinears in multiple channels: the Weiss channel which permits 
magnetic order $\langle S_i^a \rangle \!\neq\! 0$ and the hopping channel 
$\langle f^{\dagger}_{i\g} f^{\vphantom\dagger}_{j\d} 
\rangle \equiv W^{\g\d}_{ij}$ which allows parton dispersion. Since
mean field theory overestimates the tendency to order, we propose a modified parton mean field theory,
replacing
\begin{eqnarray}
\expval{S_i^{\m}S_j^{\n}}= (1-\alpha) \expval{S_i^{\m}} \expval{S_j^{\n}} 
- \frac{1}{4} \Tr(\s^{\m} W_{ij}^{*}\s^{\n} W_{ij}^{T})
\end{eqnarray}
%\begin{eqnarray}
%\expval{S_i^{\m}S_j^{\n}}&=& \frac{(1-\alpha)}{4} \Tr(\s^{\m}\chi_{ii}^T) 
%\Tr(\s^{\n}\chi_{jj}^T) \nonumber \\
%&-& \frac{1}{4} \Tr(\s^{\m}\chi_{ij}^{*}\s^{\n}\chi_{ij}^{T})
%\end{eqnarray}
while computing the mean field energy; this rescales the Weiss channel energy
by $(1-\alpha)$. When $\alpha=0$, we find that the mean field solution converges to the classical 
magnetically ordered phases with no spinon hopping ($W_{ij}\!=\!0$), reproducing
the classical phase diagram of the $J_1$-$J_3$ XY model. 
The choice of intermediate values $0 < \alpha < 1$ phenomenologically incorporates quantum fluctuation 
effects beyond parton mean field theory.
For $\alpha=1$, magnetic
orders are completely suppressed by hand and we find pure spin liquid solutions.

\begin{figure}[t]
\centering
\includegraphics[width=0.45\textwidth]{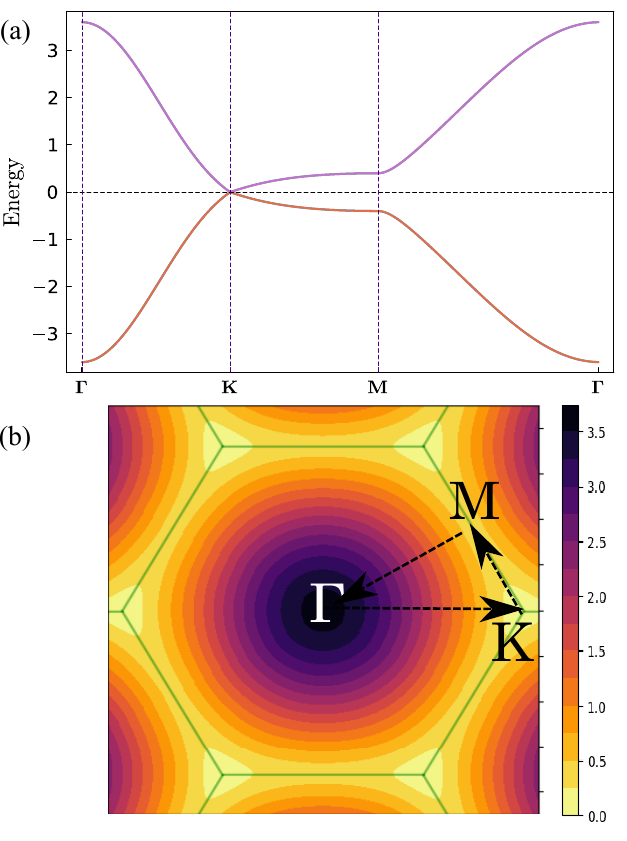}
    \caption{
   (a) Mean field parton band structure of the Dirac SL in the original BZ along a path through the $\Gamma$-K-M 
   points. Energy is measured in units of $t_1$, and we have set $t_3/t_1=0.2$ as found from our VMC optimization. 
   In self-consistent parton mean field theory, we find a slightly smaller value $t_3/t_1\!=\!0.15$.
    (b) Energy gap in units of $t_1$, shown using color scale, in the full hexagonal BZ - we find
    low energy modes along the entire edge of the BZ with vanishing gap at the Dirac K points.}
    \label{fig:diracband}
\end{figure}

For $\alpha\! = \!1$, we have done the parton mean field theory calculations,
exploring possible spatially inhomogeneous solutions directly in real space mean
field theory, as well as using momentum space mean field theory with various unit cell sizes.
Motivated
by the THz \cite{Armitage2022} and thermal conductivity \cite{kappa_NCTO_Sun_PRB_2023} 
experiments which find an abundance of low energy excitations, we have chosen to restrict our attention to
spin liquids without spinon pairing terms (which might potentially open gaps in the excitation spectrum).
In all cases examined, we find that the mean field theory converges to a 
SL phase which corresponds to a zero flux-state, described by the
spinon Hamiltonian $\mh_{\rm sp} \!=\! - \sum_{i,j}\! t^{\mu\nu}_{ij} f^\dagger_{i,\mu} f^{\vphantom\dagger}_{j,\nu}$
with $\mu,\nu$ denoting spin.
Here, the nearest-neighbor parton hopping matrix is $t_{ij}= \sigma_z t_{1}$ and third-neighbor hopping is 
$t_{ij}\!=\! \sigma_z t_{3}$ with $t_1,t_3 \!>\! 0$, such that the $\uparrow$ and $\downarrow$ spinons each see 
zero flux on  all plaquettes on the honeycomb lattice, but have opposite sign
hopping amplitudes relative to each other. This breaks the global $SU(2)$ spin rotation symmetry down to a $U(1)$ 
spin rotation symmetry, consistent with a wavefunction for an XXZ model.
The mean field parton band dispersion with $t_1,t_3$ hoppings is particle-hole symmetric, with 
$t_1 \!\approx\! 0.13 J_1$ and $t_3/t_1 \!\approx\! 0.15$, and features 
Dirac nodes at the $K,K'$ points of the hexagonal Brillouin zone (BZ) as shown in Fig.~\ref{fig:diracband}, 
so this state is a Dirac SL.
We note that a similar Dirac SL, but with
first and second neighbor
hopping, also appears as a proximate state in the $J_1$-$J_2$ honeycomb XY model \cite{j1j2_galitski2014}
although its physical properties and relevance to magnetic solids have not been explored.

\subsection{Variational Monte Carlo study}

\begin{figure}[t]
    \hspace{-0.4cm} \includegraphics[width=0.48\textwidth]{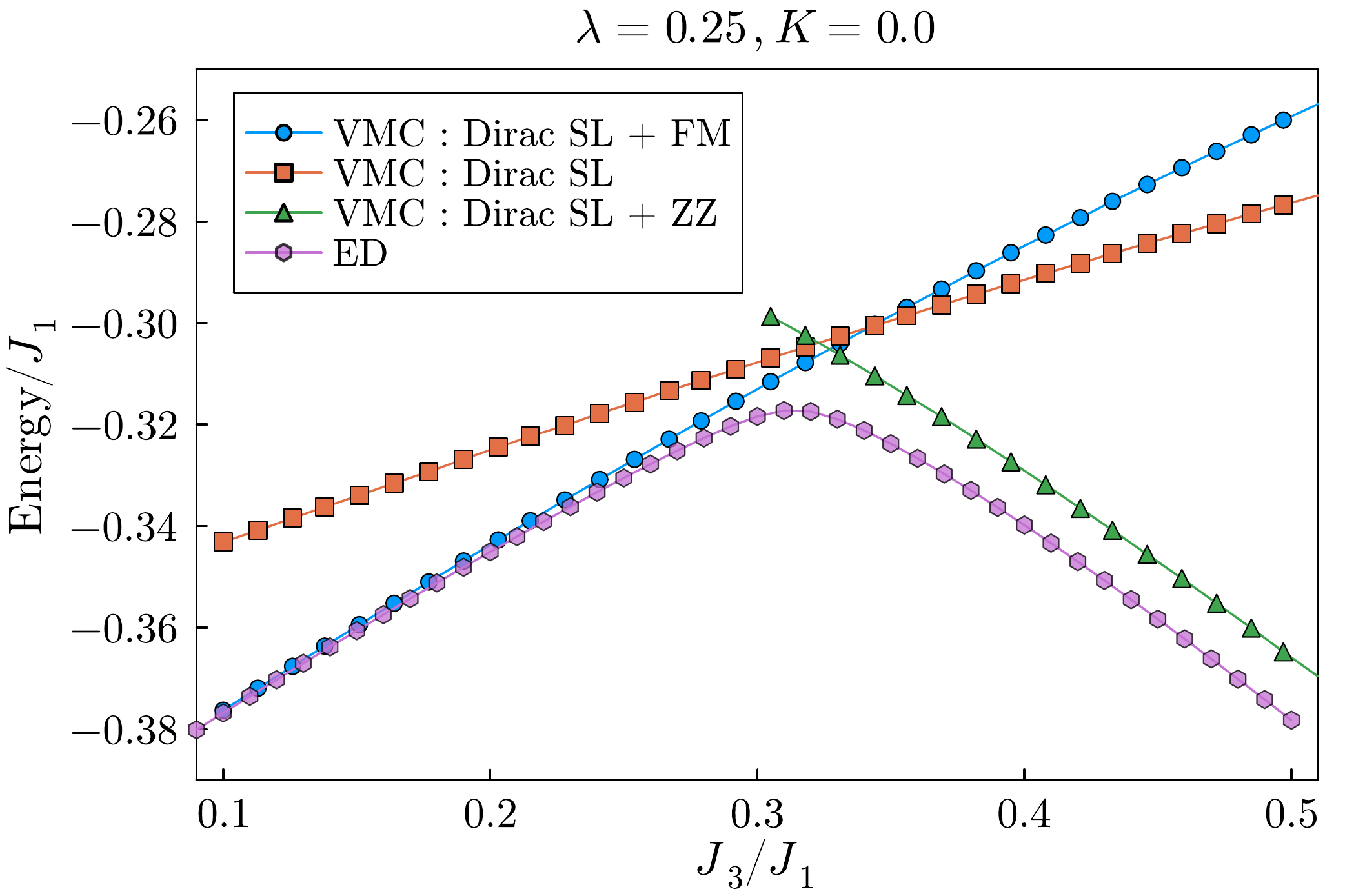}
    \caption{VMC optimized energies on \(N = 24\) sites cluster as a function of \(J_3 / J_1\), for different ansatzes, compared with ED energies on the same cluster. We have fixed $\lambda=0.25$ and $K=0$. In the FM phase, the
    VMC energy is within $0.1$-$0.5$\% of the ED. In the ZZ phase VMC is within $2$-$3$\% of ED. The largest difference of $\approx\! 
    4$\% occurs for $J_3/J_1 \!\approx \!0.32$ where the VMC energy of the Dirac SL is comparable to that of the Dirac SL with 
    coexisting FM or ZZ order.}
    \label{fig:VMC}
\end{figure}

Going beyond mean field theory, we have studied the phase diagram of the $J_1$-$J_3$ XXZ model using Gutzwiller projected parton wavefunctions. We compute the energy and correlations of the projected wavefunction through a Monte Carlo sampling of real space
configurations which obey the strict Hilbert space constraint. To obtain the best variational ground state with the lowest energy, 
and to study its correlations, we first optimize the variational parameters encoded in the parton mean field Hamiltonian. 
In our variational wavefunction, we fix
\(t_1 = 1\) in the parton Hamiltonian, and treat $t_3$, the on-site Weiss fields \(B_i\) (which allow for FM, ZZ, or $z$-N\'eel order),  
and long-range Jastrow factors as variational parameters. For optimizing these parameters, we use the
state of the art stochastic reconfiguration technique (a modified gradient descent method) \cite{becca_sorella_2017}.\par

We focus below on VMC results for the 24-site cluster which we compare with our ED results in Table~\ref{table:VMC}; 
this allows independent Jastrow factors upto third-nearest neighbors. We worked with 10,000 thermalization sweeps followed by 10,000 measurement sweeps for \emph{each} Monte Carlo run. We performed 500 such Monte Carlo runs per optimization, checking that the stochastic reconfiguration drives 
the variational parameters towards the optimum value in between runs.

\begin{table}[b]
\centering
\begin{tabular}{ |c|c|c|c|c| } 
 \hline
 \(J_3 / J_1\) & E.D. & Dirac SL+FM & Dirac SL & Dirac SL+ZZ\\
 \hline 
 0.24 & -0.33331 & -0.33114(4) & -0.31797(4) & -0.31806(4)\\
 \hline
 0.28 & -0.32265 & -0.31897(4) & -0.31110(4) & -0.31122(4)\\
 \hline
 0.32 & -0.31743 & -0.30721(4) & -0.30441(4) & -0.30302(4)\\
 \hline
 0.36 & -0.32266 & -0.29581(4) & -0.29790(4) & -0.31555(4)\\
 \hline
 0.40 & -0.33978 & -0.28481(4) & -0.29148(4) & -0.32908(4)\\
 \hline
\end{tabular}
\caption{Optimized VMC ground state energies (per spin) for the three ansatzes, 
as compared to the results from exact diagonalization on the 24-site cluster.}
\label{table:VMC}
\end{table}

As seen from Fig.~\ref{fig:VMC} and Table~\ref{table:VMC}
we find that the optimal state for $J_3/J_1 \lesssim 0.3$ corresponds to a 
Dirac SL state with imposed FM order - its energy matches with the ED energies extremely well 
(we label this state `SL + FM'). 
For $J_3/J_1 \gtrsim 0.33$, the optimal state is a Dirac SL state with 
imposed ZZ order - this exhibits a good energy 
match to within $2$-$3$\% of the ED result (we label this state `Dirac SL + ZZ')
In the intermediate regime, $0.3 \lesssim J_3/J_1 \lesssim 0.33$,
we find all ansatzes (Dirac SL, Dirac SL+FM, and Dirac SL+ZZ) 
to be very close to each other, with the largest discrepancy with the ED energies 
being $\sim \! 4$\%. These energy differences between ED and VMC are comparable to the previous best VMC results on the $J_1$-$J_2$ honeycomb XY model \cite{j1j2_galitski2014, j1j2_rigol2013}. The Gutzwiller projected Dirac SL parton state thus captures a large fraction 
$\sim\! 96\%$ of the ED energy in the {\sl} regime. The Gutzwiller projected Dirac SL has rapidly decaying $S^z$ correlations, 
while the XY spin correlations decay as $\sim 1/r^{\alpha}$, with $\alpha\approx 1.5$.

\subsection{Relation between {$\widetilde{\mathbf{SL}}$} and the Dirac SL: 
Deformations of Dirac SL}

As described in the previous section, the Gutzwiller projected Dirac SL 
captures $96$\% of the ED energy for $0.3 \lesssim J_3/J_1 \lesssim 0.32$. This suggests that
we can treat the Dirac SL as a proximate parent SL state (i.e., a SL very close by in energy), 
and the {\sl} state observed in this window
(in ED and DMRG) as a related descendant. 
In field theory language, the Dirac SL we have found involves $N_f=4$ 
flavors of massless two-component
Dirac fermions coupled to an $SU(2)$ gauge field. Large-$N_f$ versions of closely related theories have been studied in previous
work \cite{gauge_affleck1988,su2gauge_hermele_prb2007,su2gauge_karthik_prd2018,su2gauge_thomson_prx2018};
however, the low energy fate of this theory is unclear at $N_f=4$. In this framework, the
{\sl} phase might descend from the Dirac SL as a result of symmetry breaking or spinon pairing or confinement.
To explore this in VMC, we have incorporated various mass terms in our Dirac SL ansatz, 
as well as rotational symmetry breaking hoppings (nematic order).

We find that Dirac mass terms such as the (i) time-reversal preserving Kane-Mele (quantum spin Hall) mass, (ii) the
time-reversal breaking Haldane (chiral) mass, or (iii) a Kekul\'e mass which leads to 
$\sqrt{3}\times\sqrt{3}$ valence bond order, do not lower the energy of the Dirac SL. 
Motivated by related DMRG work \cite{j1j3_jiang2023quantum} and pf-FRG results 
\cite{j1j3_watanabe2022frustrated}, 
we have also considered 
(iv) a z-N\'eel mass, but also do not find it to be favored within VMC. 
%In fact, any symmetry allowed second-neighbour hopping did not seem to have any effect at all. 
We found that a rotational symmetry breaking nearest-neighbour hopping decreased the energy, but only 
in the ZZ phase and by a negligibly small value \(\approx 0.1\%\); this is too small to impact our 
comparison with ED. Given the computational complexity of studying spinon pairing states in XXZ models within
VMC (which must account for coexisting singlet and triplet pairing), we defer a study of such pairing states, as
well an exhaustive study for all Dirac mass terms, to a future
investigation.

\begin{figure}[h]
\includegraphics[width=0.5\textwidth]{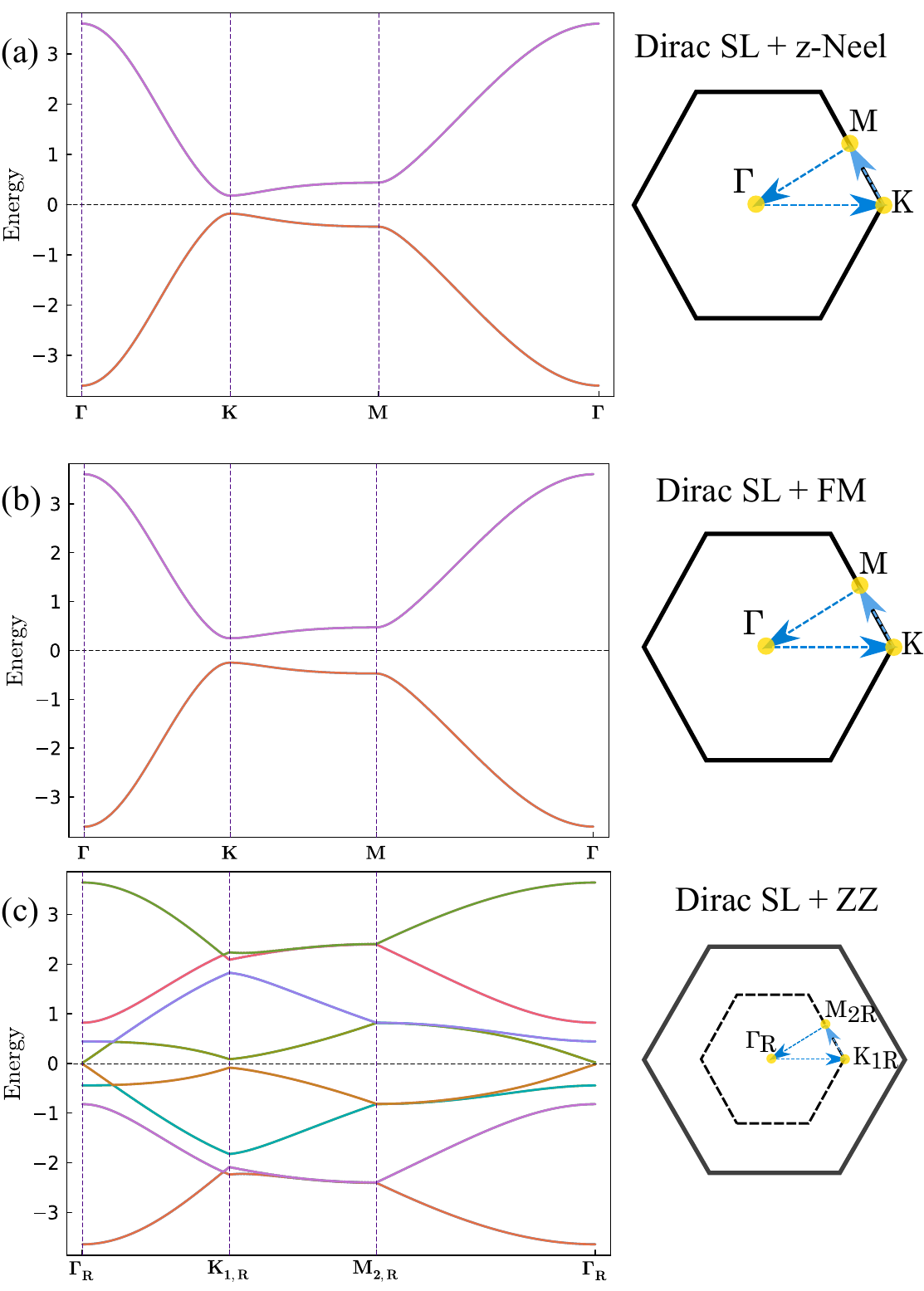}
    \caption{
   Mean field parton band structures of the perturbed Dirac SL (with energy in units of $t_1$) for $t_3/t_1\!=\!0.2$. (a) Band structure
   in the original BZ with $z$-N\'eel order parameter $m^z_{\rm N}=0.1$ showing that the Dirac nodes 
   get gapped out. (b) Band structure
   with in-plane Zeeman field $h/t_1=0.5$ showing that the Dirac nodes get gapped out. 
   (c) Band structure in the reduced BZ (dashed line) 
   for $2\times 2$ unit cell in the presence of ZZ order with order parameter $m_{\rm ZZ}=0.1$ showing
   that band touching points persist (here at $\Gamma_R$). The ZZ order is chosen with wavevector $(0,\pi)$.
   For large $m_{\rm ZZ} \gtrsim 0.25$, the dispersion is fully gapped at the chemical
   potential ($E=0$).}
    \label{fig:bands}
\end{figure}

In order to examine how some of these perturbations, specifically the FM, ZZ, and $z$-N\'eel order, impact the mean field
parton dispersion, we studied the effect of such perturbations on the single-particle band structure. 
As seen from Fig.~\ref{fig:bands}(a), including a $z$-N\'eel 
Weiss field in the Dirac SL parton ansatz gaps out the Dirac touchings at the \(K\) points.
Fig.~\ref{fig:bands}(b) shows that an in-plane FM Weiss field, which 
is equivalent to an in-plane Zeeman field, has a very similar effect.
This is an interesting point to keep in mind when later examine experimentally relevant
response functions of the Dirac SL. In particular, although {\sl} might be a weakly gapped/ordered phase 
descending from the Dirac SL, its high energy response in all these cases will look very similar to the Dirac SL. 
Moreover, weak disorder-induced broadening of the parton spectral functions on the scale of a small Dirac
mass gap might lead to apparent gapless behavior even at low energy.
 
Unlike the FM and $z$-N\'eel order, the ZZ Weiss field is not a 
simple Dirac mass term . Here, we
work with the quadrupled $2\times 2$ unit cell ($8$ site unit cell)
which accommodates the ZZ ordering. With increasing $m_{\zz}$, the Dirac points
persist as gapless band touching points (although they can shift in momentum), 
so this parton mean field state would have ZZ order coexisting with
fractionalized gapless excitations. Such a state has been conjectured to occur in
{\ncto} \cite{kappa_NCTO_Sun_PRB_2023}.
Fig.~\ref{fig:bands}(c) shows that for
$m_{\zz} \!\approx\! 0.1$ there is a band touching at the $\Gamma_R$ point of the folded BZ.
Further increase of $m_{\zz}$ leads to a quadratic band touching at $M_{2R}$ for 
$m_{\zz}\! \approx\! 0.25$, before the bands eventually get fully gapped out for $m_{\zz}\! \gtrsim\! 0.25$.

\section{Experimental implications}

With the above results on the phase diagram and the variational wavefunction in hand, we next turn to a discussion of the 
experimental data on the honeycomb cobaltates which exhibit SL like characteristics such as a broad continuum in THz 
spectroscopy and a low temperature metallic thermal conductivity. The ideal comparison between theory and experiments would
have been to explore response and transport in the {\sl} ground state. However, given our difficulty in identifying the
precise nature of the {\sl} phase, and the considerable challenge of exploring dynamics and transport in this strongly
fluctuating state, we take a slightly simpler tack here and
explore the corresponding results in the parent Dirac SL state using parton mean field theory.
Remarkably, we find that the Dirac SL, within parton mean field theory, provides
an excellent semi-quantitative description of the existing THz and metallic thermal conductivity data in these cobaltates. 
We speculate that the reason for this good agreement might be that at intermediate energy scales,
or in the presence of weak disorder or thermal fluctuations in the real materials, the balance might shift in
favor of the parent Dirac SL over the {\sl} state as a reasonable description of the spin-liquid-like behavior
found in the cobaltates.

%\subsection{Dynamical spin response}

\subsection{THz spin dynamics} 

\begin{figure}[t]
    \centering \includegraphics[width=0.45 \textwidth]{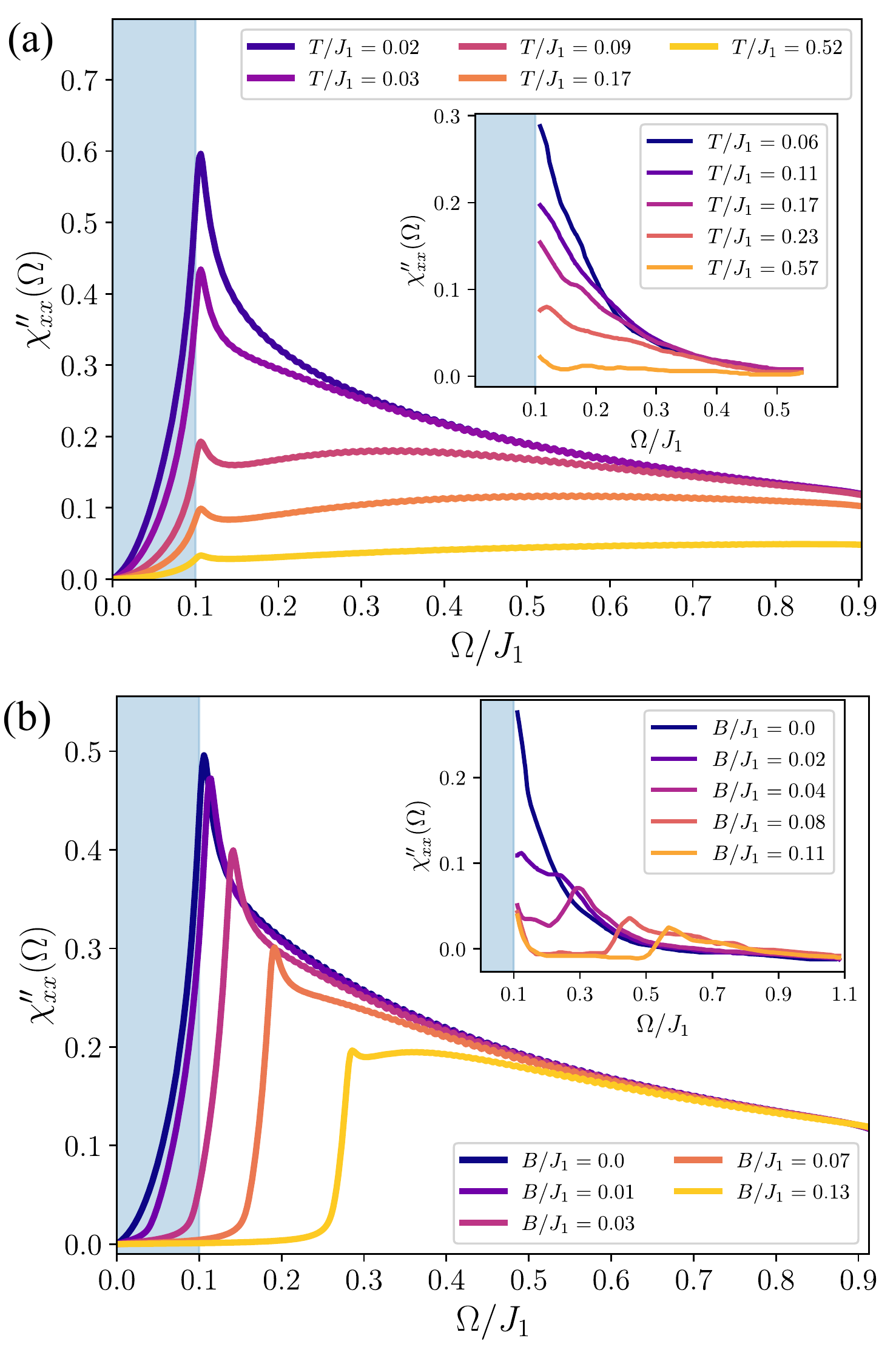}
    \caption{Imaginary part of the dynamic spin susceptibility od Dirac SL at zero momentum $\chixx(\Omega)$, {\bf a.} as a function
    of varying temperatures $T$ at $B\!=\!0$, and {\bf b.} at a fixed $T$ with varying in-plane 
    $B$ along \(\hat{x}\).
    Insets show corresponding experimental THz spectroscopy data for {\bcao} \cite{Armitage2022};
    we use $B\!=\!g \mu_B H$ with $g\!\approx\!5$ and $H$ measured in Tesla,
    and $J_1\!=\!7.6$\,meV to scale the experimental data \cite{Armitage2022,BCAO_Broholm2023}.
    The shaded regions correspond to frequency windows not explored in the THz measurements.}
    \label{fig:response}
\end{figure}

Using the Dirac SL ansatz, we have calculated the THz spin response in terms of the mean field
parton Green functions using the bubble diagram to obtain the spin susceptibility.
The zero momentum response is given by
\begin{equation}
\!\! \chi_{ab}\!(\i\Omega_n)\!\! = \!
    - \frac{1}{4}\!\! 
    \sum_{\bk,\;\omega_n} \!\! \text{Tr} [\s^{a}G^{T}\!(\Vec{k}, \i\w_n) \s^{b} G^{T}\!(\Vec{k}, \i\w_n\!+\!\i\Omega_n)]
\end{equation}
where $G_{ij}^{\a\b}(\Vec{k}, \i\w_n) = \expval{f_{i\a}^{\dagger}(\bk,\i\w_n) f_{j\b}^{\pdg}(\bk,\i\w_n)}$. We
can analytically continue $\chi_{ab}\!(\i\Omega_n \to \Omega + i0^+)$ to obtain the real frequency response.

Fig.~\ref{fig:response}(a) shows the real frequency zero wavevector THz spin response, the imaginary
component $\chixx(\Omega)$,
plotted at zero field as function of frequency $\Omega$, for various temperatures $T$, in the mean field Dirac SL
with $t_1 = 0.13 J_1$ and $t_3/t_1=0.2$. As we cool from high $T$, 
$\chixx(\Omega)$ develops a broad feature which is strongly enhanced as a low frequency `peak', before decreasing 
for $\Omega/J_1 \lesssim 0.1$. This enhancement arises from the large low energy density of states for 
spinon excitations around the entire BZ edge, as seen from Fig.~\ref{fig:diracband},
with the peak coming from states at the $M$ point of the original
honeycomb BZ. For comparison,
the inset of Fig.~\ref{fig:response}(a) shows the experimental data on {\bcao} \cite{Armitage2022}, which
highlights their striking similarity.
We note that the THz absorption data on {\bcao} \cite{Armitage2022} does not extend below $\lesssim 0.2$\,THz;
using $J_1 \!=\! 7.6$\,meV \cite{BCAO_Broholm2023},
this corresponds to $\Omega/J_1 \!\lesssim\! 0.1$.
We mark this frequency cutoff in our theoretical plot Fig.~\ref{fig:response}{\bf a} and in the
experimental data in the inset. Extrapolating the THz data to lower frequency using constraints from 
Kramers-Kr\"onig relations \cite{Armitage2022} does seem to suggest that the response decreases 
below $\lesssim 0.2$\,THz, which would be consistent with the Dirac SL. We have checked that
incorporating a small Dirac mass, for instance via weak $z$-N\'eel order, leaves the high energy continuum
unaffected while inducing a small gap at very low energies, so this would still be consistent with
the experimental THz data.

Fig.~\ref{fig:response}(b)
shows that an in-plane Zeeman field digs out a 
low energy spectral gap which increases with the field, while the high frequency response remains broad. This gap
is because the in-plane field induces the mixing and splitting of degenerate spin-$\upa$ and spin-$\dna$ parton bands.
We have set $B\!=\!g \mu_B H$ for the in-plane field, with $g\!=\!5$ \cite{Abinitio_Das2021,Armitage2022} and $H$ 
measured in Tesla. This
behavior is also qualitatively similar to the experimental data on {\bcao} \cite{Armitage2022}, which is replotted in the inset,
although the theoretical gap scale is smaller by a factor of $\sim\!2$ for comparable $B/J_1$. We attribute this
difference to parton mean field theory underestimating the internal Zeeman field generated by $J_1$, 
which can significantly boost the gap as the uniform magnetization
increases. 

At high field, the gapped experimental data may be better thought of in terms of gapped spin-$1$ magnons
around a polarized state; in the parton theory, the magnons arise most simply as
bound states of the spin-$1/2$ partons. However, the full theory of this magnon-parton response needs a more careful discussion of
confinement effects which deserves a separate investigation.
%The corresponding plots for out-of-plane Zeeman fields are shown in the SI.

\subsection{Low temperature thermal conductivity}

\begin{figure}[t]
\hspace{-0.4cm} \includegraphics[width=0.45\textwidth]{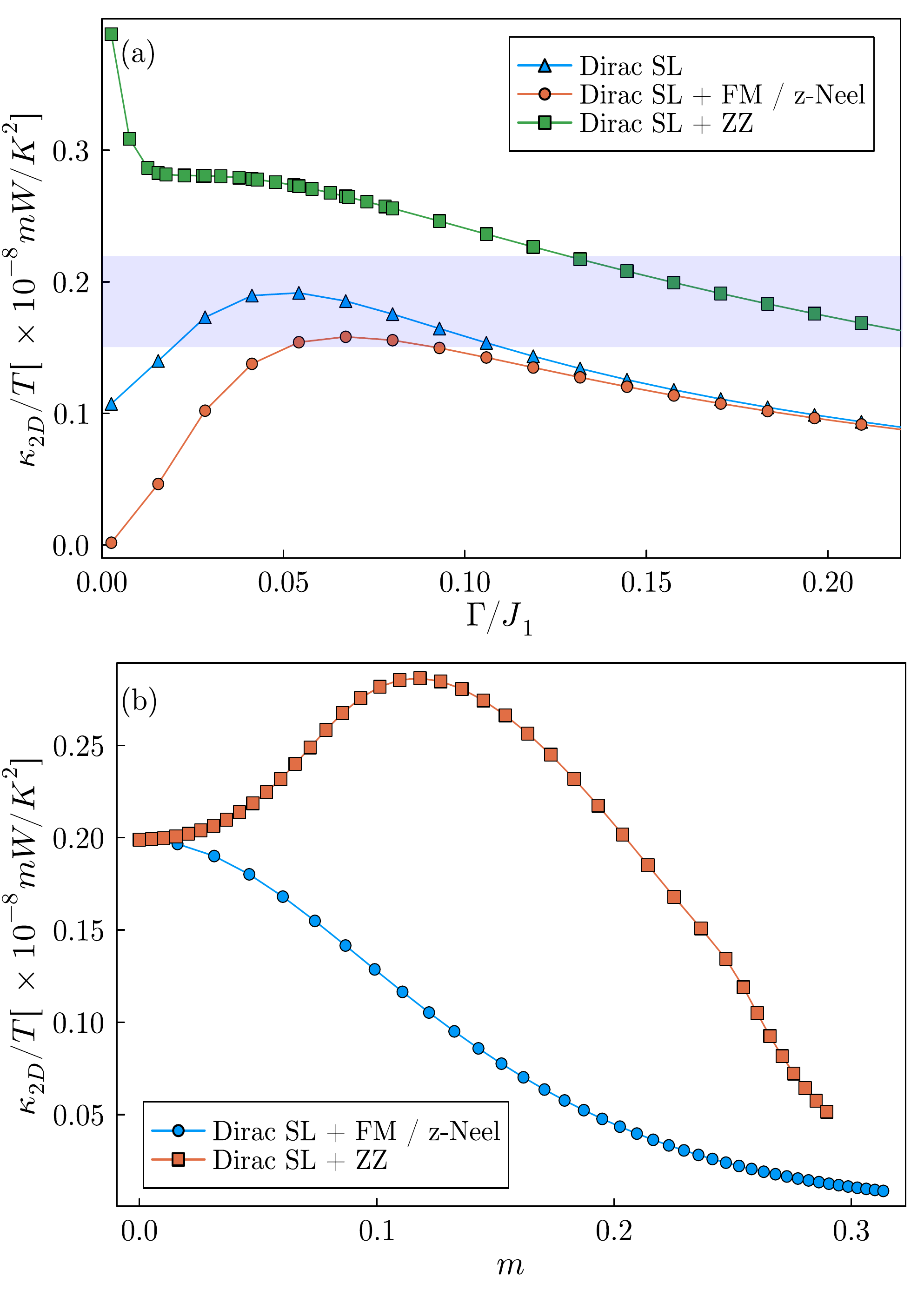}
    \caption{2D thermal conductivity $\kappa_{\rm avg}/T = (\kappa_{xx}+\kappa_{yy})/2T$ (in physical units) 
    for the Dirac SL from parton mean field theory 
    with $t_3/t_1\!=\!0.2$ and weak disorder induced broadening $\Gamma$.
    (a) $\kappa_{\rm avg}/T$ 
    as a function disorder-induced broadening \(\Gamma\) for (i) the Dirac SL, (ii) for Dirac SL with
    coexisting $z$-N\'eel or FM order with fixed order parameter of \(m = 0.1\) (both yield almost identical result), 
    and (iii) coexisting ZZ order with fixed order parameter of \(m = 0.1\). The purple band indicates
    the range of experimental values of ${\kappa}/T$ in {\ncto}.
    (b) $\kappa_{\rm avg}$ for Dirac SL with fixed disorder broadening \(\Gamma = 0.025 J_1\)
    and coexisting magnetic order ($z$-N\'eel/FM or ZZ) with varying order parameter $m$.}
    \label{fig:kappa}
\end{figure}
Within parton mean field theory, the low temperature 2D longitudinal thermal conductivity may be obtained indirectly by 
assigning a conserved `electric charge' $e_\star$ to the spinons, computing the bubble diagram for the
2D longitudinal `electrical conductivity' 
$\sigma \propto e_\star^2$, and appealing to the Wiedemann-Franz law to extract $\kappa/T \!=\! \sigma L$ 
where the Lorenz number $L\!=\!\pi^2 k_B^2/3 e_\star^2$; the charge $e_\star$ thus drops out of the final result for $\kappa/T$.
We obtain $\sigma$ using the bubble diagram with disorder-broadened Green 
functions,
\begin{equation}
    \text{Re}(\s^{\m\n}) = -\frac{e^{2}_\star}{\hbar}\sum_{\Vec{k}}\text{Tr}\left[v^{\m}(\Vec{k})A(\Vec{k},0)v^{\n}(\Vec{k})A(\Vec{k}, 0)\right]
\end{equation}
where $v^{\m}(\Vec{k}) = \partial H/\partial\bk_\m$ are velocity matrices, with $H$ being the parton mean field Hamiltonian
for the Dirac SL (or Dirac SL with coexisting ZZ or $z$-N\'eel order),
and \(A(\Vec{k}, \w)\) is the spectral function matrix.
The source of this disorder might be, for example, Na 
positional disorder in {\ncto} \cite{NCTO_samarakoon_PRB2021}. The inferred $\kappa/T$ is plotted in 
Fig.~\ref{fig:kappa}(a) as a function of the disorder broadening $\Gamma$ for the Dirac SL, as well as for
the Dirac SL with coexisting FM, or $z$-N\'eel, or ZZ order. For comparison, Fig.~\ref{fig:kappa}(a) also
shows the band of experimentally reported values in {\ncto} \cite{kappa_NCTO_Sun_PRB_2023} at zero field 
in the ZZ phase. Here, we have multiplied the measured 
3D value of $\kappa/T \! \approx\!0.03$-$0.04$\,mW/K$^2$cm by 
the honeycomb interlayer spacing $d_c\!=\!5.63 \text{\r{A}}$ for {\ncto} 
\cite{NCTO_lattice_JiangACS2019} to get the experimental result for $\kappa/T$ per 2D honeycomb plane. 
We find reasonable agreement between experiment and theory (for `Dirac SL + ZZ') over a wide range of $\Gamma/J_1$.
Metallic thermal conductivity
has also been reported in {\bcao} \cite{BCAO_thermalcond_Li2022}, but only when the
magnetic order is suppressed to reveal a window of SL. For {\bcao}, $d_c\!=\!7.83 \text{\r{A}}$ and
the metallic $\kappa/T \!\approx\! 0.03$-$0.06$\,mW K$^{-2}$cm$^{-1}$ \cite{BCAO_thermalcond_Li2022}, 
which leads to the 2D $\kappa/T \!\approx\! 0.23$-$0.46 \times 10^{-8}$\,mW K$^{-2}$ per honeycomb plane, 
in rough agreement with the theoretical estimate for the Dirac SL (over a range of $\Gamma$) in Fig.~\ref{fig:kappa}(a).

Fig.~\ref{fig:kappa}(b) shows the dependence of $\kappa/T$ per honeycomb layer on the strength of coexisting 
FM, or $z$-N\'eel, or ZZ order parameter. While the FM and $z$-N\'eel orders monotonically suppress the thermal 
conductivity due to gapping of the Dirac nodes (see Figs.~\ref{fig:bands})(a) and ~\ref{fig:bands})(b)), the
dependence on the ZZ order parameter is non-monotonic since the Dirac nodes persist, and additional quadratic
band touching points appear in the dispersion, before the spinons get fully gapped; 
cobaltates with more strong ZZ order should thus have a very strongly suppressed $\kappa/T$ at
low temperature.

\section{Discussion} 

In summary, we have used ED and DMRG calculations to show
that the spin-$1/2$ $J_1$-$J_3$ honeycomb lattice XXZ model
exhibits easy-plane FM order at small $J_3$ and ZZ order at large 
$J_3$, separated by a phase which exhibits spin-liquid-like spin correlations at intermediate $J_3/J_1$. This
intervening phase persists in the presence of weak compass anisotropy.
Using a modified parton mean field theory, supported by a Gutzwiller
wavefunction study, we have identified a candidate Dirac SL state which might be a parent of the numerically
observed {\sl}. The parent Dirac SL was shown to qualitatively capture the observed
temperature and field dependent broad continuum in THz response of BaCo$_2$(AsO$_4$)$_2$ \cite{Armitage2022}, as well as 
a linear-in-T thermal conductivity even in the presence of coexisting weak symmetry breaking orders if we account
for disorder induced broadening of the parton spectral function.
The presence of low energy excitations along the entire hexagonal BZ edge in the Dirac SL can also
allow a variety of magnetic orders to be energetically competitive and to be stabilized by 
small changes in the exchange couplings. This provides a parton theory perspective on the diversity of magnetic 
orders observed in the honeycomb cobaltates.

In future studies, it would be important to study the impact of gauge
fluctuations on the dynamical spin response and thermal conductivity presented here. In addition, it would be 
useful to explore how nonlinear spin wave theory and magnon interaction effects \cite{spinwaveXXZ_chernyshev_PRB2022}
might possibly also lead to SL type signatures in the excitation spectra and how they might also provide a description
of thermal conductivity experiments.
%Finally, going beyond mean field theory, we expect all the SL regimes with $B\!\neq\!0$, which are $U(1)$ 
%liquids with fully gapped parton excitations, to be unstable to confinement via monopole proliferation \cite{QSLgauge_XGWenBook}; the 
%nature of these confining phases remains to be clarified.
On the experimental front, looking for THz signatures of low energy excitations in {\ncto} would provide support
for the coexistence of ZZ order and Dirac partons.
Investigating these connections
would lead to a further understanding of the physics observed in these remarkable quantum magnets.

\section*{Acknowledgements}
We thank Federico Becca, Sasha Chernyshev, and Martin Klanjsek for extremely useful discussions.
This research was funded by NSERC of Canada (AB, SV, AP), DST India (MR, SS, MK, and TS-D), and
a SERB-India Vajra Fellowship VJR/2019/000076 (AP, TS-D). 
TS-D acknowledges a J.C.Bose National Fellowship (grant no. JCB/2020/000004) 
for funding. MK acknowledges support from SERB through grant no. CRG/2020/000754.
The parton and VMC computations were carried out on the Niagara supercomputer at the SciNet 
HPC Consortium and the Digital Research Alliance of Canada. AP acknowledges helpful conversations with
Ciaran Hickey, Stephan Rachel, Yasir Iqbal, Matthew Fisher, Oleg Starykh, and Cenke Xu 
during the KITP workshop ``A New Spin on Quantum Magnets'', supported in part
by the National Science Foundation under Grants No. NSF PHY-1748958 and PHY-2309135.

\bibliography{XY.bib}

\begin{thebibliography}{93}%
\makeatletter
\providecommand \@ifxundefined [1]{%
 \@ifx{#1\undefined}
}%
\providecommand \@ifnum [1]{%
 \ifnum #1\expandafter \@firstoftwo
 \else \expandafter \@secondoftwo
 \fi
}%
\providecommand \@ifx [1]{%
 \ifx #1\expandafter \@firstoftwo
 \else \expandafter \@secondoftwo
 \fi
}%
\providecommand \natexlab [1]{#1}%
\providecommand \enquote  [1]{``#1''}%
\providecommand \bibnamefont  [1]{#1}%
\providecommand \bibfnamefont [1]{#1}%
\providecommand \citenamefont [1]{#1}%
\providecommand \href@noop [0]{\@secondoftwo}%
\providecommand \href [0]{\begingroup \@sanitize@url \@href}%
\providecommand \@href[1]{\@@startlink{#1}\@@href}%
\providecommand \@@href[1]{\endgroup#1\@@endlink}%
\providecommand \@sanitize@url [0]{\catcode `\\12\catcode `\$12\catcode
  `\&12\catcode `\#12\catcode `\^12\catcode `\_12\catcode `\%12\relax}%
\providecommand \@@startlink[1]{}%
\providecommand \@@endlink[0]{}%
\providecommand \url  [0]{\begingroup\@sanitize@url \@url }%
\providecommand \@url [1]{\endgroup\@href {#1}{\urlprefix }}%
\providecommand \urlprefix  [0]{URL }%
\providecommand \Eprint [0]{\href }%
\providecommand \doibase [0]{https://doi.org/}%
\providecommand \selectlanguage [0]{\@gobble}%
\providecommand \bibinfo  [0]{\@secondoftwo}%
\providecommand \bibfield  [0]{\@secondoftwo}%
\providecommand \translation [1]{[#1]}%
\providecommand \BibitemOpen [0]{}%
\providecommand \bibitemStop [0]{}%
\providecommand \bibitemNoStop [0]{.\EOS\space}%
\providecommand \EOS [0]{\spacefactor3000\relax}%
\providecommand \BibitemShut  [1]{\csname bibitem#1\endcsname}%
\let\auto@bib@innerbib\@empty
%</preamble>
\bibitem [{\citenamefont {Wen}(2007)}]{QSLgauge_XGWenBook}%
  \BibitemOpen
  \bibfield  {author} {\bibinfo {author} {\bibfnamefont {X.~G.}\ \bibnamefont
  {Wen}},\ }\href {https://doi.org/10.1093/acprof:oso/9780199227259.001.0001}
  {\emph {\bibinfo {title} {{Quantum field theory of many-body systems: from
  the origin of sound to an origin of light and electrons}}}}\ (\bibinfo
  {publisher} {Oxford University Press},\ \bibinfo {address} {Oxford},\
  \bibinfo {year} {2007})\BibitemShut {NoStop}%
\bibitem [{\citenamefont {Lee}(2014)}]{QSLgauge_Lee2014}%
  \BibitemOpen
  \bibfield  {author} {\bibinfo {author} {\bibfnamefont {P.~A.}\ \bibnamefont
  {Lee}},\ }\bibfield  {title} {\bibinfo {title} {Quantum spin liquid: a tale
  of emergence from frustration},\ }\href
  {https://doi.org/10.1088/1742-6596/529/1/012001} {\bibfield  {journal}
  {\bibinfo  {journal} {Journal of Physics: Conference Series}\ }\textbf
  {\bibinfo {volume} {529}},\ \bibinfo {pages} {012001} (\bibinfo {year}
  {2014})}\BibitemShut {NoStop}%
\bibitem [{\citenamefont {Savary}\ and\ \citenamefont
  {Balents}(2016)}]{QSLreview_Savary_2017}%
  \BibitemOpen
  \bibfield  {author} {\bibinfo {author} {\bibfnamefont {L.}~\bibnamefont
  {Savary}}\ and\ \bibinfo {author} {\bibfnamefont {L.}~\bibnamefont
  {Balents}},\ }\bibfield  {title} {\bibinfo {title} {Quantum spin liquids: a
  review},\ }\href {https://doi.org/10.1088/0034-4885/80/1/016502} {\bibfield
  {journal} {\bibinfo  {journal} {Reports on Progress in Physics}\ }\textbf
  {\bibinfo {volume} {80}},\ \bibinfo {pages} {016502} (\bibinfo {year}
  {2016})}\BibitemShut {NoStop}%
\bibitem [{\citenamefont {Broholm}\ \emph {et~al.}(2020)\citenamefont
  {Broholm}, \citenamefont {Cava}, \citenamefont {Kivelson}, \citenamefont
  {Nocera}, \citenamefont {Norman},\ and\ \citenamefont
  {Senthil}}]{QSLreview_Broholm2020}%
  \BibitemOpen
  \bibfield  {author} {\bibinfo {author} {\bibfnamefont {C.}~\bibnamefont
  {Broholm}}, \bibinfo {author} {\bibfnamefont {R.~J.}\ \bibnamefont {Cava}},
  \bibinfo {author} {\bibfnamefont {S.~A.}\ \bibnamefont {Kivelson}}, \bibinfo
  {author} {\bibfnamefont {D.~G.}\ \bibnamefont {Nocera}}, \bibinfo {author}
  {\bibfnamefont {M.~R.}\ \bibnamefont {Norman}},\ and\ \bibinfo {author}
  {\bibfnamefont {T.}~\bibnamefont {Senthil}},\ }\bibfield  {title} {\bibinfo
  {title} {Quantum spin liquids},\ }\href
  {https://doi.org/10.1126/science.aay0668} {\bibfield  {journal} {\bibinfo
  {journal} {Science}\ }\textbf {\bibinfo {volume} {367}},\ \bibinfo {pages}
  {eaay0668} (\bibinfo {year} {2020})}\BibitemShut {NoStop}%
\bibitem [{\citenamefont {Knolle}\ and\ \citenamefont
  {Moessner}(2019)}]{QSLreview_Knolle2019}%
  \BibitemOpen
  \bibfield  {author} {\bibinfo {author} {\bibfnamefont {J.}~\bibnamefont
  {Knolle}}\ and\ \bibinfo {author} {\bibfnamefont {R.}~\bibnamefont
  {Moessner}},\ }\bibfield  {title} {\bibinfo {title} {A field guide to spin
  liquids},\ }\href {https://doi.org/10.1146/annurev-conmatphys-031218-013401}
  {\bibfield  {journal} {\bibinfo  {journal} {Annual Review of Condensed Matter
  Physics}\ }\textbf {\bibinfo {volume} {10}},\ \bibinfo {pages} {451}
  (\bibinfo {year} {2019})}\BibitemShut {NoStop}%
\bibitem [{\citenamefont {Anderson}(1973)}]{rvb_anderson1973}%
  \BibitemOpen
  \bibfield  {author} {\bibinfo {author} {\bibfnamefont {P.}~\bibnamefont
  {Anderson}},\ }\bibfield  {title} {\bibinfo {title} {Resonating valence
  bonds: A new kind of insulator?},\ }\href
  {https://doi.org/https://doi.org/10.1016/0025-5408(73)90167-0} {\bibfield
  {journal} {\bibinfo  {journal} {Materials Research Bulletin}\ }\textbf
  {\bibinfo {volume} {8}},\ \bibinfo {pages} {153} (\bibinfo {year}
  {1973})}\BibitemShut {NoStop}%
\bibitem [{\citenamefont {Moessner}\ and\ \citenamefont
  {Sondhi}(2001)}]{rvb_moessner_prl2001}%
  \BibitemOpen
  \bibfield  {author} {\bibinfo {author} {\bibfnamefont {R.}~\bibnamefont
  {Moessner}}\ and\ \bibinfo {author} {\bibfnamefont {S.~L.}\ \bibnamefont
  {Sondhi}},\ }\bibfield  {title} {\bibinfo {title} {Resonating valence bond
  phase in the triangular lattice quantum dimer model},\ }\href
  {https://doi.org/10.1103/PhysRevLett.86.1881} {\bibfield  {journal} {\bibinfo
   {journal} {Phys. Rev. Lett.}\ }\textbf {\bibinfo {volume} {86}},\ \bibinfo
  {pages} {1881} (\bibinfo {year} {2001})}\BibitemShut {NoStop}%
\bibitem [{\citenamefont {Anderson}(1987)}]{hightc_anderson_science1987}%
  \BibitemOpen
  \bibfield  {author} {\bibinfo {author} {\bibfnamefont {P.~W.}\ \bibnamefont
  {Anderson}},\ }\bibfield  {title} {\bibinfo {title} {The resonating valence
  bond state in {La$_2$CuO$_4$} and superconductivity},\ }\href
  {https://doi.org/10.1126/science.235.4793.1196} {\bibfield  {journal}
  {\bibinfo  {journal} {Science}\ }\textbf {\bibinfo {volume} {235}},\ \bibinfo
  {pages} {1196} (\bibinfo {year} {1987})}\BibitemShut {NoStop}%
\bibitem [{\citenamefont {Rokhsar}\ and\ \citenamefont
  {Kivelson}(1988)}]{rvb_rokhsarkivelson_prl1988}%
  \BibitemOpen
  \bibfield  {author} {\bibinfo {author} {\bibfnamefont {D.~S.}\ \bibnamefont
  {Rokhsar}}\ and\ \bibinfo {author} {\bibfnamefont {S.~A.}\ \bibnamefont
  {Kivelson}},\ }\bibfield  {title} {\bibinfo {title} {Superconductivity and
  the quantum hard-core dimer gas},\ }\href
  {https://doi.org/10.1103/PhysRevLett.61.2376} {\bibfield  {journal} {\bibinfo
   {journal} {Phys. Rev. Lett.}\ }\textbf {\bibinfo {volume} {61}},\ \bibinfo
  {pages} {2376} (\bibinfo {year} {1988})}\BibitemShut {NoStop}%
\bibitem [{\citenamefont {Kitaev}(2006)}]{KitaevModel_Kitaev2006}%
  \BibitemOpen
  \bibfield  {author} {\bibinfo {author} {\bibfnamefont {A.}~\bibnamefont
  {Kitaev}},\ }\bibfield  {title} {\bibinfo {title} {Anyons in an exactly
  solved model and beyond},\ }\href
  {https://doi.org/https://doi.org/10.1016/j.aop.2005.10.005} {\bibfield
  {journal} {\bibinfo  {journal} {Annals of Physics}\ }\textbf {\bibinfo
  {volume} {321}},\ \bibinfo {pages} {2} (\bibinfo {year} {2006})}\BibitemShut
  {NoStop}%
\bibitem [{\citenamefont {Helton}\ \emph {et~al.}(2007)\citenamefont {Helton},
  \citenamefont {Matan}, \citenamefont {Shores}, \citenamefont {Nytko},
  \citenamefont {Bartlett}, \citenamefont {Yoshida}, \citenamefont {Takano},
  \citenamefont {Suslov}, \citenamefont {Qiu}, \citenamefont {Chung},
  \citenamefont {Nocera},\ and\ \citenamefont {Lee}}]{kagome_neutron_Lee2007}%
  \BibitemOpen
  \bibfield  {author} {\bibinfo {author} {\bibfnamefont {J.~S.}\ \bibnamefont
  {Helton}}, \bibinfo {author} {\bibfnamefont {K.}~\bibnamefont {Matan}},
  \bibinfo {author} {\bibfnamefont {M.~P.}\ \bibnamefont {Shores}}, \bibinfo
  {author} {\bibfnamefont {E.~A.}\ \bibnamefont {Nytko}}, \bibinfo {author}
  {\bibfnamefont {B.~M.}\ \bibnamefont {Bartlett}}, \bibinfo {author}
  {\bibfnamefont {Y.}~\bibnamefont {Yoshida}}, \bibinfo {author} {\bibfnamefont
  {Y.}~\bibnamefont {Takano}}, \bibinfo {author} {\bibfnamefont
  {A.}~\bibnamefont {Suslov}}, \bibinfo {author} {\bibfnamefont
  {Y.}~\bibnamefont {Qiu}}, \bibinfo {author} {\bibfnamefont {J.-H.}\
  \bibnamefont {Chung}}, \bibinfo {author} {\bibfnamefont {D.~G.}\ \bibnamefont
  {Nocera}},\ and\ \bibinfo {author} {\bibfnamefont {Y.~S.}\ \bibnamefont
  {Lee}},\ }\bibfield  {title} {\bibinfo {title} {Spin dynamics of the
  spin-$1/2$ kagome lattice antiferromagnet
  {${\mathrm{ZnCu}}_{3}(\mathrm{OH}{)}_{6}{\mathrm{Cl}}_{2}$}},\ }\href
  {https://doi.org/10.1103/PhysRevLett.98.107204} {\bibfield  {journal}
  {\bibinfo  {journal} {Phys. Rev. Lett.}\ }\textbf {\bibinfo {volume} {98}},\
  \bibinfo {pages} {107204} (\bibinfo {year} {2007})}\BibitemShut {NoStop}%
\bibitem [{\citenamefont {Mendels}\ and\ \citenamefont
  {Bert}(2016)}]{kagome_review_mendels2016}%
  \BibitemOpen
  \bibfield  {author} {\bibinfo {author} {\bibfnamefont {P.}~\bibnamefont
  {Mendels}}\ and\ \bibinfo {author} {\bibfnamefont {F.}~\bibnamefont {Bert}},\
  }\bibfield  {title} {\bibinfo {title} {Quantum kagome frustrated
  antiferromagnets: {One} route to quantum spin liquids},\ }\href
  {https://doi.org/https://doi.org/10.1016/j.crhy.2015.12.001} {\bibfield
  {journal} {\bibinfo  {journal} {Comptes Rendus Physique}\ }\textbf {\bibinfo
  {volume} {17}},\ \bibinfo {pages} {455} (\bibinfo {year} {2016})}\BibitemShut
  {NoStop}%
\bibitem [{\citenamefont {Ran}\ \emph {et~al.}(2007)\citenamefont {Ran},
  \citenamefont {Hermele}, \citenamefont {Lee},\ and\ \citenamefont
  {Wen}}]{kagome_diracvmc_hermele2007}%
  \BibitemOpen
  \bibfield  {author} {\bibinfo {author} {\bibfnamefont {Y.}~\bibnamefont
  {Ran}}, \bibinfo {author} {\bibfnamefont {M.}~\bibnamefont {Hermele}},
  \bibinfo {author} {\bibfnamefont {P.~A.}\ \bibnamefont {Lee}},\ and\ \bibinfo
  {author} {\bibfnamefont {X.-G.}\ \bibnamefont {Wen}},\ }\bibfield  {title}
  {\bibinfo {title} {Projected-wave-function study of the spin-$1/2$
  {Heisenberg} model on the kagom\'e lattice},\ }\href
  {https://doi.org/10.1103/PhysRevLett.98.117205} {\bibfield  {journal}
  {\bibinfo  {journal} {Phys. Rev. Lett.}\ }\textbf {\bibinfo {volume} {98}},\
  \bibinfo {pages} {117205} (\bibinfo {year} {2007})}\BibitemShut {NoStop}%
\bibitem [{\citenamefont {Iqbal}\ \emph {et~al.}(2013)\citenamefont {Iqbal},
  \citenamefont {Becca}, \citenamefont {Sorella},\ and\ \citenamefont
  {Poilblanc}}]{kagome_diracvmc_iqbal2013}%
  \BibitemOpen
  \bibfield  {author} {\bibinfo {author} {\bibfnamefont {Y.}~\bibnamefont
  {Iqbal}}, \bibinfo {author} {\bibfnamefont {F.}~\bibnamefont {Becca}},
  \bibinfo {author} {\bibfnamefont {S.}~\bibnamefont {Sorella}},\ and\ \bibinfo
  {author} {\bibfnamefont {D.}~\bibnamefont {Poilblanc}},\ }\bibfield  {title}
  {\bibinfo {title} {Gapless spin-liquid phase in the kagome spin-$\frac{1}{2}$
  {Heisenberg} antiferromagnet},\ }\href
  {https://doi.org/10.1103/PhysRevB.87.060405} {\bibfield  {journal} {\bibinfo
  {journal} {Phys. Rev. B}\ }\textbf {\bibinfo {volume} {87}},\ \bibinfo
  {pages} {060405} (\bibinfo {year} {2013})}\BibitemShut {NoStop}%
\bibitem [{\citenamefont {He}\ \emph {et~al.}(2017)\citenamefont {He},
  \citenamefont {Zaletel}, \citenamefont {Oshikawa},\ and\ \citenamefont
  {Pollmann}}]{kagome_diracdmrg_pollmann2017}%
  \BibitemOpen
  \bibfield  {author} {\bibinfo {author} {\bibfnamefont {Y.-C.}\ \bibnamefont
  {He}}, \bibinfo {author} {\bibfnamefont {M.~P.}\ \bibnamefont {Zaletel}},
  \bibinfo {author} {\bibfnamefont {M.}~\bibnamefont {Oshikawa}},\ and\
  \bibinfo {author} {\bibfnamefont {F.}~\bibnamefont {Pollmann}},\ }\bibfield
  {title} {\bibinfo {title} {Signatures of {Dirac} cones in a {DMRG} study of
  the kagome {Heisenberg} model},\ }\href
  {https://doi.org/10.1103/PhysRevX.7.031020} {\bibfield  {journal} {\bibinfo
  {journal} {Phys. Rev. X}\ }\textbf {\bibinfo {volume} {7}},\ \bibinfo {pages}
  {031020} (\bibinfo {year} {2017})}\BibitemShut {NoStop}%
\bibitem [{\citenamefont {Okamoto}\ \emph {et~al.}(2007)\citenamefont
  {Okamoto}, \citenamefont {Nohara}, \citenamefont {Aruga-Katori},\ and\
  \citenamefont {Takagi}}]{hyperkagome_Takagi_PRL2007}%
  \BibitemOpen
  \bibfield  {author} {\bibinfo {author} {\bibfnamefont {Y.}~\bibnamefont
  {Okamoto}}, \bibinfo {author} {\bibfnamefont {M.}~\bibnamefont {Nohara}},
  \bibinfo {author} {\bibfnamefont {H.}~\bibnamefont {Aruga-Katori}},\ and\
  \bibinfo {author} {\bibfnamefont {H.}~\bibnamefont {Takagi}},\ }\bibfield
  {title} {\bibinfo {title} {Spin-liquid state in the $s=1/2$ hyperkagome
  antiferromagnet {${\mathrm{Na}}_{4}{\mathrm{Ir}}_{3}{\mathrm{O}}_{8}$}},\
  }\href {https://doi.org/10.1103/PhysRevLett.99.137207} {\bibfield  {journal}
  {\bibinfo  {journal} {Phys. Rev. Lett.}\ }\textbf {\bibinfo {volume} {99}},\
  \bibinfo {pages} {137207} (\bibinfo {year} {2007})}\BibitemShut {NoStop}%
\bibitem [{\citenamefont {Lawler}\ \emph {et~al.}(2008)\citenamefont {Lawler},
  \citenamefont {Paramekanti}, \citenamefont {Kim},\ and\ \citenamefont
  {Balents}}]{hyperkagome_Lawler_PRL2008}%
  \BibitemOpen
  \bibfield  {author} {\bibinfo {author} {\bibfnamefont {M.~J.}\ \bibnamefont
  {Lawler}}, \bibinfo {author} {\bibfnamefont {A.}~\bibnamefont {Paramekanti}},
  \bibinfo {author} {\bibfnamefont {Y.~B.}\ \bibnamefont {Kim}},\ and\ \bibinfo
  {author} {\bibfnamefont {L.}~\bibnamefont {Balents}},\ }\bibfield  {title}
  {\bibinfo {title} {Gapless spin liquids on the three-dimensional hyperkagome
  lattice of {${\mathrm{Na}}_{4}{\mathrm{Ir}}_{3}{\mathrm{O}}_{8}$}},\ }\href
  {https://doi.org/10.1103/PhysRevLett.101.197202} {\bibfield  {journal}
  {\bibinfo  {journal} {Phys. Rev. Lett.}\ }\textbf {\bibinfo {volume} {101}},\
  \bibinfo {pages} {197202} (\bibinfo {year} {2008})}\BibitemShut {NoStop}%
\bibitem [{\citenamefont {Zhou}\ \emph {et~al.}(2008)\citenamefont {Zhou},
  \citenamefont {Lee}, \citenamefont {Ng},\ and\ \citenamefont
  {Zhang}}]{hyperkagome_YiZhou_PRL2008}%
  \BibitemOpen
  \bibfield  {author} {\bibinfo {author} {\bibfnamefont {Y.}~\bibnamefont
  {Zhou}}, \bibinfo {author} {\bibfnamefont {P.~A.}\ \bibnamefont {Lee}},
  \bibinfo {author} {\bibfnamefont {T.-K.}\ \bibnamefont {Ng}},\ and\ \bibinfo
  {author} {\bibfnamefont {F.-C.}\ \bibnamefont {Zhang}},\ }\bibfield  {title}
  {\bibinfo {title} {{${\mathrm{Na}}_{4}{\mathrm{Ir}}_{3}{\mathrm{O}}_{8}$} as
  a {3D} spin liquid with fermionic spinons},\ }\href
  {https://doi.org/10.1103/PhysRevLett.101.197201} {\bibfield  {journal}
  {\bibinfo  {journal} {Phys. Rev. Lett.}\ }\textbf {\bibinfo {volume} {101}},\
  \bibinfo {pages} {197201} (\bibinfo {year} {2008})}\BibitemShut {NoStop}%
\bibitem [{\citenamefont {Ross}\ \emph {et~al.}(2011)\citenamefont {Ross},
  \citenamefont {Savary}, \citenamefont {Gaulin},\ and\ \citenamefont
  {Balents}}]{Pyrochlore_Ross_PRX2011}%
  \BibitemOpen
  \bibfield  {author} {\bibinfo {author} {\bibfnamefont {K.~A.}\ \bibnamefont
  {Ross}}, \bibinfo {author} {\bibfnamefont {L.}~\bibnamefont {Savary}},
  \bibinfo {author} {\bibfnamefont {B.~D.}\ \bibnamefont {Gaulin}},\ and\
  \bibinfo {author} {\bibfnamefont {L.}~\bibnamefont {Balents}},\ }\bibfield
  {title} {\bibinfo {title} {Quantum excitations in quantum spin ice},\ }\href
  {https://doi.org/10.1103/PhysRevX.1.021002} {\bibfield  {journal} {\bibinfo
  {journal} {Phys. Rev. X}\ }\textbf {\bibinfo {volume} {1}},\ \bibinfo {pages}
  {021002} (\bibinfo {year} {2011})}\BibitemShut {NoStop}%
\bibitem [{\citenamefont {Gingras}\ and\ \citenamefont
  {McClarty}(2014)}]{pyrochlore_review_gingras2014}%
  \BibitemOpen
  \bibfield  {author} {\bibinfo {author} {\bibfnamefont {M.~J.~P.}\
  \bibnamefont {Gingras}}\ and\ \bibinfo {author} {\bibfnamefont {P.~A.}\
  \bibnamefont {McClarty}},\ }\bibfield  {title} {\bibinfo {title} {Quantum
  spin ice: a search for gapless quantum spin liquids in pyrochlore magnets},\
  }\href {https://doi.org/10.1088/0034-4885/77/5/056501} {\bibfield  {journal}
  {\bibinfo  {journal} {Reports on Progress in Physics}\ }\textbf {\bibinfo
  {volume} {77}},\ \bibinfo {pages} {056501} (\bibinfo {year}
  {2014})}\BibitemShut {NoStop}%
\bibitem [{\citenamefont {Huang}\ \emph {et~al.}(2014)\citenamefont {Huang},
  \citenamefont {Chen},\ and\ \citenamefont
  {Hermele}}]{Pyrochlore_Hermele_PRL2014}%
  \BibitemOpen
  \bibfield  {author} {\bibinfo {author} {\bibfnamefont {Y.-P.}\ \bibnamefont
  {Huang}}, \bibinfo {author} {\bibfnamefont {G.}~\bibnamefont {Chen}},\ and\
  \bibinfo {author} {\bibfnamefont {M.}~\bibnamefont {Hermele}},\ }\bibfield
  {title} {\bibinfo {title} {Quantum spin ices and topological phases from
  dipolar-octupolar doublets on the pyrochlore lattice},\ }\href
  {https://doi.org/10.1103/PhysRevLett.112.167203} {\bibfield  {journal}
  {\bibinfo  {journal} {Phys. Rev. Lett.}\ }\textbf {\bibinfo {volume} {112}},\
  \bibinfo {pages} {167203} (\bibinfo {year} {2014})}\BibitemShut {NoStop}%
\bibitem [{\citenamefont {Sibille}\ \emph {et~al.}(2018)\citenamefont
  {Sibille}, \citenamefont {Gauthier}, \citenamefont {Yan}, \citenamefont
  {Ciomaga~Hatnean}, \citenamefont {Ollivier}, \citenamefont {Winn},
  \citenamefont {Filges}, \citenamefont {Balakrishnan}, \citenamefont
  {Kenzelmann}, \citenamefont {Shannon},\ and\ \citenamefont
  {Fennell}}]{Pyrochlore_Fennell_NatPhys2018}%
  \BibitemOpen
  \bibfield  {author} {\bibinfo {author} {\bibfnamefont {R.}~\bibnamefont
  {Sibille}}, \bibinfo {author} {\bibfnamefont {N.}~\bibnamefont {Gauthier}},
  \bibinfo {author} {\bibfnamefont {H.}~\bibnamefont {Yan}}, \bibinfo {author}
  {\bibfnamefont {M.}~\bibnamefont {Ciomaga~Hatnean}}, \bibinfo {author}
  {\bibfnamefont {J.}~\bibnamefont {Ollivier}}, \bibinfo {author}
  {\bibfnamefont {B.}~\bibnamefont {Winn}}, \bibinfo {author} {\bibfnamefont
  {U.}~\bibnamefont {Filges}}, \bibinfo {author} {\bibfnamefont
  {G.}~\bibnamefont {Balakrishnan}}, \bibinfo {author} {\bibfnamefont
  {M.}~\bibnamefont {Kenzelmann}}, \bibinfo {author} {\bibfnamefont
  {N.}~\bibnamefont {Shannon}},\ and\ \bibinfo {author} {\bibfnamefont
  {T.}~\bibnamefont {Fennell}},\ }\bibfield  {title} {\bibinfo {title}
  {Experimental signatures of emergent quantum electrodynamics in
  {Pr$_2$Hf$_2$O$_7$}},\ }\href {https://doi.org/10.1038/s41567-018-0116-x}
  {\bibfield  {journal} {\bibinfo  {journal} {Nature Physics}\ }\textbf
  {\bibinfo {volume} {14}},\ \bibinfo {pages} {711} (\bibinfo {year}
  {2018})}\BibitemShut {NoStop}%
\bibitem [{\citenamefont {Semeghini}\ \emph {et~al.}(2021)\citenamefont
  {Semeghini}, \citenamefont {Levine}, \citenamefont {Keesling}, \citenamefont
  {Ebadi}, \citenamefont {Wang}, \citenamefont {Bluvstein}, \citenamefont
  {Verresen}, \citenamefont {Pichler}, \citenamefont {Kalinowski},
  \citenamefont {Samajdar}, \citenamefont {Omran}, \citenamefont {Sachdev},
  \citenamefont {Vishwanath}, \citenamefont {Greiner}, \citenamefont
  {Vuletic},\ and\ \citenamefont {Lukin}}]{Rydberg_Lukin_Science2021}%
  \BibitemOpen
  \bibfield  {author} {\bibinfo {author} {\bibfnamefont {G.}~\bibnamefont
  {Semeghini}}, \bibinfo {author} {\bibfnamefont {H.}~\bibnamefont {Levine}},
  \bibinfo {author} {\bibfnamefont {A.}~\bibnamefont {Keesling}}, \bibinfo
  {author} {\bibfnamefont {S.}~\bibnamefont {Ebadi}}, \bibinfo {author}
  {\bibfnamefont {T.~T.}\ \bibnamefont {Wang}}, \bibinfo {author}
  {\bibfnamefont {D.}~\bibnamefont {Bluvstein}}, \bibinfo {author}
  {\bibfnamefont {R.}~\bibnamefont {Verresen}}, \bibinfo {author}
  {\bibfnamefont {H.}~\bibnamefont {Pichler}}, \bibinfo {author} {\bibfnamefont
  {M.}~\bibnamefont {Kalinowski}}, \bibinfo {author} {\bibfnamefont
  {R.}~\bibnamefont {Samajdar}}, \bibinfo {author} {\bibfnamefont
  {A.}~\bibnamefont {Omran}}, \bibinfo {author} {\bibfnamefont
  {S.}~\bibnamefont {Sachdev}}, \bibinfo {author} {\bibfnamefont
  {A.}~\bibnamefont {Vishwanath}}, \bibinfo {author} {\bibfnamefont
  {M.}~\bibnamefont {Greiner}}, \bibinfo {author} {\bibfnamefont
  {V.}~\bibnamefont {Vuletic}},\ and\ \bibinfo {author} {\bibfnamefont {M.~D.}\
  \bibnamefont {Lukin}},\ }\bibfield  {title} {\bibinfo {title} {Probing
  topological spin liquids on a programmable quantum simulator},\ }\href
  {https://doi.org/10.1126/science.abi8794} {\bibfield  {journal} {\bibinfo
  {journal} {Science}\ }\textbf {\bibinfo {volume} {374}},\ \bibinfo {pages}
  {1242} (\bibinfo {year} {2021})}\BibitemShut {NoStop}%
\bibitem [{\citenamefont {Hermanns}\ \emph {et~al.}(2018)\citenamefont
  {Hermanns}, \citenamefont {Kimchi},\ and\ \citenamefont
  {Knolle}}]{Kitaev_review_Hermanns2018}%
  \BibitemOpen
  \bibfield  {author} {\bibinfo {author} {\bibfnamefont {M.}~\bibnamefont
  {Hermanns}}, \bibinfo {author} {\bibfnamefont {I.}~\bibnamefont {Kimchi}},\
  and\ \bibinfo {author} {\bibfnamefont {J.}~\bibnamefont {Knolle}},\
  }\bibfield  {title} {\bibinfo {title} {Physics of the {Kitaev} model:
  Fractionalization, dynamic correlations, and material connections},\ }\href
  {https://doi.org/10.1146/annurev-conmatphys-033117-053934} {\bibfield
  {journal} {\bibinfo  {journal} {Annual Review of Condensed Matter Physics}\
  }\textbf {\bibinfo {volume} {9}},\ \bibinfo {pages} {17} (\bibinfo {year}
  {2018})}\BibitemShut {NoStop}%
\bibitem [{\citenamefont {Motome}\ and\ \citenamefont
  {Nasu}(2020)}]{Kitaev_review_Motome2020}%
  \BibitemOpen
  \bibfield  {author} {\bibinfo {author} {\bibfnamefont {Y.}~\bibnamefont
  {Motome}}\ and\ \bibinfo {author} {\bibfnamefont {J.}~\bibnamefont {Nasu}},\
  }\bibfield  {title} {\bibinfo {title} {Hunting {Majorana} fermions in
  {Kitaev} magnets},\ }\href {https://doi.org/10.7566/JPSJ.89.012002}
  {\bibfield  {journal} {\bibinfo  {journal} {Journal of the Physical Society
  of Japan}\ }\textbf {\bibinfo {volume} {89}},\ \bibinfo {pages} {012002}
  (\bibinfo {year} {2020})}\BibitemShut {NoStop}%
\bibitem [{\citenamefont {Jackeli}\ and\ \citenamefont
  {Khaliullin}(2009)}]{KitaevModel_Jackeli2009}%
  \BibitemOpen
  \bibfield  {author} {\bibinfo {author} {\bibfnamefont {G.}~\bibnamefont
  {Jackeli}}\ and\ \bibinfo {author} {\bibfnamefont {G.}~\bibnamefont
  {Khaliullin}},\ }\bibfield  {title} {\bibinfo {title} {Mott insulators in the
  strong spin-orbit coupling limit: From {Heisenberg} to a quantum compass and
  {Kitaev} models},\ }\href {https://doi.org/10.1103/PhysRevLett.102.017205}
  {\bibfield  {journal} {\bibinfo  {journal} {Phys. Rev. Lett.}\ }\textbf
  {\bibinfo {volume} {102}},\ \bibinfo {pages} {017205} (\bibinfo {year}
  {2009})}\BibitemShut {NoStop}%
\bibitem [{\citenamefont {Kimchi}\ and\ \citenamefont
  {Vishwanath}(2014)}]{iridates_kimchi2014}%
  \BibitemOpen
  \bibfield  {author} {\bibinfo {author} {\bibfnamefont {I.}~\bibnamefont
  {Kimchi}}\ and\ \bibinfo {author} {\bibfnamefont {A.}~\bibnamefont
  {Vishwanath}},\ }\bibfield  {title} {\bibinfo {title} {Kitaev-{Heisenberg}
  models for iridates on the triangular, hyperkagome, kagome, fcc, and
  pyrochlore lattices},\ }\href {https://doi.org/10.1103/PhysRevB.89.014414}
  {\bibfield  {journal} {\bibinfo  {journal} {Phys. Rev. B}\ }\textbf {\bibinfo
  {volume} {89}},\ \bibinfo {pages} {014414} (\bibinfo {year}
  {2014})}\BibitemShut {NoStop}%
\bibitem [{\citenamefont {Takagi}\ \emph {et~al.}(2019)\citenamefont {Takagi},
  \citenamefont {Takayama}, \citenamefont {Jackeli}, \citenamefont
  {Khaliullin},\ and\ \citenamefont {Nagler}}]{KitaevReview_Takagi2019}%
  \BibitemOpen
  \bibfield  {author} {\bibinfo {author} {\bibfnamefont {H.}~\bibnamefont
  {Takagi}}, \bibinfo {author} {\bibfnamefont {T.}~\bibnamefont {Takayama}},
  \bibinfo {author} {\bibfnamefont {G.}~\bibnamefont {Jackeli}}, \bibinfo
  {author} {\bibfnamefont {G.}~\bibnamefont {Khaliullin}},\ and\ \bibinfo
  {author} {\bibfnamefont {S.~E.}\ \bibnamefont {Nagler}},\ }\bibfield  {title}
  {\bibinfo {title} {Concept and realization of {Kitaev} quantum spin
  liquids},\ }\href {https://doi.org/10.1038/s42254-019-0038-2} {\bibfield
  {journal} {\bibinfo  {journal} {Nature Reviews Physics}\ }\textbf {\bibinfo
  {volume} {1}},\ \bibinfo {pages} {264} (\bibinfo {year} {2019})}\BibitemShut
  {NoStop}%
\bibitem [{\citenamefont {Trebst}\ and\ \citenamefont
  {Hickey}(2022)}]{Iridates_review_Trebst2022}%
  \BibitemOpen
  \bibfield  {author} {\bibinfo {author} {\bibfnamefont {S.}~\bibnamefont
  {Trebst}}\ and\ \bibinfo {author} {\bibfnamefont {C.}~\bibnamefont
  {Hickey}},\ }\bibfield  {title} {\bibinfo {title} {Kitaev materials},\ }\href
  {https://doi.org/https://doi.org/10.1016/j.physrep.2021.11.003} {\bibfield
  {journal} {\bibinfo  {journal} {Physics Reports}\ }\textbf {\bibinfo {volume}
  {950}},\ \bibinfo {pages} {1} (\bibinfo {year} {2022})}\BibitemShut {NoStop}%
\bibitem [{\citenamefont {Plumb}\ \emph {et~al.}(2014)\citenamefont {Plumb},
  \citenamefont {Clancy}, \citenamefont {Sandilands}, \citenamefont {Shankar},
  \citenamefont {Hu}, \citenamefont {Burch}, \citenamefont {Kee},\ and\
  \citenamefont {Kim}}]{RuCl3_SOC_Kim2014}%
  \BibitemOpen
  \bibfield  {author} {\bibinfo {author} {\bibfnamefont {K.~W.}\ \bibnamefont
  {Plumb}}, \bibinfo {author} {\bibfnamefont {J.~P.}\ \bibnamefont {Clancy}},
  \bibinfo {author} {\bibfnamefont {L.~J.}\ \bibnamefont {Sandilands}},
  \bibinfo {author} {\bibfnamefont {V.~V.}\ \bibnamefont {Shankar}}, \bibinfo
  {author} {\bibfnamefont {Y.~F.}\ \bibnamefont {Hu}}, \bibinfo {author}
  {\bibfnamefont {K.~S.}\ \bibnamefont {Burch}}, \bibinfo {author}
  {\bibfnamefont {H.-Y.}\ \bibnamefont {Kee}},\ and\ \bibinfo {author}
  {\bibfnamefont {Y.-J.}\ \bibnamefont {Kim}},\ }\bibfield  {title} {\bibinfo
  {title} {{$\alpha$-RuCl$_3$}: A spin-orbit assisted {Mott} insulator on a
  honeycomb lattice},\ }\href {https://doi.org/10.1103/PhysRevB.90.041112}
  {\bibfield  {journal} {\bibinfo  {journal} {Phys. Rev. B}\ }\textbf {\bibinfo
  {volume} {90}},\ \bibinfo {pages} {041112} (\bibinfo {year}
  {2014})}\BibitemShut {NoStop}%
\bibitem [{\citenamefont {Sandilands}\ \emph {et~al.}(2015)\citenamefont
  {Sandilands}, \citenamefont {Tian}, \citenamefont {Plumb}, \citenamefont
  {Kim},\ and\ \citenamefont {Burch}}]{RuCl3_Raman_Burch2015}%
  \BibitemOpen
  \bibfield  {author} {\bibinfo {author} {\bibfnamefont {L.~J.}\ \bibnamefont
  {Sandilands}}, \bibinfo {author} {\bibfnamefont {Y.}~\bibnamefont {Tian}},
  \bibinfo {author} {\bibfnamefont {K.~W.}\ \bibnamefont {Plumb}}, \bibinfo
  {author} {\bibfnamefont {Y.-J.}\ \bibnamefont {Kim}},\ and\ \bibinfo {author}
  {\bibfnamefont {K.~S.}\ \bibnamefont {Burch}},\ }\bibfield  {title} {\bibinfo
  {title} {Scattering continuum and possible fractionalized excitations in
  {$\alpha$-RuCl$_3$}},\ }\href
  {https://doi.org/10.1103/PhysRevLett.114.147201} {\bibfield  {journal}
  {\bibinfo  {journal} {Phys. Rev. Lett.}\ }\textbf {\bibinfo {volume} {114}},\
  \bibinfo {pages} {147201} (\bibinfo {year} {2015})}\BibitemShut {NoStop}%
\bibitem [{\citenamefont {Banerjee}\ \emph {et~al.}(2017)\citenamefont
  {Banerjee}, \citenamefont {Yan}, \citenamefont {Knolle}, \citenamefont
  {Bridges}, \citenamefont {Stone}, \citenamefont {Lumsden}, \citenamefont
  {Mandrus}, \citenamefont {Tennant}, \citenamefont {Moessner},\ and\
  \citenamefont {Nagler}}]{RuCl3_neutron_Nagler2017}%
  \BibitemOpen
  \bibfield  {author} {\bibinfo {author} {\bibfnamefont {A.}~\bibnamefont
  {Banerjee}}, \bibinfo {author} {\bibfnamefont {J.}~\bibnamefont {Yan}},
  \bibinfo {author} {\bibfnamefont {J.}~\bibnamefont {Knolle}}, \bibinfo
  {author} {\bibfnamefont {C.~A.}\ \bibnamefont {Bridges}}, \bibinfo {author}
  {\bibfnamefont {M.~B.}\ \bibnamefont {Stone}}, \bibinfo {author}
  {\bibfnamefont {M.~D.}\ \bibnamefont {Lumsden}}, \bibinfo {author}
  {\bibfnamefont {D.~G.}\ \bibnamefont {Mandrus}}, \bibinfo {author}
  {\bibfnamefont {D.~A.}\ \bibnamefont {Tennant}}, \bibinfo {author}
  {\bibfnamefont {R.}~\bibnamefont {Moessner}},\ and\ \bibinfo {author}
  {\bibfnamefont {S.~E.}\ \bibnamefont {Nagler}},\ }\bibfield  {title}
  {\bibinfo {title} {Neutron scattering in the proximate quantum spin liquid
  {$\alpha$-RuCl$_3$}},\ }\href {https://doi.org/10.1126/science.aah6015}
  {\bibfield  {journal} {\bibinfo  {journal} {Science}\ }\textbf {\bibinfo
  {volume} {356}},\ \bibinfo {pages} {1055} (\bibinfo {year}
  {2017})}\BibitemShut {NoStop}%
\bibitem [{\citenamefont {Baskaran}\ \emph {et~al.}(2008)\citenamefont
  {Baskaran}, \citenamefont {Sen},\ and\ \citenamefont
  {Shankar}}]{KitaevSpinS_Baskaran2008}%
  \BibitemOpen
  \bibfield  {author} {\bibinfo {author} {\bibfnamefont {G.}~\bibnamefont
  {Baskaran}}, \bibinfo {author} {\bibfnamefont {D.}~\bibnamefont {Sen}},\ and\
  \bibinfo {author} {\bibfnamefont {R.}~\bibnamefont {Shankar}},\ }\bibfield
  {title} {\bibinfo {title} {Spin-{$S$} {Kitaev} model: {Classical} ground
  states, order from disorder, and exact correlation functions},\ }\href
  {https://doi.org/10.1103/PhysRevB.78.115116} {\bibfield  {journal} {\bibinfo
  {journal} {Phys. Rev. B}\ }\textbf {\bibinfo {volume} {78}},\ \bibinfo
  {pages} {115116} (\bibinfo {year} {2008})}\BibitemShut {NoStop}%
\bibitem [{\citenamefont {Stavropoulos}\ \emph {et~al.}(2019)\citenamefont
  {Stavropoulos}, \citenamefont {Pereira},\ and\ \citenamefont
  {Kee}}]{KitaevSpinS_Kee2019}%
  \BibitemOpen
  \bibfield  {author} {\bibinfo {author} {\bibfnamefont {P.~P.}\ \bibnamefont
  {Stavropoulos}}, \bibinfo {author} {\bibfnamefont {D.}~\bibnamefont
  {Pereira}},\ and\ \bibinfo {author} {\bibfnamefont {H.-Y.}\ \bibnamefont
  {Kee}},\ }\bibfield  {title} {\bibinfo {title} {Microscopic mechanism for a
  higher-spin {Kitaev} model},\ }\href
  {https://doi.org/10.1103/PhysRevLett.123.037203} {\bibfield  {journal}
  {\bibinfo  {journal} {Phys. Rev. Lett.}\ }\textbf {\bibinfo {volume} {123}},\
  \bibinfo {pages} {037203} (\bibinfo {year} {2019})}\BibitemShut {NoStop}%
\bibitem [{\citenamefont {Liu}\ and\ \citenamefont
  {Khaliullin}(2018)}]{CoKitaev_Liu2018}%
  \BibitemOpen
  \bibfield  {author} {\bibinfo {author} {\bibfnamefont {H.}~\bibnamefont
  {Liu}}\ and\ \bibinfo {author} {\bibfnamefont {G.}~\bibnamefont
  {Khaliullin}},\ }\bibfield  {title} {\bibinfo {title} {Pseudospin exchange
  interactions in ${d}^{7}$ cobalt compounds: Possible realization of the
  {Kitaev} model},\ }\href {https://doi.org/10.1103/PhysRevB.97.014407}
  {\bibfield  {journal} {\bibinfo  {journal} {Phys. Rev. B}\ }\textbf {\bibinfo
  {volume} {97}},\ \bibinfo {pages} {014407} (\bibinfo {year}
  {2018})}\BibitemShut {NoStop}%
\bibitem [{\citenamefont {Liu}\ \emph {et~al.}(2020)\citenamefont {Liu},
  \citenamefont {Chaloupka},\ and\ \citenamefont
  {Khaliullin}}]{CoKitaev_Liu2020}%
  \BibitemOpen
  \bibfield  {author} {\bibinfo {author} {\bibfnamefont {H.}~\bibnamefont
  {Liu}}, \bibinfo {author} {\bibfnamefont {J.~c.~v.}\ \bibnamefont
  {Chaloupka}},\ and\ \bibinfo {author} {\bibfnamefont {G.}~\bibnamefont
  {Khaliullin}},\ }\bibfield  {title} {\bibinfo {title} {Kitaev spin liquid in
  $3d$ transition metal compounds},\ }\href
  {https://doi.org/10.1103/PhysRevLett.125.047201} {\bibfield  {journal}
  {\bibinfo  {journal} {Phys. Rev. Lett.}\ }\textbf {\bibinfo {volume} {125}},\
  \bibinfo {pages} {047201} (\bibinfo {year} {2020})}\BibitemShut {NoStop}%
\bibitem [{\citenamefont {Regnault}\ \emph {et~al.}(1977)\citenamefont
  {Regnault}, \citenamefont {Burlet},\ and\ \citenamefont
  {Rossat-Mignod}}]{BCAO_regnault1977}%
  \BibitemOpen
  \bibfield  {author} {\bibinfo {author} {\bibfnamefont {L.}~\bibnamefont
  {Regnault}}, \bibinfo {author} {\bibfnamefont {P.}~\bibnamefont {Burlet}},\
  and\ \bibinfo {author} {\bibfnamefont {J.}~\bibnamefont {Rossat-Mignod}},\
  }\bibfield  {title} {\bibinfo {title} {Magnetic ordering in a planar {XY}
  model: {BaCo$_2$(AsO$_4$)$_2$}},\ }\href
  {https://doi.org/https://doi.org/10.1016/0378-4363(77)90635-0} {\bibfield
  {journal} {\bibinfo  {journal} {Physica B+C}\ }\textbf {\bibinfo {volume}
  {86-88}},\ \bibinfo {pages} {660} (\bibinfo {year} {1977})}\BibitemShut
  {NoStop}%
\bibitem [{\citenamefont {Regnault}\ \emph {et~al.}(2018)\citenamefont
  {Regnault}, \citenamefont {Boullier},\ and\ \citenamefont
  {Lorenzo}}]{BCAO_Regnault2018}%
  \BibitemOpen
  \bibfield  {author} {\bibinfo {author} {\bibfnamefont {L.~P.}\ \bibnamefont
  {Regnault}}, \bibinfo {author} {\bibfnamefont {C.}~\bibnamefont {Boullier}},\
  and\ \bibinfo {author} {\bibfnamefont {J.~E.}\ \bibnamefont {Lorenzo}},\
  }\bibfield  {title} {\bibinfo {title} {Polarized-neutron investigation of
  magnetic ordering and spin dynamics in {BaCo$_{2}$(AsO$_{4}$)$_{2}$}
  frustrated honeycomb-lattice magnet},\ }\href
  {https://doi.org/10.1016/j.heliyon.2018.e00507} {\bibfield  {journal}
  {\bibinfo  {journal} {Heliyon}\ }\textbf {\bibinfo {volume} {4}},\ \bibinfo
  {pages} {E00507} (\bibinfo {year} {2018})}\BibitemShut {NoStop}%
\bibitem [{\citenamefont {Zhong}\ \emph {et~al.}(2019)\citenamefont {Zhong},
  \citenamefont {Gao}, \citenamefont {Ong},\ and\ \citenamefont
  {Cava}}]{BCAO_Zhong2019}%
  \BibitemOpen
  \bibfield  {author} {\bibinfo {author} {\bibfnamefont {R.}~\bibnamefont
  {Zhong}}, \bibinfo {author} {\bibfnamefont {T.}~\bibnamefont {Gao}}, \bibinfo
  {author} {\bibfnamefont {N.~P.}\ \bibnamefont {Ong}},\ and\ \bibinfo {author}
  {\bibfnamefont {R.~J.}\ \bibnamefont {Cava}},\ }\bibfield  {title} {\bibinfo
  {title} {Weak-field induced nonmagnetic state in a {Co}-based honeycomb},\
  }\href {https://doi.org/10.1126/sciadv.aay6953} {\bibfield  {journal}
  {\bibinfo  {journal} {Science Advances}\ }\textbf {\bibinfo {volume} {6
  (4)}},\ \bibinfo {pages} {aay6953} (\bibinfo {year} {2019})}\BibitemShut
  {NoStop}%
\bibitem [{\citenamefont {Zhang}\ \emph
  {et~al.}(2023{\natexlab{a}})\citenamefont {Zhang}, \citenamefont {Xu},
  \citenamefont {Halloran}, \citenamefont {Zhong}, \citenamefont {Broholm},
  \citenamefont {Cava}, \citenamefont {Drichko},\ and\ \citenamefont
  {Armitage}}]{Armitage2022}%
  \BibitemOpen
  \bibfield  {author} {\bibinfo {author} {\bibfnamefont {X.}~\bibnamefont
  {Zhang}}, \bibinfo {author} {\bibfnamefont {Y.}~\bibnamefont {Xu}}, \bibinfo
  {author} {\bibfnamefont {T.}~\bibnamefont {Halloran}}, \bibinfo {author}
  {\bibfnamefont {R.}~\bibnamefont {Zhong}}, \bibinfo {author} {\bibfnamefont
  {C.}~\bibnamefont {Broholm}}, \bibinfo {author} {\bibfnamefont {R.~J.}\
  \bibnamefont {Cava}}, \bibinfo {author} {\bibfnamefont {N.}~\bibnamefont
  {Drichko}},\ and\ \bibinfo {author} {\bibfnamefont {N.~P.}\ \bibnamefont
  {Armitage}},\ }\bibfield  {title} {\bibinfo {title} {A magnetic continuum in
  the cobalt-based honeycomb magnet {BaCo$_2$(AsO$_4$)$_2$}},\ }\href
  {https://doi.org/10.1038/s41563-022-01403-1} {\bibfield  {journal} {\bibinfo
  {journal} {Nature Materials}\ }\textbf {\bibinfo {volume} {22}},\ \bibinfo
  {pages} {58} (\bibinfo {year} {2023}{\natexlab{a}})}\BibitemShut {NoStop}%
\bibitem [{\citenamefont {Halloran}\ \emph {et~al.}(2023)\citenamefont
  {Halloran}, \citenamefont {Desrochers}, \citenamefont {Zhang}, \citenamefont
  {Chen}, \citenamefont {Chern}, \citenamefont {Xu}, \citenamefont {Winn},
  \citenamefont {Graves-Brook}, \citenamefont {Stone}, \citenamefont
  {Kolesnikov}, \citenamefont {Qiu}, \citenamefont {Zhong}, \citenamefont
  {Cava}, \citenamefont {Kim},\ and\ \citenamefont
  {Broholm}}]{BCAO_Broholm2023}%
  \BibitemOpen
  \bibfield  {author} {\bibinfo {author} {\bibfnamefont {T.}~\bibnamefont
  {Halloran}}, \bibinfo {author} {\bibfnamefont {F.}~\bibnamefont
  {Desrochers}}, \bibinfo {author} {\bibfnamefont {E.~Z.}\ \bibnamefont
  {Zhang}}, \bibinfo {author} {\bibfnamefont {T.}~\bibnamefont {Chen}},
  \bibinfo {author} {\bibfnamefont {L.~E.}\ \bibnamefont {Chern}}, \bibinfo
  {author} {\bibfnamefont {Z.}~\bibnamefont {Xu}}, \bibinfo {author}
  {\bibfnamefont {B.}~\bibnamefont {Winn}}, \bibinfo {author} {\bibfnamefont
  {M.}~\bibnamefont {Graves-Brook}}, \bibinfo {author} {\bibfnamefont {M.~B.}\
  \bibnamefont {Stone}}, \bibinfo {author} {\bibfnamefont {A.~I.}\ \bibnamefont
  {Kolesnikov}}, \bibinfo {author} {\bibfnamefont {Y.}~\bibnamefont {Qiu}},
  \bibinfo {author} {\bibfnamefont {R.}~\bibnamefont {Zhong}}, \bibinfo
  {author} {\bibfnamefont {R.}~\bibnamefont {Cava}}, \bibinfo {author}
  {\bibfnamefont {Y.~B.}\ \bibnamefont {Kim}},\ and\ \bibinfo {author}
  {\bibfnamefont {C.}~\bibnamefont {Broholm}},\ }\bibfield  {title} {\bibinfo
  {title} {Geometrical frustration versus {Kitaev} interactions in
  {BaCo$_2$(AsO$_4$)$_2$}},\ }\href {https://doi.org/10.1073/pnas.2215509119}
  {\bibfield  {journal} {\bibinfo  {journal} {Proceedings of the National
  Academy of Sciences}\ }\textbf {\bibinfo {volume} {120}},\ \bibinfo {pages}
  {e2215509119} (\bibinfo {year} {2023})}\BibitemShut {NoStop}%
\bibitem [{\citenamefont {{Tu}}\ \emph {et~al.}(2022)\citenamefont {{Tu}},
  \citenamefont {{Dai}}, \citenamefont {{Zhang}}, \citenamefont {{Zhao}},
  \citenamefont {{Jin}}, \citenamefont {{Gao}}, \citenamefont {{Dai}},\ and\
  \citenamefont {{Li}}}]{BCAO_thermalcond_Li2022}%
  \BibitemOpen
  \bibfield  {author} {\bibinfo {author} {\bibfnamefont {C.}~\bibnamefont
  {{Tu}}}, \bibinfo {author} {\bibfnamefont {D.}~\bibnamefont {{Dai}}},
  \bibinfo {author} {\bibfnamefont {X.}~\bibnamefont {{Zhang}}}, \bibinfo
  {author} {\bibfnamefont {C.}~\bibnamefont {{Zhao}}}, \bibinfo {author}
  {\bibfnamefont {X.}~\bibnamefont {{Jin}}}, \bibinfo {author} {\bibfnamefont
  {B.}~\bibnamefont {{Gao}}}, \bibinfo {author} {\bibfnamefont
  {P.}~\bibnamefont {{Dai}}},\ and\ \bibinfo {author} {\bibfnamefont
  {S.}~\bibnamefont {{Li}}},\ }\bibfield  {title} {\bibinfo {title} {{Evidence
  for gapless quantum spin liquid in a honeycomb lattice}},\ }\href@noop {}
  {\bibfield  {journal} {\bibinfo  {journal} {arXiv e-prints}\ ,\ \bibinfo
  {eid} {arXiv:2212.07322}} (\bibinfo {year} {2022})},\ \Eprint
  {https://arxiv.org/abs/2212.07322} {arXiv:2212.07322 [cond-mat.str-el]}
  \BibitemShut {NoStop}%
\bibitem [{\citenamefont {Nair}\ \emph {et~al.}(2018)\citenamefont {Nair},
  \citenamefont {Brown}, \citenamefont {Coldren}, \citenamefont {Hester},
  \citenamefont {Gelfand}, \citenamefont {Podlesnyak}, \citenamefont {Huang},\
  and\ \citenamefont {Ross}}]{BCPO_Nair2018}%
  \BibitemOpen
  \bibfield  {author} {\bibinfo {author} {\bibfnamefont {H.~S.}\ \bibnamefont
  {Nair}}, \bibinfo {author} {\bibfnamefont {J.~M.}\ \bibnamefont {Brown}},
  \bibinfo {author} {\bibfnamefont {E.}~\bibnamefont {Coldren}}, \bibinfo
  {author} {\bibfnamefont {G.}~\bibnamefont {Hester}}, \bibinfo {author}
  {\bibfnamefont {M.~P.}\ \bibnamefont {Gelfand}}, \bibinfo {author}
  {\bibfnamefont {A.}~\bibnamefont {Podlesnyak}}, \bibinfo {author}
  {\bibfnamefont {Q.}~\bibnamefont {Huang}},\ and\ \bibinfo {author}
  {\bibfnamefont {K.~A.}\ \bibnamefont {Ross}},\ }\bibfield  {title} {\bibinfo
  {title} {Short-range order in the quantum {XXZ} honeycomb lattice material
  {${\mathrm{BaCo}}_{2}{({\mathrm{PO}}_{4})}_{2}$}},\ }\href
  {https://doi.org/10.1103/PhysRevB.97.134409} {\bibfield  {journal} {\bibinfo
  {journal} {Phys. Rev. B}\ }\textbf {\bibinfo {volume} {97}},\ \bibinfo
  {pages} {134409} (\bibinfo {year} {2018})}\BibitemShut {NoStop}%
\bibitem [{\citenamefont {Lefrancois}\ \emph {et~al.}(2016)\citenamefont
  {Lefrancois}, \citenamefont {Songvilay}, \citenamefont {Robert},
  \citenamefont {Nataf}, \citenamefont {Jordan}, \citenamefont {Chaix},
  \citenamefont {Colin}, \citenamefont {Lejay}, \citenamefont {Hadj-Azzem},
  \citenamefont {Ballou},\ and\ \citenamefont {Simonet}}]{ncto_simonet2016}%
  \BibitemOpen
  \bibfield  {author} {\bibinfo {author} {\bibfnamefont {E.}~\bibnamefont
  {Lefrancois}}, \bibinfo {author} {\bibfnamefont {M.}~\bibnamefont
  {Songvilay}}, \bibinfo {author} {\bibfnamefont {J.}~\bibnamefont {Robert}},
  \bibinfo {author} {\bibfnamefont {G.}~\bibnamefont {Nataf}}, \bibinfo
  {author} {\bibfnamefont {E.}~\bibnamefont {Jordan}}, \bibinfo {author}
  {\bibfnamefont {L.}~\bibnamefont {Chaix}}, \bibinfo {author} {\bibfnamefont
  {C.~V.}\ \bibnamefont {Colin}}, \bibinfo {author} {\bibfnamefont
  {P.}~\bibnamefont {Lejay}}, \bibinfo {author} {\bibfnamefont
  {A.}~\bibnamefont {Hadj-Azzem}}, \bibinfo {author} {\bibfnamefont
  {R.}~\bibnamefont {Ballou}},\ and\ \bibinfo {author} {\bibfnamefont
  {V.}~\bibnamefont {Simonet}},\ }\bibfield  {title} {\bibinfo {title}
  {Magnetic properties of the honeycomb oxide {Na$_3$Co$_2$TeO$_6$}},\ }\href
  {https://doi.org/10.1103/PhysRevB.94.214416} {\bibfield  {journal} {\bibinfo
  {journal} {Phys. Rev. B}\ }\textbf {\bibinfo {volume} {94}},\ \bibinfo
  {pages} {214416} (\bibinfo {year} {2016})}\BibitemShut {NoStop}%
\bibitem [{\citenamefont {Songvilay}\ \emph {et~al.}(2020)\citenamefont
  {Songvilay}, \citenamefont {Robert}, \citenamefont {Petit}, \citenamefont
  {Rodriguez-Rivera}, \citenamefont {Ratcliff}, \citenamefont {Damay},
  \citenamefont {Bal\'edent}, \citenamefont {Jim\'enez-Ruiz}, \citenamefont
  {Lejay}, \citenamefont {Pachoud}, \citenamefont {Hadj-Azzem}, \citenamefont
  {Simonet},\ and\ \citenamefont {Stock}}]{ncto_ncso_stock2020}%
  \BibitemOpen
  \bibfield  {author} {\bibinfo {author} {\bibfnamefont {M.}~\bibnamefont
  {Songvilay}}, \bibinfo {author} {\bibfnamefont {J.}~\bibnamefont {Robert}},
  \bibinfo {author} {\bibfnamefont {S.}~\bibnamefont {Petit}}, \bibinfo
  {author} {\bibfnamefont {J.~A.}\ \bibnamefont {Rodriguez-Rivera}}, \bibinfo
  {author} {\bibfnamefont {W.~D.}\ \bibnamefont {Ratcliff}}, \bibinfo {author}
  {\bibfnamefont {F.}~\bibnamefont {Damay}}, \bibinfo {author} {\bibfnamefont
  {V.}~\bibnamefont {Bal\'edent}}, \bibinfo {author} {\bibfnamefont
  {M.}~\bibnamefont {Jim\'enez-Ruiz}}, \bibinfo {author} {\bibfnamefont
  {P.}~\bibnamefont {Lejay}}, \bibinfo {author} {\bibfnamefont
  {E.}~\bibnamefont {Pachoud}}, \bibinfo {author} {\bibfnamefont
  {A.}~\bibnamefont {Hadj-Azzem}}, \bibinfo {author} {\bibfnamefont
  {V.}~\bibnamefont {Simonet}},\ and\ \bibinfo {author} {\bibfnamefont
  {C.}~\bibnamefont {Stock}},\ }\bibfield  {title} {\bibinfo {title} {Kitaev
  interactions in the {Co} honeycomb antiferromagnets {Na$_3$Co$_2$SbO$_6$} and
  {Na$_3$Co$_2$TeO$_6$}},\ }\href {https://doi.org/10.1103/PhysRevB.102.224429}
  {\bibfield  {journal} {\bibinfo  {journal} {Phys. Rev. B}\ }\textbf {\bibinfo
  {volume} {102}},\ \bibinfo {pages} {224429} (\bibinfo {year}
  {2020})}\BibitemShut {NoStop}%
\bibitem [{\citenamefont {Yang}\ \emph {et~al.}(2022)\citenamefont {Yang},
  \citenamefont {Kim}, \citenamefont {Choi}, \citenamefont {Lee}, \citenamefont
  {Lin}, \citenamefont {Ma}, \citenamefont {Kratochv\'{\i}lov\'a},
  \citenamefont {Proschek}, \citenamefont {Moon}, \citenamefont {Lee},
  \citenamefont {Oh},\ and\ \citenamefont {Park}}]{Kappaxy_NCoTeO_Park2022}%
  \BibitemOpen
  \bibfield  {author} {\bibinfo {author} {\bibfnamefont {H.}~\bibnamefont
  {Yang}}, \bibinfo {author} {\bibfnamefont {C.}~\bibnamefont {Kim}}, \bibinfo
  {author} {\bibfnamefont {Y.}~\bibnamefont {Choi}}, \bibinfo {author}
  {\bibfnamefont {J.~H.}\ \bibnamefont {Lee}}, \bibinfo {author} {\bibfnamefont
  {G.}~\bibnamefont {Lin}}, \bibinfo {author} {\bibfnamefont {J.}~\bibnamefont
  {Ma}}, \bibinfo {author} {\bibfnamefont {M.}~\bibnamefont
  {Kratochv\'{\i}lov\'a}}, \bibinfo {author} {\bibfnamefont {P.}~\bibnamefont
  {Proschek}}, \bibinfo {author} {\bibfnamefont {E.-G.}\ \bibnamefont {Moon}},
  \bibinfo {author} {\bibfnamefont {K.~H.}\ \bibnamefont {Lee}}, \bibinfo
  {author} {\bibfnamefont {Y.~S.}\ \bibnamefont {Oh}},\ and\ \bibinfo {author}
  {\bibfnamefont {J.-G.}\ \bibnamefont {Park}},\ }\bibfield  {title} {\bibinfo
  {title} {Significant thermal {Hall} effect in the $3d$ cobalt {Kitaev} system
  {${\mathrm{Na}}_{2}{\mathrm{Co}}_{2}\mathrm{Te}{\mathrm{O}}_{6}$}},\ }\href
  {https://doi.org/10.1103/PhysRevB.106.L081116} {\bibfield  {journal}
  {\bibinfo  {journal} {Phys. Rev. B}\ }\textbf {\bibinfo {volume} {106}},\
  \bibinfo {pages} {L081116} (\bibinfo {year} {2022})}\BibitemShut {NoStop}%
\bibitem [{\citenamefont {Zhang}\ \emph
  {et~al.}(2023{\natexlab{b}})\citenamefont {Zhang}, \citenamefont {Lee},
  \citenamefont {Woods}, \citenamefont {Thomas}, \citenamefont {Movshovich},
  \citenamefont {Brosha}, \citenamefont {Huang}, \citenamefont {Zhou},
  \citenamefont {Zapf},\ and\ \citenamefont {Lee}}]{LANL_ncto2022}%
  \BibitemOpen
  \bibfield  {author} {\bibinfo {author} {\bibfnamefont {S.}~\bibnamefont
  {Zhang}}, \bibinfo {author} {\bibfnamefont {S.}~\bibnamefont {Lee}}, \bibinfo
  {author} {\bibfnamefont {A.~J.}\ \bibnamefont {Woods}}, \bibinfo {author}
  {\bibfnamefont {S.~M.}\ \bibnamefont {Thomas}}, \bibinfo {author}
  {\bibfnamefont {R.}~\bibnamefont {Movshovich}}, \bibinfo {author}
  {\bibfnamefont {E.}~\bibnamefont {Brosha}}, \bibinfo {author} {\bibfnamefont
  {Q.}~\bibnamefont {Huang}}, \bibinfo {author} {\bibfnamefont
  {H.}~\bibnamefont {Zhou}}, \bibinfo {author} {\bibfnamefont {V.~S.}\
  \bibnamefont {Zapf}},\ and\ \bibinfo {author} {\bibfnamefont
  {M.}~\bibnamefont {Lee}},\ }\href@noop {} {\bibinfo {title} {Electronic and
  magnetic phase diagrams of {Kitaev} quantum spin liquid candidate
  {Na$_2$Co$_2$TeO$_6$}}} (\bibinfo {year} {2023}{\natexlab{b}}),\ \Eprint
  {https://arxiv.org/abs/2212.03849} {arXiv:2212.03849 [cond-mat.str-el]}
  \BibitemShut {NoStop}%
\bibitem [{\citenamefont {Guang}\ \emph {et~al.}(2023)\citenamefont {Guang},
  \citenamefont {Li}, \citenamefont {Luo}, \citenamefont {Huang}, \citenamefont
  {Wang}, \citenamefont {Yue}, \citenamefont {Xia}, \citenamefont {Li},
  \citenamefont {Zhao}, \citenamefont {Chen}, \citenamefont {Zhou},\ and\
  \citenamefont {Sun}}]{kappa_NCTO_Sun_PRB_2023}%
  \BibitemOpen
  \bibfield  {author} {\bibinfo {author} {\bibfnamefont {S.}~\bibnamefont
  {Guang}}, \bibinfo {author} {\bibfnamefont {N.}~\bibnamefont {Li}}, \bibinfo
  {author} {\bibfnamefont {R.~L.}\ \bibnamefont {Luo}}, \bibinfo {author}
  {\bibfnamefont {Q.}~\bibnamefont {Huang}}, \bibinfo {author} {\bibfnamefont
  {Y.}~\bibnamefont {Wang}}, \bibinfo {author} {\bibfnamefont {X.}~\bibnamefont
  {Yue}}, \bibinfo {author} {\bibfnamefont {K.}~\bibnamefont {Xia}}, \bibinfo
  {author} {\bibfnamefont {Q.}~\bibnamefont {Li}}, \bibinfo {author}
  {\bibfnamefont {X.}~\bibnamefont {Zhao}}, \bibinfo {author} {\bibfnamefont
  {G.}~\bibnamefont {Chen}}, \bibinfo {author} {\bibfnamefont {H.}~\bibnamefont
  {Zhou}},\ and\ \bibinfo {author} {\bibfnamefont {X.}~\bibnamefont {Sun}},\
  }\bibfield  {title} {\bibinfo {title} {Thermal transport of fractionalized
  antiferromagnetic and field-induced states in the {Kitaev material}
  {${\mathrm{Na}}_{2}{\mathrm{Co}}_{2}{\mathrm{TeO}}_{6}$}},\ }\href
  {https://doi.org/10.1103/PhysRevB.107.184423} {\bibfield  {journal} {\bibinfo
   {journal} {Phys. Rev. B}\ }\textbf {\bibinfo {volume} {107}},\ \bibinfo
  {pages} {184423} (\bibinfo {year} {2023})}\BibitemShut {NoStop}%
\bibitem [{\citenamefont {Yan}\ \emph {et~al.}(2019)\citenamefont {Yan},
  \citenamefont {Okamoto}, \citenamefont {Wu}, \citenamefont {Zheng},
  \citenamefont {Zhou}, \citenamefont {Cao},\ and\ \citenamefont
  {McGuire}}]{ncso_mcguire2019}%
  \BibitemOpen
  \bibfield  {author} {\bibinfo {author} {\bibfnamefont {J.-Q.}\ \bibnamefont
  {Yan}}, \bibinfo {author} {\bibfnamefont {S.}~\bibnamefont {Okamoto}},
  \bibinfo {author} {\bibfnamefont {Y.}~\bibnamefont {Wu}}, \bibinfo {author}
  {\bibfnamefont {Q.}~\bibnamefont {Zheng}}, \bibinfo {author} {\bibfnamefont
  {H.~D.}\ \bibnamefont {Zhou}}, \bibinfo {author} {\bibfnamefont {H.~B.}\
  \bibnamefont {Cao}},\ and\ \bibinfo {author} {\bibfnamefont {M.~A.}\
  \bibnamefont {McGuire}},\ }\bibfield  {title} {\bibinfo {title} {Magnetic
  order in single crystals of
  {${\mathrm{Na}}_{3}{\mathrm{Co}}_{2}{\mathrm{SbO}}_{6}$} with a honeycomb
  arrangement of
  {${3\mathrm{d}}^{7}\phantom{\rule{0.28em}{0ex}}{\mathrm{Co}}^{2+}$} ions},\
  }\href {https://doi.org/10.1103/PhysRevMaterials.3.074405} {\bibfield
  {journal} {\bibinfo  {journal} {Phys. Rev. Materials}\ }\textbf {\bibinfo
  {volume} {3}},\ \bibinfo {pages} {074405} (\bibinfo {year}
  {2019})}\BibitemShut {NoStop}%
\bibitem [{\citenamefont {Yuan}\ \emph {et~al.}(2020)\citenamefont {Yuan},
  \citenamefont {Khait}, \citenamefont {Shu}, \citenamefont {Chou},
  \citenamefont {Stone}, \citenamefont {Clancy}, \citenamefont {Paramekanti},\
  and\ \citenamefont {Kim}}]{CTO_Yuan2020}%
  \BibitemOpen
  \bibfield  {author} {\bibinfo {author} {\bibfnamefont {B.}~\bibnamefont
  {Yuan}}, \bibinfo {author} {\bibfnamefont {I.}~\bibnamefont {Khait}},
  \bibinfo {author} {\bibfnamefont {G.-J.}\ \bibnamefont {Shu}}, \bibinfo
  {author} {\bibfnamefont {F.~C.}\ \bibnamefont {Chou}}, \bibinfo {author}
  {\bibfnamefont {M.~B.}\ \bibnamefont {Stone}}, \bibinfo {author}
  {\bibfnamefont {J.~P.}\ \bibnamefont {Clancy}}, \bibinfo {author}
  {\bibfnamefont {A.}~\bibnamefont {Paramekanti}},\ and\ \bibinfo {author}
  {\bibfnamefont {Y.-J.}\ \bibnamefont {Kim}},\ }\bibfield  {title} {\bibinfo
  {title} {Dirac magnons in a honeycomb lattice quantum {$XY$} magnet
  {CoTiO$_{3}$}},\ }\href {https://doi.org/10.1103/PhysRevX.10.011062}
  {\bibfield  {journal} {\bibinfo  {journal} {Phys. Rev. X}\ }\textbf {\bibinfo
  {volume} {10}},\ \bibinfo {pages} {011062} (\bibinfo {year}
  {2020})}\BibitemShut {NoStop}%
\bibitem [{\citenamefont {Elliot}\ \emph {et~al.}(2021)\citenamefont {Elliot},
  \citenamefont {McClarty}, \citenamefont {Prabhakaran}, \citenamefont
  {Johnson}, \citenamefont {Walker}, \citenamefont {Manuel},\ and\
  \citenamefont {Coldea}}]{CTO_Coldea_Ncomm2021}%
  \BibitemOpen
  \bibfield  {author} {\bibinfo {author} {\bibfnamefont {M.}~\bibnamefont
  {Elliot}}, \bibinfo {author} {\bibfnamefont {P.~A.}\ \bibnamefont
  {McClarty}}, \bibinfo {author} {\bibfnamefont {D.}~\bibnamefont
  {Prabhakaran}}, \bibinfo {author} {\bibfnamefont {R.~D.}\ \bibnamefont
  {Johnson}}, \bibinfo {author} {\bibfnamefont {H.~C.}\ \bibnamefont {Walker}},
  \bibinfo {author} {\bibfnamefont {P.}~\bibnamefont {Manuel}},\ and\ \bibinfo
  {author} {\bibfnamefont {R.}~\bibnamefont {Coldea}},\ }\bibfield  {title}
  {\bibinfo {title} {Order-by-disorder from bond-dependent exchange and
  intensity signature of nodal quasiparticles in a honeycomb cobaltate},\
  }\href {https://doi.org/10.1038/s41467-021-23851-0} {\bibfield  {journal}
  {\bibinfo  {journal} {Nature Communications}\ }\textbf {\bibinfo {volume}
  {12}},\ \bibinfo {pages} {3936} (\bibinfo {year} {2021})}\BibitemShut
  {NoStop}%
\bibitem [{\citenamefont {Li}\ \emph {et~al.}(2022)\citenamefont {Li},
  \citenamefont {Gu}, \citenamefont {Chen}, \citenamefont {Garlea},
  \citenamefont {Iida}, \citenamefont {Kamazawa}, \citenamefont {Li},
  \citenamefont {Deng}, \citenamefont {Xiao}, \citenamefont {Zheng},
  \citenamefont {Ye}, \citenamefont {Peng}, \citenamefont {Zaliznyak},
  \citenamefont {Tranquada},\ and\ \citenamefont
  {Li}}]{Anisotropy_NCoSb_Yuan2022}%
  \BibitemOpen
  \bibfield  {author} {\bibinfo {author} {\bibfnamefont {X.}~\bibnamefont
  {Li}}, \bibinfo {author} {\bibfnamefont {Y.}~\bibnamefont {Gu}}, \bibinfo
  {author} {\bibfnamefont {Y.}~\bibnamefont {Chen}}, \bibinfo {author}
  {\bibfnamefont {V.~O.}\ \bibnamefont {Garlea}}, \bibinfo {author}
  {\bibfnamefont {K.}~\bibnamefont {Iida}}, \bibinfo {author} {\bibfnamefont
  {K.}~\bibnamefont {Kamazawa}}, \bibinfo {author} {\bibfnamefont
  {Y.}~\bibnamefont {Li}}, \bibinfo {author} {\bibfnamefont {G.}~\bibnamefont
  {Deng}}, \bibinfo {author} {\bibfnamefont {Q.}~\bibnamefont {Xiao}}, \bibinfo
  {author} {\bibfnamefont {X.}~\bibnamefont {Zheng}}, \bibinfo {author}
  {\bibfnamefont {Z.}~\bibnamefont {Ye}}, \bibinfo {author} {\bibfnamefont
  {Y.}~\bibnamefont {Peng}}, \bibinfo {author} {\bibfnamefont {I.~A.}\
  \bibnamefont {Zaliznyak}}, \bibinfo {author} {\bibfnamefont {J.~M.}\
  \bibnamefont {Tranquada}},\ and\ \bibinfo {author} {\bibfnamefont
  {Y.}~\bibnamefont {Li}},\ }\bibfield  {title} {\bibinfo {title} {Giant
  magnetic in-plane anisotropy and competing instabilities in
  {${\mathrm{Na}}_{3}{\mathrm{Co}}_{2}{\mathrm{SbO}}_{6}$}},\ }\href
  {https://doi.org/10.1103/PhysRevX.12.041024} {\bibfield  {journal} {\bibinfo
  {journal} {Phys. Rev. X}\ }\textbf {\bibinfo {volume} {12}},\ \bibinfo
  {pages} {041024} (\bibinfo {year} {2022})}\BibitemShut {NoStop}%
\bibitem [{\citenamefont {Kr\"uger}\ \emph {et~al.}(2022)\citenamefont
  {Kr\"uger}, \citenamefont {Chen}, \citenamefont {Jin}, \citenamefont {Li},\
  and\ \citenamefont {Janssen}}]{tripleQ_Janssen2022}%
  \BibitemOpen
  \bibfield  {author} {\bibinfo {author} {\bibfnamefont {W.~G.~F.}\
  \bibnamefont {Kr\"uger}}, \bibinfo {author} {\bibfnamefont {W.}~\bibnamefont
  {Chen}}, \bibinfo {author} {\bibfnamefont {X.}~\bibnamefont {Jin}}, \bibinfo
  {author} {\bibfnamefont {Y.}~\bibnamefont {Li}},\ and\ \bibinfo {author}
  {\bibfnamefont {L.}~\bibnamefont {Janssen}},\ }\href@noop {} {\bibinfo
  {title} {Triple-q order in {Na$_2$Co$_2$TeO$_6$} from proximity to
  hidden-{SU(2)}-symmetric point}} (\bibinfo {year} {2022}),\ \Eprint
  {https://arxiv.org/abs/2211.16957} {arXiv:2211.16957 [cond-mat.str-el]}
  \BibitemShut {NoStop}%
\bibitem [{\citenamefont {Hong}\ \emph {et~al.}(2023)\citenamefont {Hong},
  \citenamefont {Gillig}, \citenamefont {Yao}, \citenamefont {Janssen},
  \citenamefont {Kocsis}, \citenamefont {Gass}, \citenamefont {Li},
  \citenamefont {Wolter}, \citenamefont {Büchner},\ and\ \citenamefont
  {Hess}}]{hong2023phonon}%
  \BibitemOpen
  \bibfield  {author} {\bibinfo {author} {\bibfnamefont {X.}~\bibnamefont
  {Hong}}, \bibinfo {author} {\bibfnamefont {M.}~\bibnamefont {Gillig}},
  \bibinfo {author} {\bibfnamefont {W.}~\bibnamefont {Yao}}, \bibinfo {author}
  {\bibfnamefont {L.}~\bibnamefont {Janssen}}, \bibinfo {author} {\bibfnamefont
  {V.}~\bibnamefont {Kocsis}}, \bibinfo {author} {\bibfnamefont
  {S.}~\bibnamefont {Gass}}, \bibinfo {author} {\bibfnamefont {Y.}~\bibnamefont
  {Li}}, \bibinfo {author} {\bibfnamefont {A.~U.~B.}\ \bibnamefont {Wolter}},
  \bibinfo {author} {\bibfnamefont {B.}~\bibnamefont {Büchner}},\ and\
  \bibinfo {author} {\bibfnamefont {C.}~\bibnamefont {Hess}},\ }\href@noop {}
  {\bibinfo {title} {Phonon thermal transport shaped by strong spin-phonon
  scattering in a {Kitaev} material {Na$_2$Co$_2$TeO$_6$}}} (\bibinfo {year}
  {2023}),\ \Eprint {https://arxiv.org/abs/2306.16963} {arXiv:2306.16963
  [cond-mat.str-el]} \BibitemShut {NoStop}%
\bibitem [{\citenamefont {Sanders}\ \emph {et~al.}(2022)\citenamefont
  {Sanders}, \citenamefont {Mole}, \citenamefont {Liu}, \citenamefont {Brown},
  \citenamefont {Yu}, \citenamefont {Ling},\ and\ \citenamefont
  {Rachel}}]{Anisotropy_NCoSb_Rachel2022}%
  \BibitemOpen
  \bibfield  {author} {\bibinfo {author} {\bibfnamefont {A.~L.}\ \bibnamefont
  {Sanders}}, \bibinfo {author} {\bibfnamefont {R.~A.}\ \bibnamefont {Mole}},
  \bibinfo {author} {\bibfnamefont {J.}~\bibnamefont {Liu}}, \bibinfo {author}
  {\bibfnamefont {A.~J.}\ \bibnamefont {Brown}}, \bibinfo {author}
  {\bibfnamefont {D.}~\bibnamefont {Yu}}, \bibinfo {author} {\bibfnamefont
  {C.~D.}\ \bibnamefont {Ling}},\ and\ \bibinfo {author} {\bibfnamefont
  {S.}~\bibnamefont {Rachel}},\ }\bibfield  {title} {\bibinfo {title} {Dominant
  kitaev interactions in the honeycomb materials
  {${\mathrm{Na}}_{3}{\mathrm{Co}}_{2}{\mathrm{SbO}}_{6}$} and
  {${\mathrm{Na}}_{2}{\mathrm{Co}}_{2}{\mathrm{TeO}}_{6}$}},\ }\href
  {https://doi.org/10.1103/PhysRevB.106.014413} {\bibfield  {journal} {\bibinfo
   {journal} {Phys. Rev. B}\ }\textbf {\bibinfo {volume} {106}},\ \bibinfo
  {pages} {014413} (\bibinfo {year} {2022})}\BibitemShut {NoStop}%
\bibitem [{\citenamefont {Das}\ \emph {et~al.}(2021)\citenamefont {Das},
  \citenamefont {Voleti}, \citenamefont {Saha-Dasgupta},\ and\ \citenamefont
  {Paramekanti}}]{Abinitio_Das2021}%
  \BibitemOpen
  \bibfield  {author} {\bibinfo {author} {\bibfnamefont {S.}~\bibnamefont
  {Das}}, \bibinfo {author} {\bibfnamefont {S.}~\bibnamefont {Voleti}},
  \bibinfo {author} {\bibfnamefont {T.}~\bibnamefont {Saha-Dasgupta}},\ and\
  \bibinfo {author} {\bibfnamefont {A.}~\bibnamefont {Paramekanti}},\
  }\bibfield  {title} {\bibinfo {title} {Xy magnetism, {Kitaev} exchange, and
  long-range frustration in the ${J}_{\mathrm{eff}}=\frac{1}{2}$ honeycomb
  cobaltates},\ }\href {https://doi.org/10.1103/PhysRevB.104.134425} {\bibfield
   {journal} {\bibinfo  {journal} {Phys. Rev. B}\ }\textbf {\bibinfo {volume}
  {104}},\ \bibinfo {pages} {134425} (\bibinfo {year} {2021})}\BibitemShut
  {NoStop}%
\bibitem [{\citenamefont {Maksimov}\ \emph {et~al.}(2022)\citenamefont
  {Maksimov}, \citenamefont {Ushakov}, \citenamefont {Pchelkina}, \citenamefont
  {Li}, \citenamefont {Winter},\ and\ \citenamefont
  {Streltsov}}]{Abinitio_Streltsov2022}%
  \BibitemOpen
  \bibfield  {author} {\bibinfo {author} {\bibfnamefont {P.~A.}\ \bibnamefont
  {Maksimov}}, \bibinfo {author} {\bibfnamefont {A.~V.}\ \bibnamefont
  {Ushakov}}, \bibinfo {author} {\bibfnamefont {Z.~V.}\ \bibnamefont
  {Pchelkina}}, \bibinfo {author} {\bibfnamefont {Y.}~\bibnamefont {Li}},
  \bibinfo {author} {\bibfnamefont {S.~M.}\ \bibnamefont {Winter}},\ and\
  \bibinfo {author} {\bibfnamefont {S.~V.}\ \bibnamefont {Streltsov}},\
  }\bibfield  {title} {\bibinfo {title} {Ab initio guided minimal model for the
  ``{Kitaev}'' material {${\mathrm{BaCo}}_{2}$(${\mathrm{AsO}}_{4}{)}_{2}$}:
  {Importance} of direct hopping, third-neighbor exchange, and quantum
  fluctuations},\ }\href {https://doi.org/10.1103/PhysRevB.106.165131}
  {\bibfield  {journal} {\bibinfo  {journal} {Phys. Rev. B}\ }\textbf {\bibinfo
  {volume} {106}},\ \bibinfo {pages} {165131} (\bibinfo {year}
  {2022})}\BibitemShut {NoStop}%
\bibitem [{\citenamefont {Samanta}\ \emph {et~al.}(2022)\citenamefont
  {Samanta}, \citenamefont {Detrattanawichai}, \citenamefont {Na-Phattalung},\
  and\ \citenamefont {Kim}}]{Abinitio_HSKim2022}%
  \BibitemOpen
  \bibfield  {author} {\bibinfo {author} {\bibfnamefont {S.}~\bibnamefont
  {Samanta}}, \bibinfo {author} {\bibfnamefont {P.}~\bibnamefont
  {Detrattanawichai}}, \bibinfo {author} {\bibfnamefont {S.}~\bibnamefont
  {Na-Phattalung}},\ and\ \bibinfo {author} {\bibfnamefont {H.-S.}\
  \bibnamefont {Kim}},\ }\bibfield  {title} {\bibinfo {title} {Active orbital
  degree of freedom and potential spin-orbit-entangled moments in the {Kitaev}
  magnet candidate {${\mathrm{BaCo}}_{2}{({\mathrm{AsO}}_{4})}_{2}$}},\ }\href
  {https://doi.org/10.1103/PhysRevB.106.195136} {\bibfield  {journal} {\bibinfo
   {journal} {Phys. Rev. B}\ }\textbf {\bibinfo {volume} {106}},\ \bibinfo
  {pages} {195136} (\bibinfo {year} {2022})}\BibitemShut {NoStop}%
\bibitem [{\citenamefont {Rastelli}\ \emph {et~al.}(1979)\citenamefont
  {Rastelli}, \citenamefont {Tassi},\ and\ \citenamefont
  {Reatto}}]{j1j2j3_rastelli1979}%
  \BibitemOpen
  \bibfield  {author} {\bibinfo {author} {\bibfnamefont {E.}~\bibnamefont
  {Rastelli}}, \bibinfo {author} {\bibfnamefont {A.}~\bibnamefont {Tassi}},\
  and\ \bibinfo {author} {\bibfnamefont {L.}~\bibnamefont {Reatto}},\
  }\bibfield  {title} {\bibinfo {title} {Non-simple magnetic order for simple
  {Hamiltonians}},\ }\href
  {https://doi.org/https://doi.org/10.1016/0378-4363(79)90002-0} {\bibfield
  {journal} {\bibinfo  {journal} {Physica B+C}\ }\textbf {\bibinfo {volume}
  {97}},\ \bibinfo {pages} {1} (\bibinfo {year} {1979})}\BibitemShut {NoStop}%
\bibitem [{\citenamefont {Fouet}\ \emph {et~al.}(2001)\citenamefont {Fouet},
  \citenamefont {Sindzingre},\ and\ \citenamefont
  {Lhuillier}}]{j1j2j3_Fouet2001}%
  \BibitemOpen
  \bibfield  {author} {\bibinfo {author} {\bibfnamefont {J.~B.}\ \bibnamefont
  {Fouet}}, \bibinfo {author} {\bibfnamefont {P.}~\bibnamefont {Sindzingre}},\
  and\ \bibinfo {author} {\bibfnamefont {C.}~\bibnamefont {Lhuillier}},\
  }\bibfield  {title} {\bibinfo {title} {An investigation of the quantum
  {J1-J2-J3} model on the honeycomb lattice},\ }\href
  {https://doi.org/10.1007/s100510170273} {\bibfield  {journal} {\bibinfo
  {journal} {The European Physical Journal B - Condensed Matter and Complex
  Systems}\ }\textbf {\bibinfo {volume} {20}},\ \bibinfo {pages} {241}
  (\bibinfo {year} {2001})}\BibitemShut {NoStop}%
\bibitem [{\citenamefont {Mulder}\ \emph {et~al.}(2010)\citenamefont {Mulder},
  \citenamefont {Ganesh}, \citenamefont {Capriotti},\ and\ \citenamefont
  {Paramekanti}}]{j1j2_ganesh2010}%
  \BibitemOpen
  \bibfield  {author} {\bibinfo {author} {\bibfnamefont {A.}~\bibnamefont
  {Mulder}}, \bibinfo {author} {\bibfnamefont {R.}~\bibnamefont {Ganesh}},
  \bibinfo {author} {\bibfnamefont {L.}~\bibnamefont {Capriotti}},\ and\
  \bibinfo {author} {\bibfnamefont {A.}~\bibnamefont {Paramekanti}},\
  }\bibfield  {title} {\bibinfo {title} {Spiral order by disorder and lattice
  nematic order in a frustrated {Heisenberg} antiferromagnet on the honeycomb
  lattice},\ }\href {https://doi.org/10.1103/PhysRevB.81.214419} {\bibfield
  {journal} {\bibinfo  {journal} {Phys. Rev. B}\ }\textbf {\bibinfo {volume}
  {81}},\ \bibinfo {pages} {214419} (\bibinfo {year} {2010})}\BibitemShut
  {NoStop}%
\bibitem [{\citenamefont {Ganesh}\ \emph {et~al.}(2011)\citenamefont {Ganesh},
  \citenamefont {Sheng}, \citenamefont {Kim},\ and\ \citenamefont
  {Paramekanti}}]{j1j2_ganesh2011}%
  \BibitemOpen
  \bibfield  {author} {\bibinfo {author} {\bibfnamefont {R.}~\bibnamefont
  {Ganesh}}, \bibinfo {author} {\bibfnamefont {D.~N.}\ \bibnamefont {Sheng}},
  \bibinfo {author} {\bibfnamefont {Y.-J.}\ \bibnamefont {Kim}},\ and\ \bibinfo
  {author} {\bibfnamefont {A.}~\bibnamefont {Paramekanti}},\ }\bibfield
  {title} {\bibinfo {title} {Quantum paramagnetic ground states on the
  honeycomb lattice and field-induced {N\'eel} order},\ }\href
  {https://doi.org/10.1103/PhysRevB.83.144414} {\bibfield  {journal} {\bibinfo
  {journal} {Phys. Rev. B}\ }\textbf {\bibinfo {volume} {83}},\ \bibinfo
  {pages} {144414} (\bibinfo {year} {2011})}\BibitemShut {NoStop}%
\bibitem [{\citenamefont {Mezzacapo}\ and\ \citenamefont
  {Boninsegni}(2012)}]{j1j2_peps_Boninsegni2012}%
  \BibitemOpen
  \bibfield  {author} {\bibinfo {author} {\bibfnamefont {F.}~\bibnamefont
  {Mezzacapo}}\ and\ \bibinfo {author} {\bibfnamefont {M.}~\bibnamefont
  {Boninsegni}},\ }\bibfield  {title} {\bibinfo {title} {Ground-state phase
  diagram of the quantum {${J}_{1}\ensuremath{-}{J}_{2}$} model on the
  honeycomb lattice},\ }\href {https://doi.org/10.1103/PhysRevB.85.060402}
  {\bibfield  {journal} {\bibinfo  {journal} {Phys. Rev. B}\ }\textbf {\bibinfo
  {volume} {85}},\ \bibinfo {pages} {060402} (\bibinfo {year}
  {2012})}\BibitemShut {NoStop}%
\bibitem [{\citenamefont {Ganesh}\ \emph {et~al.}(2013)\citenamefont {Ganesh},
  \citenamefont {van~den Brink},\ and\ \citenamefont
  {Nishimoto}}]{j1j2_DQCP_ganesh2013}%
  \BibitemOpen
  \bibfield  {author} {\bibinfo {author} {\bibfnamefont {R.}~\bibnamefont
  {Ganesh}}, \bibinfo {author} {\bibfnamefont {J.}~\bibnamefont {van~den
  Brink}},\ and\ \bibinfo {author} {\bibfnamefont {S.}~\bibnamefont
  {Nishimoto}},\ }\bibfield  {title} {\bibinfo {title} {Deconfined criticality
  in the frustrated {Heisenberg} honeycomb antiferromagnet},\ }\href
  {https://doi.org/10.1103/PhysRevLett.110.127203} {\bibfield  {journal}
  {\bibinfo  {journal} {Phys. Rev. Lett.}\ }\textbf {\bibinfo {volume} {110}},\
  \bibinfo {pages} {127203} (\bibinfo {year} {2013})}\BibitemShut {NoStop}%
\bibitem [{\citenamefont {Pujari}\ \emph {et~al.}(2013)\citenamefont {Pujari},
  \citenamefont {Damle},\ and\ \citenamefont {Alet}}]{j1j2_DQCP_pujari2013}%
  \BibitemOpen
  \bibfield  {author} {\bibinfo {author} {\bibfnamefont {S.}~\bibnamefont
  {Pujari}}, \bibinfo {author} {\bibfnamefont {K.}~\bibnamefont {Damle}},\ and\
  \bibinfo {author} {\bibfnamefont {F.}~\bibnamefont {Alet}},\ }\bibfield
  {title} {\bibinfo {title} {N\'eel-state to valence-bond-solid transition on
  the honeycomb lattice: Evidence for deconfined criticality},\ }\href
  {https://doi.org/10.1103/PhysRevLett.111.087203} {\bibfield  {journal}
  {\bibinfo  {journal} {Phys. Rev. Lett.}\ }\textbf {\bibinfo {volume} {111}},\
  \bibinfo {pages} {087203} (\bibinfo {year} {2013})}\BibitemShut {NoStop}%
\bibitem [{\citenamefont {Gong}\ \emph {et~al.}(2013)\citenamefont {Gong},
  \citenamefont {Sheng}, \citenamefont {Motrunich},\ and\ \citenamefont
  {Fisher}}]{j1j2_dmrg_sheng2013}%
  \BibitemOpen
  \bibfield  {author} {\bibinfo {author} {\bibfnamefont {S.-S.}\ \bibnamefont
  {Gong}}, \bibinfo {author} {\bibfnamefont {D.~N.}\ \bibnamefont {Sheng}},
  \bibinfo {author} {\bibfnamefont {O.~I.}\ \bibnamefont {Motrunich}},\ and\
  \bibinfo {author} {\bibfnamefont {M.~P.~A.}\ \bibnamefont {Fisher}},\
  }\bibfield  {title} {\bibinfo {title} {Phase diagram of the
  spin-$\frac{1}{2}$ {${J}_{1}$-${J}_{2}$} {Heisenberg} model on a honeycomb
  lattice},\ }\href {https://doi.org/10.1103/PhysRevB.88.165138} {\bibfield
  {journal} {\bibinfo  {journal} {Phys. Rev. B}\ }\textbf {\bibinfo {volume}
  {88}},\ \bibinfo {pages} {165138} (\bibinfo {year} {2013})}\BibitemShut
  {NoStop}%
\bibitem [{\citenamefont {Carrasquilla}\ \emph {et~al.}(2013)\citenamefont
  {Carrasquilla}, \citenamefont {Ciolo}, \citenamefont {Becca}, \citenamefont
  {Galitski},\ and\ \citenamefont {Rigol}}]{j1j2_rigol2013}%
  \BibitemOpen
  \bibfield  {author} {\bibinfo {author} {\bibfnamefont {J.}~\bibnamefont
  {Carrasquilla}}, \bibinfo {author} {\bibfnamefont {A.~D.}\ \bibnamefont
  {Ciolo}}, \bibinfo {author} {\bibfnamefont {F.}~\bibnamefont {Becca}},
  \bibinfo {author} {\bibfnamefont {V.}~\bibnamefont {Galitski}},\ and\
  \bibinfo {author} {\bibfnamefont {M.}~\bibnamefont {Rigol}},\ }\bibfield
  {title} {\bibinfo {title} {Nature of the phases in the frustrated {$XY$}
  model on the honeycomb lattice},\ }\href
  {https://doi.org/10.1103/PhysRevB.88.241109} {\bibfield  {journal} {\bibinfo
  {journal} {Phys. Rev. B}\ }\textbf {\bibinfo {volume} {88}},\ \bibinfo
  {pages} {241109} (\bibinfo {year} {2013})}\BibitemShut {NoStop}%
\bibitem [{\citenamefont {Di~Ciolo}\ \emph {et~al.}(2014)\citenamefont
  {Di~Ciolo}, \citenamefont {Carrasquilla}, \citenamefont {Becca},
  \citenamefont {Rigol},\ and\ \citenamefont {Galitski}}]{j1j2_galitski2014}%
  \BibitemOpen
  \bibfield  {author} {\bibinfo {author} {\bibfnamefont {A.}~\bibnamefont
  {Di~Ciolo}}, \bibinfo {author} {\bibfnamefont {J.}~\bibnamefont
  {Carrasquilla}}, \bibinfo {author} {\bibfnamefont {F.}~\bibnamefont {Becca}},
  \bibinfo {author} {\bibfnamefont {M.}~\bibnamefont {Rigol}},\ and\ \bibinfo
  {author} {\bibfnamefont {V.}~\bibnamefont {Galitski}},\ }\bibfield  {title}
  {\bibinfo {title} {Spiral antiferromagnets beyond the spin-wave
  approximation: Frustrated {$XY$} and {Heisenberg} models on the honeycomb
  lattice},\ }\href {https://doi.org/10.1103/PhysRevB.89.094413} {\bibfield
  {journal} {\bibinfo  {journal} {Phys. Rev. B}\ }\textbf {\bibinfo {volume}
  {89}},\ \bibinfo {pages} {094413} (\bibinfo {year} {2014})}\BibitemShut
  {NoStop}%
\bibitem [{\citenamefont {Huang}\ \emph {et~al.}(2021)\citenamefont {Huang},
  \citenamefont {Dong}, \citenamefont {Sheng},\ and\ \citenamefont
  {Ting}}]{j1j2_Ting2021}%
  \BibitemOpen
  \bibfield  {author} {\bibinfo {author} {\bibfnamefont {Y.}~\bibnamefont
  {Huang}}, \bibinfo {author} {\bibfnamefont {X.-Y.}\ \bibnamefont {Dong}},
  \bibinfo {author} {\bibfnamefont {D.~N.}\ \bibnamefont {Sheng}},\ and\
  \bibinfo {author} {\bibfnamefont {C.~S.}\ \bibnamefont {Ting}},\ }\bibfield
  {title} {\bibinfo {title} {Quantum phase diagram and chiral spin liquid in
  the extended spin-$\frac{1}{2}$ honeycomb {XY} model},\ }\href
  {https://doi.org/10.1103/PhysRevB.103.L041108} {\bibfield  {journal}
  {\bibinfo  {journal} {Phys. Rev. B}\ }\textbf {\bibinfo {volume} {103}},\
  \bibinfo {pages} {L041108} (\bibinfo {year} {2021})}\BibitemShut {NoStop}%
\bibitem [{\citenamefont {Maksimov}\ and\ \citenamefont
  {Chernyshev}(2022)}]{spinwaveXXZ_chernyshev_PRB2022}%
  \BibitemOpen
  \bibfield  {author} {\bibinfo {author} {\bibfnamefont {P.~A.}\ \bibnamefont
  {Maksimov}}\ and\ \bibinfo {author} {\bibfnamefont {A.~L.}\ \bibnamefont
  {Chernyshev}},\ }\bibfield  {title} {\bibinfo {title} {Easy-plane
  anisotropic-exchange magnets on a honeycomb lattice: Quantum effects and
  dealing with them},\ }\href {https://doi.org/10.1103/PhysRevB.106.214411}
  {\bibfield  {journal} {\bibinfo  {journal} {Phys. Rev. B}\ }\textbf {\bibinfo
  {volume} {106}},\ \bibinfo {pages} {214411} (\bibinfo {year}
  {2022})}\BibitemShut {NoStop}%
\bibitem [{\citenamefont {Davidson}(1975)}]{DAVIDSON197587}%
  \BibitemOpen
  \bibfield  {author} {\bibinfo {author} {\bibfnamefont {E.~R.}\ \bibnamefont
  {Davidson}},\ }\bibfield  {title} {\bibinfo {title} {The iterative
  calculation of a few of the lowest eigenvalues and corresponding eigenvectors
  of large real-symmetric matrices},\ }\href
  {https://doi.org/https://doi.org/10.1016/0021-9991(75)90065-0} {\bibfield
  {journal} {\bibinfo  {journal} {Journal of Computational Physics}\ }\textbf
  {\bibinfo {volume} {17}},\ \bibinfo {pages} {87} (\bibinfo {year}
  {1975})}\BibitemShut {NoStop}%
\bibitem [{\citenamefont {Davidson}\ and\ \citenamefont
  {Thompson}(1993)}]{davidson1993monster}%
  \BibitemOpen
  \bibfield  {author} {\bibinfo {author} {\bibfnamefont {E.~R.}\ \bibnamefont
  {Davidson}}\ and\ \bibinfo {author} {\bibfnamefont {W.~J.}\ \bibnamefont
  {Thompson}},\ }\bibfield  {title} {\bibinfo {title} {Monster matrices: their
  eigenvalues and eigenvectors},\ }\href
  {https://doi.org/https://doi.org/10.1063/1.4823212} {\bibfield  {journal}
  {\bibinfo  {journal} {Computers in Physics}\ }\textbf {\bibinfo {volume}
  {7}},\ \bibinfo {pages} {519} (\bibinfo {year} {1993})}\BibitemShut {NoStop}%
\bibitem [{\citenamefont {Murray}\ \emph {et~al.}(1992)\citenamefont {Murray},
  \citenamefont {Racine},\ and\ \citenamefont {Davidson}}]{MURRAY1992382}%
  \BibitemOpen
  \bibfield  {author} {\bibinfo {author} {\bibfnamefont {C.~W.}\ \bibnamefont
  {Murray}}, \bibinfo {author} {\bibfnamefont {S.~C.}\ \bibnamefont {Racine}},\
  and\ \bibinfo {author} {\bibfnamefont {E.~R.}\ \bibnamefont {Davidson}},\
  }\bibfield  {title} {\bibinfo {title} {Improved algorithms for the lowest few
  eigenvalues and associated eigenvectors of large matrices},\ }\href
  {https://doi.org/https://doi.org/10.1016/0021-9991(92)90409-R} {\bibfield
  {journal} {\bibinfo  {journal} {Journal of Computational Physics}\ }\textbf
  {\bibinfo {volume} {103}},\ \bibinfo {pages} {382} (\bibinfo {year}
  {1992})}\BibitemShut {NoStop}%
\bibitem [{sup()}]{suppmat}%
  \BibitemOpen
  \href@noop {} {}\bibinfo {note} {See Supplemental Material for details about
  (i) ED clusters and additional ED data and (ii) supporting DMRG
  data.}\BibitemShut {Stop}%
\bibitem [{\citenamefont {White}(1992)}]{white1992density}%
  \BibitemOpen
  \bibfield  {author} {\bibinfo {author} {\bibfnamefont {S.~R.}\ \bibnamefont
  {White}},\ }\bibfield  {title} {\bibinfo {title} {Density matrix formulation
  for quantum renormalization groups},\ }\href
  {https://doi.org/10.1103/PhysRevLett.69.2863} {\bibfield  {journal} {\bibinfo
   {journal} {Phys. Rev. Lett.}\ }\textbf {\bibinfo {volume} {69}},\ \bibinfo
  {pages} {2863} (\bibinfo {year} {1992})}\BibitemShut {NoStop}%
\bibitem [{\citenamefont {White}(1993)}]{white1993density}%
  \BibitemOpen
  \bibfield  {author} {\bibinfo {author} {\bibfnamefont {S.~R.}\ \bibnamefont
  {White}},\ }\bibfield  {title} {\bibinfo {title} {Density-matrix algorithms
  for quantum renormalization groups},\ }\href
  {https://doi.org/10.1103/PhysRevB.48.10345} {\bibfield  {journal} {\bibinfo
  {journal} {Phys. Rev. B}\ }\textbf {\bibinfo {volume} {48}},\ \bibinfo
  {pages} {10345} (\bibinfo {year} {1993})}\BibitemShut {NoStop}%
\bibitem [{\citenamefont {Kumar}\ \emph {et~al.}(2010)\citenamefont {Kumar},
  \citenamefont {Soos}, \citenamefont {Sen},\ and\ \citenamefont
  {Ramasesha}}]{Mkumar2010}%
  \BibitemOpen
  \bibfield  {author} {\bibinfo {author} {\bibfnamefont {M.}~\bibnamefont
  {Kumar}}, \bibinfo {author} {\bibfnamefont {Z.~G.}\ \bibnamefont {Soos}},
  \bibinfo {author} {\bibfnamefont {D.}~\bibnamefont {Sen}},\ and\ \bibinfo
  {author} {\bibfnamefont {S.}~\bibnamefont {Ramasesha}},\ }\bibfield  {title}
  {\bibinfo {title} {Modified density matrix renormalization group algorithm
  for the zigzag spin-$\frac{1}{2}$ chain with frustrated antiferromagnetic
  exchange: Comparison with field theory at large {${J}_{2}/{J}_{1}$}},\ }\href
  {https://doi.org/10.1103/PhysRevB.81.104406} {\bibfield  {journal} {\bibinfo
  {journal} {Phys. Rev. B}\ }\textbf {\bibinfo {volume} {81}},\ \bibinfo
  {pages} {104406} (\bibinfo {year} {2010})}\BibitemShut {NoStop}%
\bibitem [{\citenamefont {Schollw\"ock}(2005)}]{schollwock2005density}%
  \BibitemOpen
  \bibfield  {author} {\bibinfo {author} {\bibfnamefont {U.}~\bibnamefont
  {Schollw\"ock}},\ }\bibfield  {title} {\bibinfo {title} {The density-matrix
  renormalization group},\ }\href {https://doi.org/10.1103/RevModPhys.77.259}
  {\bibfield  {journal} {\bibinfo  {journal} {Rev. Mod. Phys.}\ }\textbf
  {\bibinfo {volume} {77}},\ \bibinfo {pages} {259} (\bibinfo {year}
  {2005})}\BibitemShut {NoStop}%
\bibitem [{\citenamefont {Emery}\ \emph {et~al.}(2000)\citenamefont {Emery},
  \citenamefont {Fradkin}, \citenamefont {Kivelson},\ and\ \citenamefont
  {Lubensky}}]{SLL_Emery_PRL2000}%
  \BibitemOpen
  \bibfield  {author} {\bibinfo {author} {\bibfnamefont {V.~J.}\ \bibnamefont
  {Emery}}, \bibinfo {author} {\bibfnamefont {E.}~\bibnamefont {Fradkin}},
  \bibinfo {author} {\bibfnamefont {S.~A.}\ \bibnamefont {Kivelson}},\ and\
  \bibinfo {author} {\bibfnamefont {T.~C.}\ \bibnamefont {Lubensky}},\
  }\bibfield  {title} {\bibinfo {title} {Quantum theory of the smectic metal
  state in stripe phases},\ }\href
  {https://doi.org/10.1103/PhysRevLett.85.2160} {\bibfield  {journal} {\bibinfo
   {journal} {Phys. Rev. Lett.}\ }\textbf {\bibinfo {volume} {85}},\ \bibinfo
  {pages} {2160} (\bibinfo {year} {2000})}\BibitemShut {NoStop}%
\bibitem [{\citenamefont {Vishwanath}\ and\ \citenamefont
  {Carpentier}(2001)}]{SLL_Vishwanath_PRL2001}%
  \BibitemOpen
  \bibfield  {author} {\bibinfo {author} {\bibfnamefont {A.}~\bibnamefont
  {Vishwanath}}\ and\ \bibinfo {author} {\bibfnamefont {D.}~\bibnamefont
  {Carpentier}},\ }\bibfield  {title} {\bibinfo {title} {Two-dimensional
  anisotropic non-{Fermi}-liquid phase of coupled {Luttinger} liquids},\ }\href
  {https://doi.org/10.1103/PhysRevLett.86.676} {\bibfield  {journal} {\bibinfo
  {journal} {Phys. Rev. Lett.}\ }\textbf {\bibinfo {volume} {86}},\ \bibinfo
  {pages} {676} (\bibinfo {year} {2001})}\BibitemShut {NoStop}%
\bibitem [{\citenamefont {Mukhopadhyay}\ \emph {et~al.}(2001)\citenamefont
  {Mukhopadhyay}, \citenamefont {Kane},\ and\ \citenamefont
  {Lubensky}}]{CSLL_Mukhopadhyay_PRB2001}%
  \BibitemOpen
  \bibfield  {author} {\bibinfo {author} {\bibfnamefont {R.}~\bibnamefont
  {Mukhopadhyay}}, \bibinfo {author} {\bibfnamefont {C.~L.}\ \bibnamefont
  {Kane}},\ and\ \bibinfo {author} {\bibfnamefont {T.~C.}\ \bibnamefont
  {Lubensky}},\ }\bibfield  {title} {\bibinfo {title} {Sliding {Luttinger}
  liquid phases},\ }\href {https://doi.org/10.1103/PhysRevB.64.045120}
  {\bibfield  {journal} {\bibinfo  {journal} {Phys. Rev. B}\ }\textbf {\bibinfo
  {volume} {64}},\ \bibinfo {pages} {045120} (\bibinfo {year}
  {2001})}\BibitemShut {NoStop}%
\bibitem [{\citenamefont {Starykh}\ \emph {et~al.}(2002)\citenamefont
  {Starykh}, \citenamefont {Singh},\ and\ \citenamefont
  {Levine}}]{CSLL_Starykh_PRL2002}%
  \BibitemOpen
  \bibfield  {author} {\bibinfo {author} {\bibfnamefont {O.~A.}\ \bibnamefont
  {Starykh}}, \bibinfo {author} {\bibfnamefont {R.~R.~P.}\ \bibnamefont
  {Singh}},\ and\ \bibinfo {author} {\bibfnamefont {G.~C.}\ \bibnamefont
  {Levine}},\ }\bibfield  {title} {\bibinfo {title} {Spinons in a
  crossed-chains model of a 2d spin liquid},\ }\href
  {https://doi.org/10.1103/PhysRevLett.88.167203} {\bibfield  {journal}
  {\bibinfo  {journal} {Phys. Rev. Lett.}\ }\textbf {\bibinfo {volume} {88}},\
  \bibinfo {pages} {167203} (\bibinfo {year} {2002})}\BibitemShut {NoStop}%
\bibitem [{\citenamefont {Sindzingre}\ \emph {et~al.}(2002)\citenamefont
  {Sindzingre}, \citenamefont {Fouet},\ and\ \citenamefont
  {Lhuillier}}]{CSLL_Sindzingre_PRB2002}%
  \BibitemOpen
  \bibfield  {author} {\bibinfo {author} {\bibfnamefont {P.}~\bibnamefont
  {Sindzingre}}, \bibinfo {author} {\bibfnamefont {J.-B.}\ \bibnamefont
  {Fouet}},\ and\ \bibinfo {author} {\bibfnamefont {C.}~\bibnamefont
  {Lhuillier}},\ }\bibfield  {title} {\bibinfo {title} {One-dimensional
  behavior and sliding {Luttinger} liquid phase in a frustrated
  spin-$\frac{1}{2}$ crossed chain model: Contribution of exact
  diagonalizations},\ }\href {https://doi.org/10.1103/PhysRevB.66.174424}
  {\bibfield  {journal} {\bibinfo  {journal} {Phys. Rev. B}\ }\textbf {\bibinfo
  {volume} {66}},\ \bibinfo {pages} {174424} (\bibinfo {year}
  {2002})}\BibitemShut {NoStop}%
\bibitem [{\citenamefont {Bose}\ \emph {et~al.}(2023)\citenamefont {Bose},
  \citenamefont {Routh}, \citenamefont {Voleti}, \citenamefont {Saha},
  \citenamefont {Kumar}, \citenamefont {Saha-Dasgupta},\ and\ \citenamefont
  {Paramekanti}}]{bose2023proximate}%
  \BibitemOpen
  \bibfield  {author} {\bibinfo {author} {\bibfnamefont {A.}~\bibnamefont
  {Bose}}, \bibinfo {author} {\bibfnamefont {M.}~\bibnamefont {Routh}},
  \bibinfo {author} {\bibfnamefont {S.}~\bibnamefont {Voleti}}, \bibinfo
  {author} {\bibfnamefont {S.~K.}\ \bibnamefont {Saha}}, \bibinfo {author}
  {\bibfnamefont {M.}~\bibnamefont {Kumar}}, \bibinfo {author} {\bibfnamefont
  {T.}~\bibnamefont {Saha-Dasgupta}},\ and\ \bibinfo {author} {\bibfnamefont
  {A.}~\bibnamefont {Paramekanti}},\ }\href@noop {} {\bibinfo {title}
  {Proximate dirac spin liquid in the j1-j3 xxz model for honeycomb
  cobaltates}} (\bibinfo {year} {2023}),\ \Eprint
  {https://arxiv.org/abs/2212.13271} {arXiv:2212.13271 [cond-mat.str-el]}
  \BibitemShut {NoStop}%
\bibitem [{\citenamefont {Jiang}\ \emph {et~al.}(2023)\citenamefont {Jiang},
  \citenamefont {White},\ and\ \citenamefont
  {Chernyshev}}]{j1j3_jiang2023quantum}%
  \BibitemOpen
  \bibfield  {author} {\bibinfo {author} {\bibfnamefont {S.}~\bibnamefont
  {Jiang}}, \bibinfo {author} {\bibfnamefont {S.~R.}\ \bibnamefont {White}},\
  and\ \bibinfo {author} {\bibfnamefont {A.~L.}\ \bibnamefont {Chernyshev}},\
  }\href@noop {} {\bibinfo {title} {Quantum phases in the honeycomb-lattice
  {$J_1$--$J_3$} ferro-antiferromagnetic model}} (\bibinfo {year} {2023}),\
  \Eprint {https://arxiv.org/abs/2304.06062} {arXiv:2304.06062
  [cond-mat.str-el]} \BibitemShut {NoStop}%
\bibitem [{\citenamefont {Watanabe}\ \emph {et~al.}(2022)\citenamefont
  {Watanabe}, \citenamefont {Trebst},\ and\ \citenamefont
  {Hickey}}]{j1j3_watanabe2022frustrated}%
  \BibitemOpen
  \bibfield  {author} {\bibinfo {author} {\bibfnamefont {Y.}~\bibnamefont
  {Watanabe}}, \bibinfo {author} {\bibfnamefont {S.}~\bibnamefont {Trebst}},\
  and\ \bibinfo {author} {\bibfnamefont {C.}~\bibnamefont {Hickey}},\
  }\href@noop {} {\bibinfo {title} {Frustrated ferromagnetism of honeycomb
  cobaltates: Incommensurate spirals, quantum disordered phases, and
  out-of-plane {Ising} order}} (\bibinfo {year} {2022}),\ \Eprint
  {https://arxiv.org/abs/2212.14053} {arXiv:2212.14053 [cond-mat.str-el]}
  \BibitemShut {NoStop}%
\bibitem [{\citenamefont {Becca}\ and\ \citenamefont
  {Sorella}(2017)}]{becca_sorella_2017}%
  \BibitemOpen
  \bibfield  {author} {\bibinfo {author} {\bibfnamefont {F.}~\bibnamefont
  {Becca}}\ and\ \bibinfo {author} {\bibfnamefont {S.}~\bibnamefont
  {Sorella}},\ }\href {https://doi.org/10.1017/9781316417041} {\emph {\bibinfo
  {title} {Quantum Monte Carlo Approaches for Correlated Systems}}}\ (\bibinfo
  {publisher} {Cambridge University Press},\ \bibinfo {year}
  {2017})\BibitemShut {NoStop}%
\bibitem [{\citenamefont {Affleck}\ \emph {et~al.}(1988)\citenamefont
  {Affleck}, \citenamefont {Zou}, \citenamefont {Hsu},\ and\ \citenamefont
  {Anderson}}]{gauge_affleck1988}%
  \BibitemOpen
  \bibfield  {author} {\bibinfo {author} {\bibfnamefont {I.}~\bibnamefont
  {Affleck}}, \bibinfo {author} {\bibfnamefont {Z.}~\bibnamefont {Zou}},
  \bibinfo {author} {\bibfnamefont {T.}~\bibnamefont {Hsu}},\ and\ \bibinfo
  {author} {\bibfnamefont {P.~W.}\ \bibnamefont {Anderson}},\ }\bibfield
  {title} {\bibinfo {title} {{SU(2)} gauge symmetry of the large-{$U$} limit of
  the {Hubbard} model},\ }\href {https://doi.org/10.1103/PhysRevB.38.745}
  {\bibfield  {journal} {\bibinfo  {journal} {Phys. Rev. B}\ }\textbf {\bibinfo
  {volume} {38}},\ \bibinfo {pages} {745} (\bibinfo {year} {1988})}\BibitemShut
  {NoStop}%
\bibitem [{\citenamefont {Hermele}(2007)}]{su2gauge_hermele_prb2007}%
  \BibitemOpen
  \bibfield  {author} {\bibinfo {author} {\bibfnamefont {M.}~\bibnamefont
  {Hermele}},\ }\bibfield  {title} {\bibinfo {title} {{SU(2)} gauge theory of
  the {Hubbard} model and application to the honeycomb lattice},\ }\href
  {https://doi.org/10.1103/PhysRevB.76.035125} {\bibfield  {journal} {\bibinfo
  {journal} {Phys. Rev. B}\ }\textbf {\bibinfo {volume} {76}},\ \bibinfo
  {pages} {035125} (\bibinfo {year} {2007})}\BibitemShut {NoStop}%
\bibitem [{\citenamefont {Karthik}\ and\ \citenamefont
  {Narayanan}(2018)}]{su2gauge_karthik_prd2018}%
  \BibitemOpen
  \bibfield  {author} {\bibinfo {author} {\bibfnamefont {N.}~\bibnamefont
  {Karthik}}\ and\ \bibinfo {author} {\bibfnamefont {R.}~\bibnamefont
  {Narayanan}},\ }\bibfield  {title} {\bibinfo {title} {Scale-invariance and
  scale-breaking in parity-invariant three-dimensional {QCD}},\ }\href
  {https://doi.org/10.1103/PhysRevD.97.054510} {\bibfield  {journal} {\bibinfo
  {journal} {Phys. Rev. D}\ }\textbf {\bibinfo {volume} {97}},\ \bibinfo
  {pages} {054510} (\bibinfo {year} {2018})}\BibitemShut {NoStop}%
\bibitem [{\citenamefont {Thomson}\ and\ \citenamefont
  {Sachdev}(2018)}]{su2gauge_thomson_prx2018}%
  \BibitemOpen
  \bibfield  {author} {\bibinfo {author} {\bibfnamefont {A.}~\bibnamefont
  {Thomson}}\ and\ \bibinfo {author} {\bibfnamefont {S.}~\bibnamefont
  {Sachdev}},\ }\bibfield  {title} {\bibinfo {title} {Fermionic spinon theory
  of square lattice spin liquids near the {N\'eel} state},\ }\href
  {https://doi.org/10.1103/PhysRevX.8.011012} {\bibfield  {journal} {\bibinfo
  {journal} {Phys. Rev. X}\ }\textbf {\bibinfo {volume} {8}},\ \bibinfo {pages}
  {011012} (\bibinfo {year} {2018})}\BibitemShut {NoStop}%
\bibitem [{\citenamefont {Samarakoon}\ \emph {et~al.}(2021)\citenamefont
  {Samarakoon}, \citenamefont {Chen}, \citenamefont {Zhou},\ and\ \citenamefont
  {Garlea}}]{NCTO_samarakoon_PRB2021}%
  \BibitemOpen
  \bibfield  {author} {\bibinfo {author} {\bibfnamefont {A.~M.}\ \bibnamefont
  {Samarakoon}}, \bibinfo {author} {\bibfnamefont {Q.}~\bibnamefont {Chen}},
  \bibinfo {author} {\bibfnamefont {H.}~\bibnamefont {Zhou}},\ and\ \bibinfo
  {author} {\bibfnamefont {V.~O.}\ \bibnamefont {Garlea}},\ }\bibfield  {title}
  {\bibinfo {title} {Static and dynamic magnetic properties of honeycomb
  lattice antiferromagnets {${\mathrm{Na}}_{2}{M}_{2}{\mathrm{TeO}}_{6}$},
  {$M=\mathrm{Co}$ and Ni}},\ }\href
  {https://doi.org/10.1103/PhysRevB.104.184415} {\bibfield  {journal} {\bibinfo
   {journal} {Phys. Rev. B}\ }\textbf {\bibinfo {volume} {104}},\ \bibinfo
  {pages} {184415} (\bibinfo {year} {2021})}\BibitemShut {NoStop}%
\bibitem [{\citenamefont {Xiao}\ \emph {et~al.}(2019)\citenamefont {Xiao},
  \citenamefont {Xia}, \citenamefont {Zhang}, \citenamefont {Yue},
  \citenamefont {Huang}, \citenamefont {Zhang}, \citenamefont {Yang},
  \citenamefont {Song}, \citenamefont {Wei}, \citenamefont {Deng},\ and\
  \citenamefont {Jiang}}]{NCTO_lattice_JiangACS2019}%
  \BibitemOpen
  \bibfield  {author} {\bibinfo {author} {\bibfnamefont {G.}~\bibnamefont
  {Xiao}}, \bibinfo {author} {\bibfnamefont {Z.}~\bibnamefont {Xia}}, \bibinfo
  {author} {\bibfnamefont {W.}~\bibnamefont {Zhang}}, \bibinfo {author}
  {\bibfnamefont {X.}~\bibnamefont {Yue}}, \bibinfo {author} {\bibfnamefont
  {S.}~\bibnamefont {Huang}}, \bibinfo {author} {\bibfnamefont
  {X.}~\bibnamefont {Zhang}}, \bibinfo {author} {\bibfnamefont
  {F.}~\bibnamefont {Yang}}, \bibinfo {author} {\bibfnamefont {Y.}~\bibnamefont
  {Song}}, \bibinfo {author} {\bibfnamefont {M.}~\bibnamefont {Wei}}, \bibinfo
  {author} {\bibfnamefont {H.}~\bibnamefont {Deng}},\ and\ \bibinfo {author}
  {\bibfnamefont {D.}~\bibnamefont {Jiang}},\ }\bibfield  {title} {\bibinfo
  {title} {Crystal growth and the magnetic properties of {Na$_2$Co$_2$TeO$_6$}
  with quasi-two-dimensional honeycomb lattice},\ }\href
  {https://doi.org/10.1021/acs.cgd.8b01770} {\bibfield  {journal} {\bibinfo
  {journal} {Crystal Growth \& Design}\ }\textbf {\bibinfo {volume} {19}},\
  \bibinfo {pages} {2658} (\bibinfo {year} {2019})}\BibitemShut {NoStop}%
\end{thebibliography}%

\end{document}